\documentclass[11pt]{article}

\usepackage[final]{acl}

\usepackage{times}
\usepackage{latexsym}

\usepackage[T1]{fontenc}

\usepackage[utf8]{inputenc}

\usepackage{amsmath,amsfonts,bm}
\usepackage{booktabs}    
\usepackage{amssymb}     
\usepackage{pifont}      
\usepackage{natbib}      
\usepackage{graphicx}    
\usepackage{array}       
\usepackage{caption}     
\usepackage{booktabs}    
\usepackage{amssymb}     
\usepackage{multirow}    
\usepackage{array}       
\usepackage{caption}     
\usepackage{natbib}      
\usepackage{pifont}
\usepackage{fontawesome5}
\usepackage{url}
\usepackage{xurl}
\usepackage{hyperref}
\usepackage{makecell}
\usepackage{tabularx}
\usepackage{tablefootnote}
\usepackage{wasysym}     
\newcommand{\cmark}{\textcolor{red}{\ding{51}}}  
\newcommand{\xmark}{\textcolor{green!60!black}{\ding{55}}} 

\usepackage[table]{xcolor}

\usepackage{ragged2e}
\newcolumntype{Y}{>{\RaggedRight\arraybackslash}X} 
\newcolumntype{C}{>{\Centering\arraybackslash}X}  
 
\definecolor{GrayBG}{gray}{0.95}
\newcommand{\pubheader}[1]{\rowcolor{GrayBG}[0pt][0pt]\multicolumn{2}{@{}l@{}}{\hspace{1em}\bfseries\scshape #1}}
\newcolumntype{L}{>{\raggedright\arraybackslash}X}

\def\eqref#1{equation~\ref{#1}}

\def\1{\bm{1}}

\DeclareMathAlphabet{\mathsfit}{\encodingdefault}{\sfdefault}{m}{sl}
\SetMathAlphabet{\mathsfit}{bold}{\encodingdefault}{\sfdefault}{bx}{n}

\usepackage{microtype}
\usepackage{threeparttable}
\usepackage{hyperref}
\usepackage{url}
\usepackage[skins]{tcolorbox}
\tcbuselibrary{listings,breakable}
\usepackage{pifont}
\usepackage{amsthm}
\usepackage{alltt}
\usepackage{booktabs}
\usepackage{siunitx}
\usepackage{cleveref}
\usepackage{kantlipsum}
\usepackage{setspace}
\usepackage{tablefootnote}

\usepackage{xspace}

\newtheorem{definition}{Definition}

\newtcolorbox{PromptBox}[1][]{
  title=Prompt,
  enhanced,
  breakable,
  colback=blue!5,
  colframe=blue!40,  
  coltitle=black,
  fonttitle=\bfseries,
  listing only,
  listing options={basicstyle=\ttfamily\footnotesize, breaklines=true},
  #1
}

\usepackage{inconsolata}
\usepackage{hyperref}
\usepackage{graphicx}
\usepackage{listings}
\usepackage{xcolor} 
\definecolor{lightergray}{rgb}{0.95, 0.95, 0.95}
\usepackage{tcolorbox}
\usepackage{enumitem}
\usepackage{wrapfig}
\usepackage{longtable}
\definecolor{orange}{RGB}{255,198,140}
\newcommand{\titleicon}{
  \raisebox{-0.2em}{\includegraphics[height=1.1em]{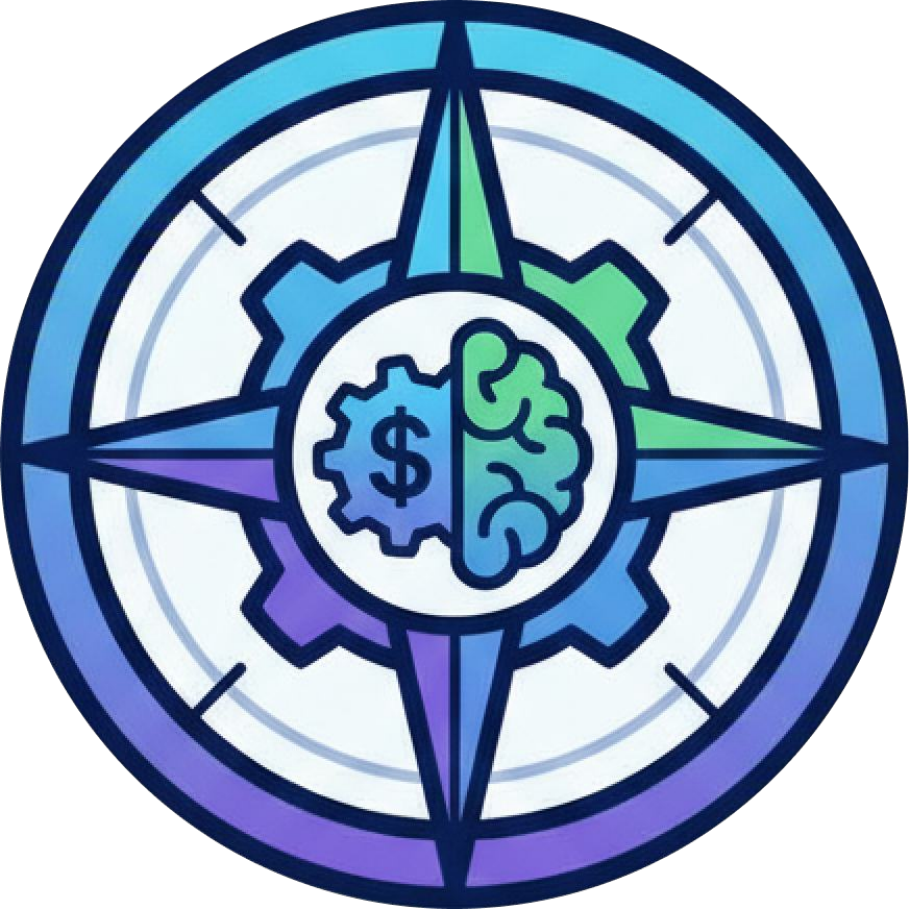}}
}

\newcommand{\eg}{\emph{e.g.}\xspace}

\definecolor{darkgreen}{RGB}{0,100,0}
\newcommand{\gain}[1]{\textcolor{red}{(+#1)}}
\newcommand{\loss}[1]{\textcolor{darkgreen}{(-#1)}}
\usepackage[para]{footmisc}
\title{\titleicon BizCompass: Benchmarking the Reasoning Capabilities of LLMs in Business Knowledge and Applications}

\author{
    \textbf{Jianing Hao}\textsuperscript{1}\thanks{Equal contribution.}, 
    \textbf{Yuhe Wu}\textsuperscript{1}\footnotemark[1],                 
    \textbf{Yuanjian Xu}\textsuperscript{1}\footnotemark[1],                 
    \textbf{Shichang Meng}\textsuperscript{2}\thanks{Work done while the author was a Research Assistant at HKUST-GZ.},
    \\
    \textbf{Shuai Yuan}\textsuperscript{3}\footnotemark[2],                 
    \textbf{Wei Zeng}\textsuperscript{1},
    \textbf{Zixuan Wang}\textsuperscript{1},
    \textbf{Guang Zhang}\textsuperscript{1}\thanks{Correspondence: \href{mailto:guangzhang@hkust-gz.edu.cn}{guangzhang@hkust-gz.edu.cn}} 
    \\
    \\
    \textsuperscript{1}The Hong Kong University of Science and Technology (Guangzhou), Guangzhou, China\\
    \textsuperscript{2}City University of Hong Kong, Hong Kong, China\\
    \textsuperscript{3}Peking University, Beijing, China
}
\begin{document}
\maketitle

\begin{abstract}
Large language models (LLMs) hold great promise for business applications, yet business analysis remains inherently complex, demanding rigorous reasoning and the integration of diverse knowledge sources. Existing benchmarks typically target narrow tasks and thus leave a fundamental question unanswered: \emph{how can LLMs be reliably applied in business, and how are these applications grounded in underlying theoretical capabilities?}  To address this gap, we introduce \textbf{BizCompass}, a benchmark explicitly designed to connect theoretical foundations with practical business knowledge and applications.
At the knowledge level, BizCompass covers four core domains—finance, economics, statistics, and operations management.
At the application level, it structures tasks around three representative roles: the \emph{analyst}, the \emph{trader}, and the \emph{consultant}.
This dual-axis design not only exposes performance differences across realistic scenarios but also diagnoses which foundational capabilities enable or constrain success.
We systematically evaluate both open-source and commercial LLMs, revealing how theoretical knowledge translates into practical performance in business.
The results provide actionable insights for model selection and training optimization in real-world business contexts.
All datasets and evaluation code are publicly released to support reproducibility and future research: \url{https://bizcompass.dev.ypemc.com}.
\end{abstract}

\section{Introduction}

Large language models (LLMs) have rapidly become a focal point of AI research \citep{llm_brown2020language,llm_openai2023gpt4,llm_deepseek2025r1}, and their potential to support business analysis and decision-making is increasingly regarded as a critical direction for exploration \citep{llm_in_bus_zhou2024can,llm_in_bus_chen2024large,llm_in_bus_de2025chatgpt,llm_in_bus_siano2025news}. From financial statement analysis and portfolio optimization to strategic consulting and market forecasting, both academia and industry expect LLMs to serve as a key driver of intelligent enterprise transformation. However, business analysis itself is inherently complex: it requires not only broad interdisciplinary knowledge but also rigorous reasoning and decision-making in uncertain and dynamic environments \citep{inter_eisenhardt1992strategic,inter_vahed2025interdisciplinarity}. For example, when advising a client on entering a new market, an analyst must combine financial assessment, economic analysis, statistical evidence from surveys, and operations management methods to ensure both strategic soundness and operational feasibility.
This complexity distinguishes business reasoning from general reasoning: it requires integrating multi-disciplinary knowledge within strict domain-specific constraints.
For example, trading requires reconciling statistical data with economic rules, while consulting involves tracking how operational changes propagate through a firm.
Existing benchmarks have made significant progress; however, they still fail to address a fundamental question for reliable LLM deployment in business:

\begin{table*}[t]
\centering
\small
\begin{tabular}{l@{}|cccc|ccc}
\toprule
\multirow{2}{*}{\textbf{Benchmark}} & \multicolumn{4}{c}{\textbf{Knowledge Coverage}} & \multicolumn{3}{c}{\textbf{Application Dimensions}} \\
\cmidrule(lr){2-5} \cmidrule(lr){6-8}
 & Finance & Economics & Statistics & OM & Analysis & Trading & Consulting \\
\midrule
FinQA \citep{FinQA_2021} & \cmark & \xmark & \xmark & \xmark & \cmark & \xmark & \xmark \\
CovFinQA \citep{convfinqa_2022} & \cmark & \xmark & \xmark & \xmark & \cmark & \xmark & \xmark \\
FinanceBench \citep{financebench_2023} & \cmark & \xmark & \xmark & \xmark & \cmark & \xmark & \xmark \\
BizBench \citep{bizbench_2024} & \cmark & \cmark & \xmark & \xmark & \cmark & \xmark & \xmark \\
FinBen \citep{Finben_2024} & \cmark & \cmark & \cmark & \xmark & \cmark & \cmark & \xmark \\
FinMaster \citep{finmaster_2025} & \cmark & \cmark & \cmark & \cmark & \cmark & \xmark & \cmark \\
\midrule
\rowcolor{gray!15}
\textbf{BizCompass (ours)} & \textbf{\cmark} & \textbf{\cmark} & \textbf{\cmark} & \textbf{\cmark} & \textbf{\cmark} & \textbf{\cmark} & \textbf{\cmark} \\
\bottomrule
\end{tabular}
\caption{A sample comparison between BizCompass and existing benchmarks.
Red \cmark\ indicates full coverage and green \xmark\ indicates lack of coverage. 
A detailed comparison is shown in Appendix~\ref{app:comparison}.}
\label{tab:dual_axis_comparison}
\end{table*} 

\begin{tcolorbox}[colback=blue!5,colframe=blue!40,
  leftrule=1.2mm, left=3mm]
\faQuestionCircle \quad 
\emph{Which business scenarios can LLMs reliably support, and what foundational capabilities are required to enable their effective deployment?}
\end{tcolorbox}
Despite growing interest in applying LLMs to business decision-making, existing benchmarks remain unable to answer the fundamental question of what LLMs can reliably do in real business contexts and which foundational capabilities support such applications.
While these benchmarks represent meaningful progress beyond generic NLP evaluations, they suffer from two critical limitations. First, they provide only a narrow view of real business scenarios.
For example, FinQA \citep{FinQA_2021} focuses on multi-step numerical reasoning over financial reports, but its scope is largely limited to tabular calculations and does not reflect the diverse tasks that financial institutions prioritize.
While FinanceBench \citep{financebench_2023} moves a step closer to practice by evaluating open-domain QA over SEC filings, its problem formulation remains narrow.
It primarily tests single-turn, isolated questions, which falls short of the complex information synthesis required in real-world professional workflows. 
Second, current benchmarks lack a systematic bridge between theoretical capabilities and practical applications. Some remain overly abstract and detached from real-world practice, while others emphasize narrow applications without clarifying the foundational skills required.
For instance, although FinanceQA \citep{FinanceQA_2025} simulates professional investment workflows, it does not evaluate the underlying foundational knowledge.
Consequently, when mainstream models fail on nearly 60\% of cases, it is impossible to diagnose which specific capability gaps caused the failure, underscoring the absence of a clear mapping from theoretical knowledge to applied performance.
As a result, prior efforts offer fragmented insights rather than a coherent framework, leaving the connection between disciplinary foundations and real business applications unresolved.
An abstract comparison is shown in Table~\ref{tab:dual_axis_comparison} to demonstrate our advantages. The detailed comparison between BizCompass and other past 5-year representative benchmarks is provided in Table~\ref{tab:com_comparison} in Appendix~\ref{app:comparison}.

To fill this critical gap and enable the reliable deployment of LLMs in business, we introduce \textbf{BizCompass}, the first benchmark that systematically bridges disciplinary foundations with realistic business applications. BizCompass offers an integrated evaluation framework tailored to high-stakes business contexts. Its design is guided by two central principles. First, rigorous evaluation must be grounded in \emph{realistic business scenarios}: to this end, we engaged in extensive discussions with over ten leading financial institutions to identify the application settings they regard as most critical. Second, meaningful assessment requires \emph{linking these scenarios back to foundational capabilities}: we therefore constructed tasks that capture both disciplinary knowledge and its applied manifestations.  
Our contributions go substantially beyond prior efforts and can be summarized as follows:
\begin{itemize}[leftmargin=*,labelsep=5pt]
    \item \textbf{Systematic coverage:} BizCompass is the first benchmark to integrate four foundational disciplines—finance, economics, statistics, and operations management—together with role-aware tasks spanning analysis, trading, and consulting. This structured design ensures that both theoretical underpinnings and practical applications are comprehensively represented.  
    \item \textbf{Explanatory evaluation:} BizCompass moves beyond surface-level task accuracy by explicitly linking performance to underlying theoretical capabilities. This enables not only the measurement of task success or failure, but also the diagnosis of which foundational skills drive or hinder performance in real-world business contexts.  
    \item \textbf{Reliable and rigorous data:} All tasks were co-developed with domain experts and subjected to a systematic, multi-stage review process. Coupled with a substantial volume of carefully designed questions, this guarantees realism, statistical robustness, and minimal evaluation bias, thereby establishing a new standard for business-oriented LLM benchmarks.  
\end{itemize}

Through this \textbf{carefully constructed framework}, BizCompass establishes the first systematic mapping between foundational capabilities and applied business scenarios. In doing so, it uncovers critical performance gaps of current LLMs, while providing actionable guidance for model selection, training strategies, and real-world deployment. 

\section{Benchmark Construction}

\begin{figure*}[ht]
    \centering
    \includegraphics[width=0.98\linewidth]{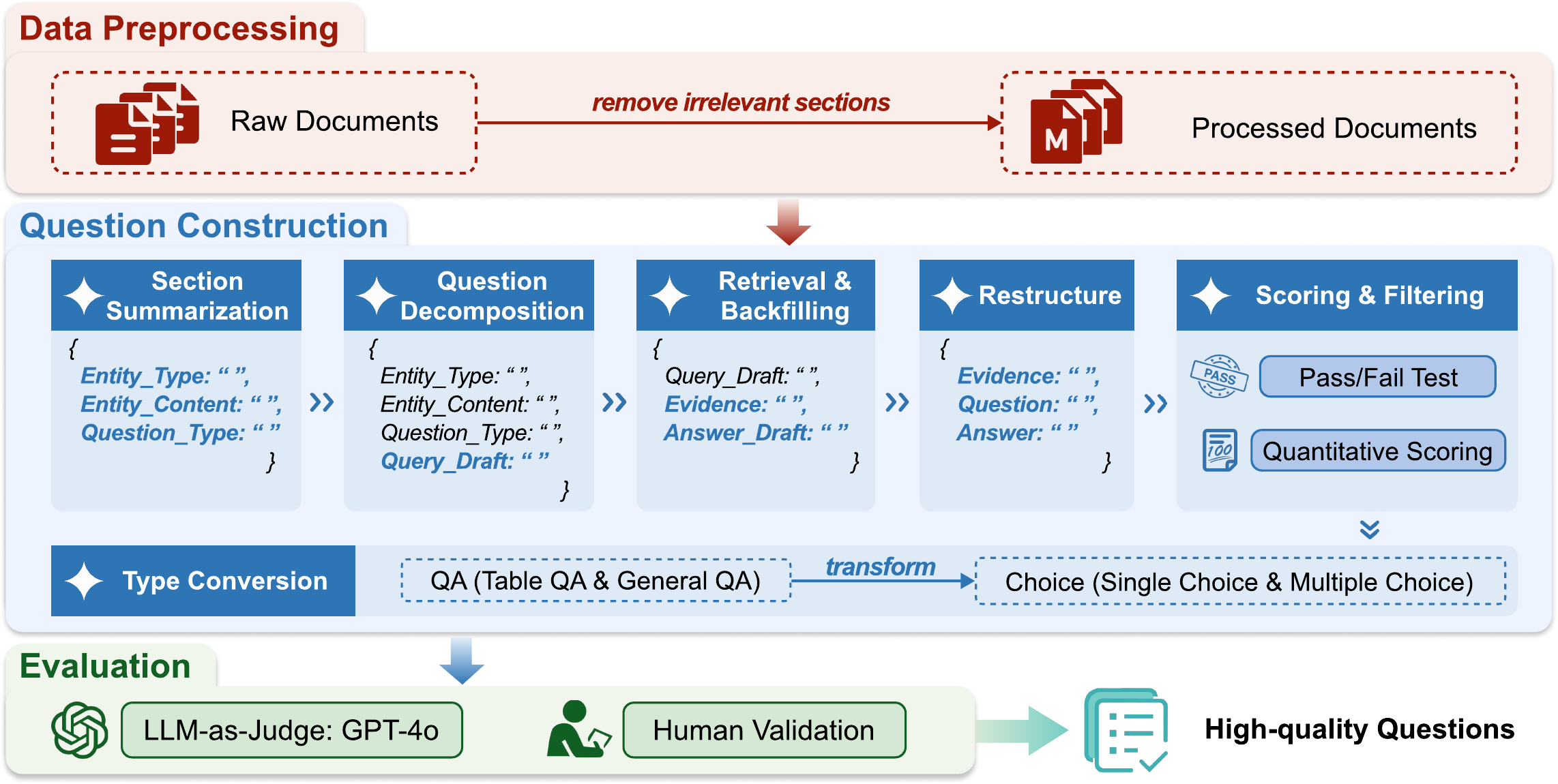}
    \caption{The three-phase pipeline of BizCompass benchmark construction.}
    \label{fig:pipeline}
\end{figure*}

In evaluating domain knowledge for business applications, we focus on Finance, Economics, Statistics, and Operations Management, which jointly provide a coherent analytical framework. Their relevance is supported by both the curricular structures of leading business schools and prior research~\citep{Biehl2006}. To bridge this disciplinary foundation and practical applications, we draw on the O*NET occupational classification system\footnote{National Center for O*NET Development. ONET-SOC Taxonomy. \url{https://www.onetcenter.org/taxonomy.html}} to identify relevant professional categories. We find that analysis, trading, and consulting collectively span over 80\% of business scenarios.
Further details are provided in the Appendix~\ref{sec:sta_ana_of_bus}.

\subsection{Data Sources}

To bridge theory and practice, we curate a high-quality corpus to construct BizCompass and organize it into two categories: one for constructing knowledge-based tasks and the other for designing application-based scenarios.

\paragraph{Data Sources for Knowledge-Based Tasks.} For knowledge-based tasks, we rely on academic journal articles, motivated by two main considerations. First, their peer-reviewed nature makes them reliable sources of knowledge. Second, they combine methodological rigor with both established and emerging theories, providing a solid foundation for testing whether LLMs can function as domain experts. We select 82 representative journals across economics, finance, operations management, and statistics, referencing recommended lists from top universities and prior research~\citep{ham2021new}. The corpus spans 1954–2025, thus covering both classic theories—often only briefly covered in textbooks—and frontier research. In terms of quantity, it comprises 152,077 articles, sufficient to ensure a comprehensive and unbiased evaluation. Further details are provided in Appendix~\ref{app:raw-data}.

\paragraph{Data Sources for Application-Based Tasks.}
Before developing application-based tasks, we systematically reviewed critical business scenarios and existing benchmarks. This analysis shows that current benchmarks remain limited, with coverage restricted to a few tasks, exemplified by CCFraud \citep{feng2023empowering}. To address the lack of benchmarks for other tasks, we established three principles for material selection. Authenticity requires materials to mirror real-world contexts, in line with financial analysis practices of grounding conclusions in public documents \citep{LoughranMcDonald2011}. Utility requires materials to include step-by-step operational details, reflecting the practical need in financial machine learning to turn theory into deployable strategies \citep{LopezDePrado2018}. Governance requires clear source traceability and usage rights, in line with standards such as Datasheets for Datasets \citep{gebru2021datasheets} and Model Cards \citep{Mitchell2019}. Based on the above principles, we primarily rely on two types of source materials:  

\begin{itemize}[leftmargin=*,labelsep=5pt]
    \item \textbf{Practitioner-Oriented Textbooks:} Practitioner-oriented textbooks serve as our primary source for practical knowledge \citep{LopezDePrado2018, Chan2013}. We convert their stepwise workflows, methods, and constraints into multi-step application tasks. For instance, the chapter of the Sharpe ratio—including its definition, formula, and interpretation—is converted into a composite task where a model must calculate the ratio from given asset returns, analyze its implications for risk-adjusted performance, and provide a final investment recommendation.

    \item \textbf{Real-World Business Documents:} Real-world business documents, in turn, provide the raw business context and factual grounding for our evaluation tasks \citep{LoughranMcDonald2011}. These materials include corporate disclosures (10-Ks, 10-Qs)\footnote{The 10-K is an annual report and the 10-Q is a quarterly report filed by public companies with the U.S. Securities and Exchange Commission (SEC). Available via the SEC's EDGAR database: \url{https://www.sec.gov/edgar/searchedgar/companysearch}} for fundamental analysis, third-party structured datasets (Morningstar)\footnote{The morningstar is a global financial services firm that provides extensive financial data and investment research on a wide range of securities: \url{https://www.morningstar.com/}} for quantitative reasoning, and newspaper archives for event-driven analysis.

\end{itemize}

\subsection{Construction Pipeline}
We design a three-phase pipeline comprising data preprocessing, question construction, and evaluation, as illustrated in Figure \ref{fig:pipeline}. In the preprocessing phase, raw PDFs are converted and cleaned into an analysis-ready corpus. The construction phase follows six modular steps: (i) section summarization extracts key knowledge units such as equations, propositions, and tables; (ii) question decomposition generates candidate queries; (iii) retrieval and backfilling ground each query in verbatim textual evidence; (iv) restructuring organizes evidence and queries into coherent question–answer pairs; (v) scoring and filtering ensure consistency and quality; and (vi) type conversion adapts each item into QA or choice format. This stepwise design makes the process auditable and reduces error propagation. The final evaluation phase adopts a dual-track expert review, where specialists assess knowledge-based tasks and practitioners assess application-based ones.
These iterative reviews focused on clarity, representativeness, realism, and difficulty.
We adopted a \textit{``Review-Discuss-Resolve''} protocol to handle disagreements.
Convergence was defined as achieving unanimous consent; if consensus could not be reached after two rounds of discussion, the item was strictly discarded.
Detailed descriptions of each step are provided in Appendix~\ref{app:construct}.

\subsection{BizCompass Statistics}
\begin{figure}[h]
    \centering
    \includegraphics[width=\linewidth]{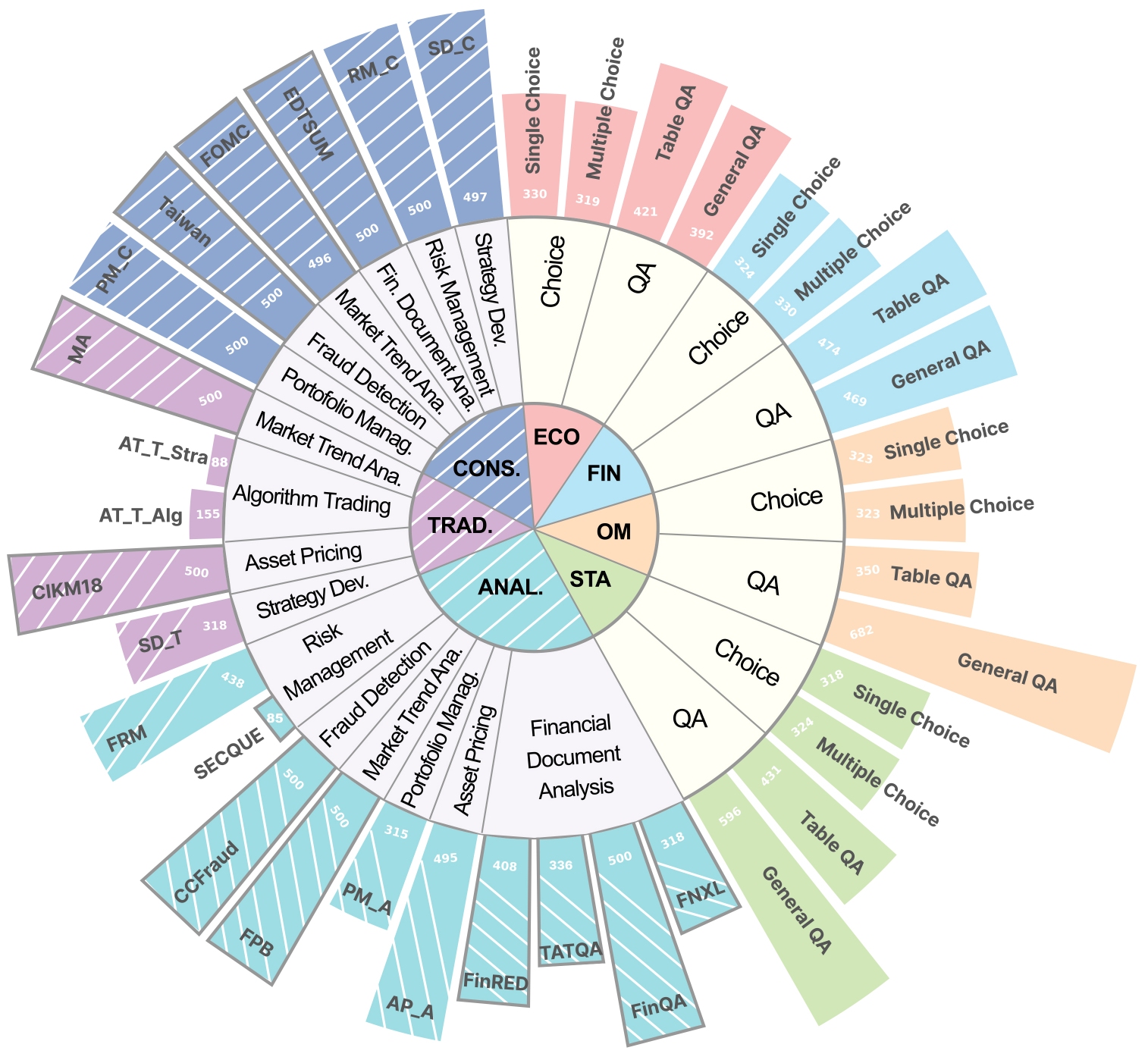}
    \caption{BizCompass's statistics. Bordered bars indicate subsets derived from existing datasets, while non-bordered bars represent newly created original subsets by BizCompass. For consistency, we define the abbreviations as follows:
\textbf{AP} = Asset Pricing,
\textbf{MTA} = Market Trend Analysis,
\textbf{RM} = Risk Management,
\textbf{SD} = Strategy Development,
\textbf{PM} = Portfolio Management,
\textbf{AT} = Algorithmic Trading (with suffixes \textbf{Stra} = Strategy and \textbf{Alg} = Algorithm),
\textbf{FD} = Fraud Detection,
and \textbf{FDA} = Financial Document Analysis.
\textbf{A}, \textbf{T}, and \textbf{C} denote the three major application domains:
Analysis, Trading, and Consulting, respectively.
}
    \vspace{-0.5em}
    \label{fig:bizcompass_stat}
\end{figure}

BizCompass contains a total of 14,855 questions. Among them, 6,406 are knowledge-based questions, and 8,449 are application-based questions. The distribution of the data is illustrated in Figure \ref{fig:bizcompass_stat}. The inner circle of the figure represents four categories of knowledge-based questions and three major application domains, namely Consulting, Trading, and Analysis. The middle and outer layers further expand into specific sub-tasks and question formats, such as question answering, single choice, and multiple choice. 

\begin{figure*}[ht]
    \centering
    \includegraphics[width=0.99\linewidth]{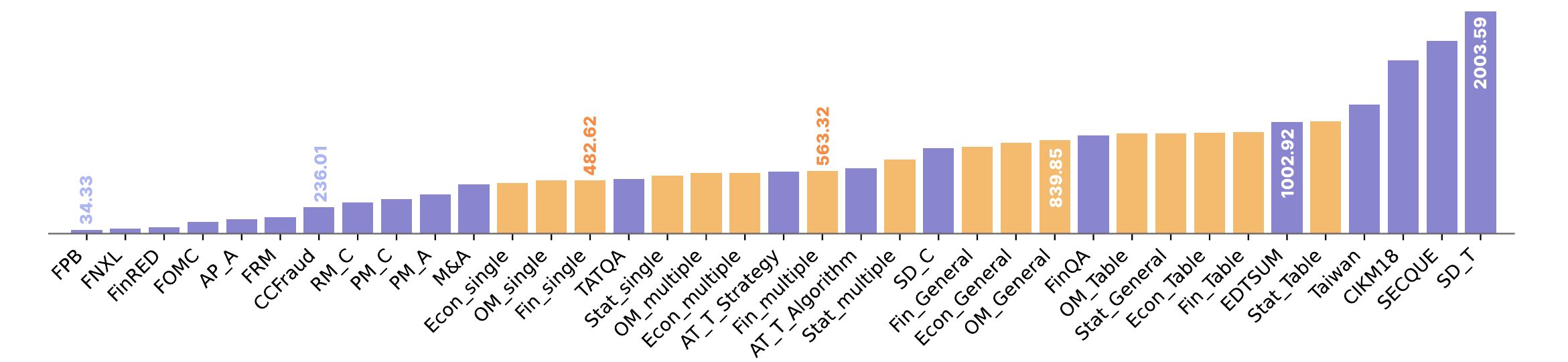}
    \caption{The average token length of each subset in BizCompass.}
    \label{fig:len_distribution}
\end{figure*}

The design of BizCompass carefully considers the varying contextual length requirements across different task types, as shown in Figure~\ref{fig:len_distribution}. Some tasks naturally require long-context analytical capabilities, involving cross-sentence reasoning, information extraction, or document-level comprehension. In contrast, other interactive scenarios rely on shorter contexts with more frequent exchanges rather than lengthy narratives. This balanced design ensures that BizCompass aligns with real-world business communication patterns, thereby enabling more accurate evaluation of LLMs.

\section{Experiments}
BizCompass is designed to support model selection for business applications and to provide guidance on how LLMs can be further optimized.
To meet these objectives, we structure our experiments around three research questions that establish performance baselines, explain underlying causes, and identify pathways for improvement:
\begin{itemize}[leftmargin=*,labelsep=5pt]
    \item \textbf{RQ1 (What):} What is the overall performance of LLMs on BizCompass, and what is the relationship between application scenarios and knowledge domains in shaping performance?  
    \item \textbf{RQ2 (Why):} What factors underlie performance differences, and how are these reflected in the reasoning paths of models when solving complex business problems?  
    \item \textbf{RQ3 (How):} How can appropriate fine-tuning methods be employed to enhance the performance of LLMs in business scenarios?  
\end{itemize}
To provide a solid foundation for investigating these research questions, we begin by outlining the experimental setup, focusing on model selection and fairness control.
\subsection{Evaluation Setup}
\paragraph{Model Selection.} We systematically evaluate 23 representative LLMs under the few-shot setting.
To ensure comprehensive coverage, the evaluated LLM series includes both open-source and proprietary models, such as GPT~\cite{gpt_2024}, Gemini~\cite{gemini_2025}, Llama~\cite{llama_2025}, Claude~\cite{claude_2025}, DeepSeek~\cite{llm_deepseek2025r1}, Qwen~\cite{qwen_2024}, and Grok~\cite{grok_2025}.
Our evaluation also includes distilled models (\eg, `DeepSeek-R1-Distill-Llama-70B'~\cite{deepdistill_2025}), and models with integrated thinking mechanisms for efficient comparisons.
To facilitate a fair and standardized comparison, we consistently employ the official default configurations for models with adjustable reasoning capacities.

\paragraph{Evaluation Metrics.}
We employ a diverse set of evaluation metrics tailored to the specific nature of each task.
\begin{itemize}[leftmargin=*,labelsep=5pt]
    \item \textbf{Accuracy.} For single-choice questions and tasks with label-based outputs, we use Accuracy to measure the proportion of correct predictions, which provides a straightforward assessment of model performance on classification tasks.
    \item \textbf{F1.} For multi-choice questions, we complement Accuracy with F1 score to better capture both precision and recall, providing a more balanced evaluation of model performance on multi-label classification tasks.
    \item \textbf{ROUGE.} For tasks requiring text generation, we employ ROUGE-1, ROUGE-2, and ROUGE-L metrics and compute their average to provide a comprehensive assessment of text similarity~\citep{lin2004rouge}. This averaging approach ensures a more balanced evaluation across different levels of n-gram matching and longest common subsequence matching.
    \item \textbf{GPT-Eval.} Considering the dynamic nature of answers and the need for understanding the logic of answers, we adopt LLM-as-a-Judge~\citep{judging_2023} for evaluation, name the metric GPT-Eval and select GPT-4o as our evaluator. To actively mitigate potential evaluation biases, we design structured, dimension-specific scoring criteria inspired by G-Eval~\citep{Geval_2023} (The prompts are detailed in Appendix~\ref{app:llmeval}). Recent studies, such as JudgeBench~\citep{judgebench_2025}, have validated GPT-4o's strong alignment with human judges, thereby justifying its selection as a reliable evaluator.
\end{itemize}
\paragraph{Fairness control.}
To ensure fair and comparable evaluations across diverse models, we standardize the hyperparameters via a comprehensive grid search on the Llama3-8B model, exploring temperature values in \{0, 0.2, 0.4, 0.6, 0.8\} and top-p values in \{0.6, 0.8, 0.9, 0.95\}.
For multiple-choice questions in knowledge-based tasks, accuracy is maximized at a temperature of 0.8 and a top-p of 0.95.
For general QA benchmarks, where GPT-4o is employed as the evaluator following the protocol in Appendix~\ref{app:llmeval}, the optimal configuration shifts to a temperature of 0.8 paired with a top-p of 0.6. Detailed outcomes of the grid search are provided in Appendix~\ref{app:hyper}, specifically in Table~\ref{tab:hyper}.

\subsection{Evaluation Results}
As shown in Tables~\ref{tab:eva_knowledge} and~\ref{tab:eva_application}, proprietary LLMs (e.g., GPT, Gemini, Claude) outperform open-source models in both domain knowledge and practical application tasks. In knowledge-based evaluation, proprietary models show higher consistency across finance, economics, statistics, and operations management, suggesting broader coverage of professional corpora. In contrast, the performance gap widens in application-based tasks that require contextual reasoning and multi-step decision making. Larger model scales do not necessarily lead to higher accuracy, as DeepSeek-R1 performs worse than some smaller proprietary LLMs, indicating that scaling alone is not sufficient. Likewise, incorporating a chain-of-thought mechanism~\citep{wei2022chain} does not guarantee improvement, implying that reasoning effectiveness depends more on data quality and alignment than on explicit reasoning traces. We can conclude that domain knowledge, contextual reasoning, model scale, and thinking mechanisms contribute differently to business performance, and their effects are not additive.
\begin{table*}[h]
    \centering
    \resizebox{\linewidth}{!}{
    \begin{tabular}{lcccccccccc}
    \toprule
    \multirow{2}{*}{Model} & \multirow{2}{*}{Size} & \multirow{2}{*}{Thinking} & \multicolumn{2}{c}{FIN} & \multicolumn{2}{c}{ECON} & \multicolumn{2}{c}{OM} & \multicolumn{2}{c}{STAT} \\
    \cmidrule(lr){4-5} \cmidrule(lr){6-7} \cmidrule(lr){8-9} \cmidrule(lr){10-11}
        &  &  & Acc. & GPT-Eval & Acc. & GPT-Eval & Acc. & GPT-Eval & Acc. & GPT-Eval \\
    \midrule
    \multicolumn{11}{l}{\emph{\textbf{Proprietary LLMs}}} \\
    GPT & N/D & \cmark & 80.43\% & 4.98\cellcolor{orange} & 83.03\% & \cellcolor{orange}4.98 & 79.26\% & \cellcolor{orange}4.98 & 83.80\% & \cellcolor{orange}4.99 \\
    Gemini & N/D & \cmark & \cellcolor{orange}82.13\% & \cellcolor{orange!50}4.95 & \cellcolor{orange}87.77\% & \cellcolor{orange!50}4.95 & \cellcolor{orange}82.66\% & \cellcolor{orange!50}4.93 & \cellcolor{orange}85.67\% & \cellcolor{orange!50}4.97 \\
    Claude & N/D & \cmark & \cellcolor{orange!50}81.82\% & 4.90 & \cellcolor{orange!50}85.78\% & 4.92 & \cellcolor{orange!50}80.18\% & 4.89 & \cellcolor{orange!50}84.58\% & 4.93 \\
    Grok & N/D  & \cmark & 76.58\% & 4.94 & 83.03\% & 4.93 &  76.63\% & 4.90 & 78.51\% & 4.96 \\
    \midrule
    \multicolumn{11}{l}{\emph{\textbf{Open-source LLMs}}} \\
    DeepSeek & 671B & \cmark & 73.81\% & 4.91 & 81.65\% & 4.91 & 71.05\% & 4.89 & 70.87\% & 4.95\\
    Llama & 70B & \xmark & 52.59\% & 4.66 & 62.84\% & 4.74 & 60.52\% & 4.61 & 57.79\% & 4.71 \\
    Qwen & 235B & \cmark & 78.58\% & 4.86 & 81.65\% & 4.87 & 80.03\% & 4.87 & 82.09\% & 4.91 \\
    \midrule
    \multicolumn{11}{l}{\emph{\textbf{Distilled LLMs}}} \\
    DeepSeek-R1-Distill & 70B & \cmark & 70.11\% & 4.56 & 79.97\% & 4.68 & 72.13\% & 4.54 & 71.96\% & 4.70 \\
    \bottomrule
    \end{tabular}
    }
    \caption{Performance of different models on knowledge-based tasks of BizCompass. For each model series, we report the model with highest average accuracy and GPT-Eval scores. N/D means the parameter size is not disclosed. The complete results of all selected LLMs are presented in Table~\ref{tab:full-knowledge-result} in Appendix~\ref{app:eva_result}.}
    \label{tab:eva_knowledge}
\end{table*}
\begin{table*}[h]
    \centering
    \resizebox{0.9\linewidth}{!}{
    \begin{tabular}{lcccccc}
    \toprule
    \multirow{2}{*}{Model} & \multirow{2}{*}{Size} & \multirow{2}{*}{Thinking} & 8-task Avg. & 3-task Avg. & 4-task Avg. & 6-task Avg. \\
    \cmidrule(lr){4-4} \cmidrule(lr){5-5} \cmidrule(lr){6-6} \cmidrule(lr){7-7}
        &  &  & Accuracy & ROUGE & GPT-Eval (Label) & GPT-Eval (Score) \\
    \midrule
    \multicolumn{7}{l}{\emph{\textbf{Proprietary LLMs}}} \\
    GPT & N/D & \cmark & \cellcolor{orange}79.94\% & 24.30\% & \cellcolor{orange!50}52.35\% & \cellcolor{orange}4.82 \\ 
    Claude & N/D & \cmark & 75.50\% & \cellcolor{orange}30.73\% & \cellcolor{orange}56.86\% & 4.43 \\
    Gemini & N/D & \cmark & 77.43\% & 23.68\% & 50.92\% & \cellcolor{orange!50}4.47 \\
    Grok & N/D  & \cmark & \cellcolor{orange!50}78.40\% & 21.09\% & 50.13\% & 4.25 \\
    \midrule
    \multicolumn{7}{l}{\emph{\textbf{Open-source LLMs}}} \\
    DeepSeek & 671B & \xmark & 71.26\% & 19.58\% & 44.97\% & 4.31 \\
    Llama & 70B & \xmark & 60.24\% & \cellcolor{orange!50}25.15\% & 41.93\% & 4.01 \\
    Qwen & 235B & \cmark & 64.78\% & 14.12\% & 49.56\% & 4.37 \\
    \midrule
    \multicolumn{7}{l}{\emph{\textbf{Distilled LLMs}}} \\
    Deepseek-R1-Distill & 32B & \cmark & 62.30\% & 24.32\% & 45.06\% & 4.11 \\
    \bottomrule
    \end{tabular}
    }
    \caption{Performance of different models on application-based tasks of BizCompass. For each model series, we report the one with the best performance. It is computed as the weighted average of metrics across tasks using the same metric. The complete results of all selected LLMs are presented in Table~\ref{tab:full-application-result} in Appendix~\ref{app:eva_result}.}
    \label{tab:eva_application}
\end{table*}

\section{Discussion}
Our results highlight a relationship between knowledge, reasoning, and application performance. To better understand these findings, our discussion first addresses RQ2 to diagnose the factors driving performance differences. Building on this diagnostic insight, we then explore potential remedies by addressing RQ3.
\subsection{Correlation Analysis}
\begin{figure}
    \centering
    \includegraphics[width=\linewidth]{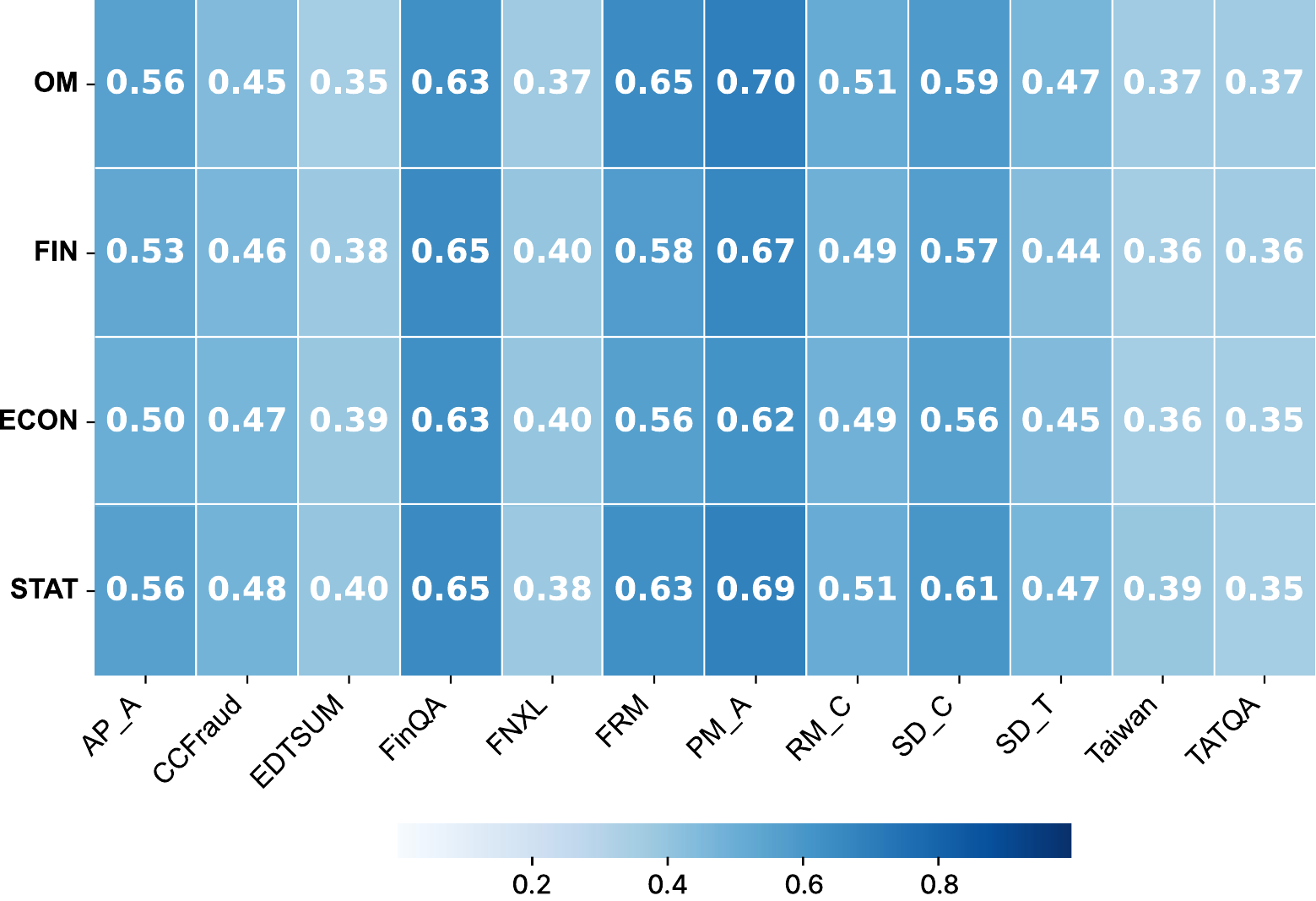}
    \caption{Correlation between application-based tasks and the 4 knowledge domains.}
    \label{fig:heatmap}
\end{figure}

\paragraph{Cross-domain Correlation.}  
As shown in Figure~\ref{fig:heatmap}, most tasks are influenced by domain differences, indicating that domain representations systematically shape task features. Some tasks show stronger domain dependence: FRM and PM\_A are more correlated with OM and STAT, suggesting shared structural patterns between statistical modeling and operational decision-making. In contrast, FIN and ECON exhibit weaker and more uniform correlations, implying looser structural connections. From the task perspective, analytical and quantitative tasks display higher cross-domain consistency, while text and consulting tasks show lower correlations, reflecting their stronger reliance on semantic and contextual information.

\begin{figure*}[ht]
    \centering
    \includegraphics[width=0.99\linewidth]{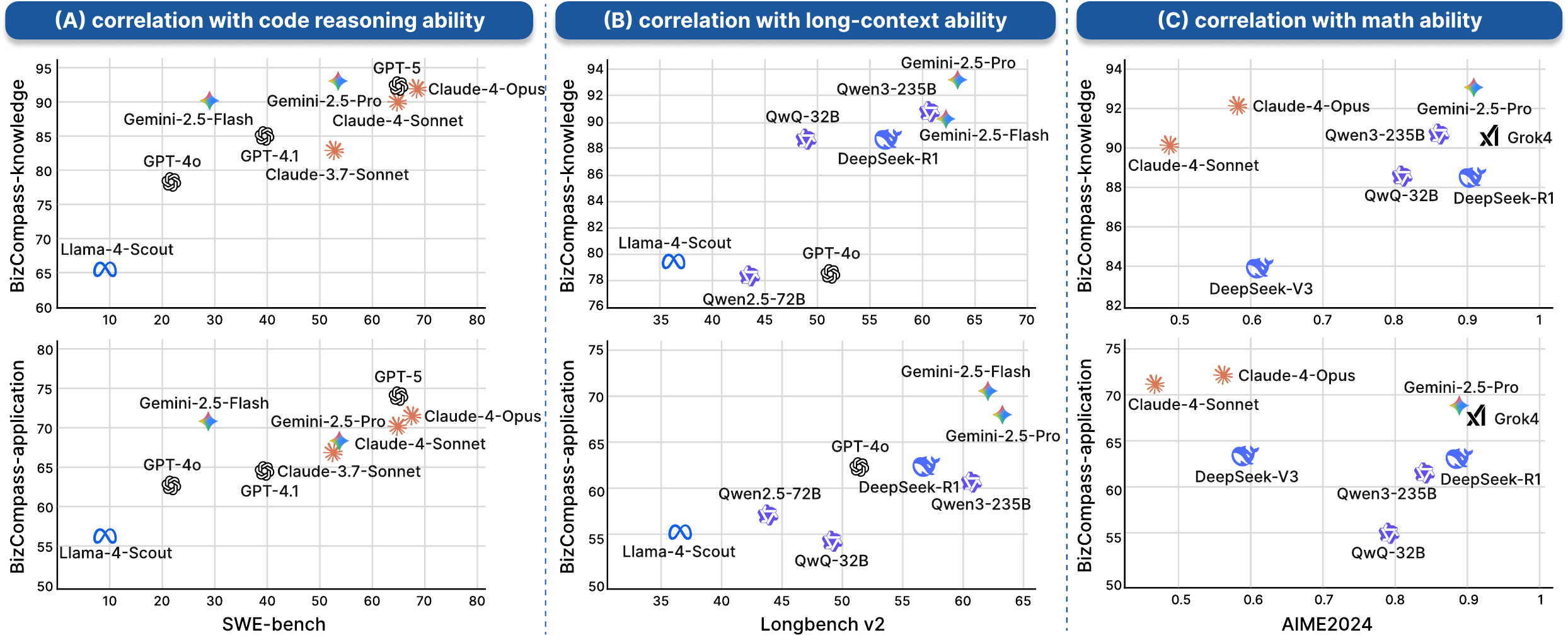}
    \caption{Correlation of model performance on (A) SWE-bench (resolved \%), (B) LongBench v2 (accuracy \%), and (C) AIME2024 (10× accuracy rate \%) with weighted scores from BizCompass knowledge-based and application-based evaluations.}
    \label{fig:corr-knowledge-actual}
\end{figure*}

\paragraph{Related to Code Reasoning Ability.}
We evaluate the relationship between models' ability to solve complex code reasoning problems and their performance on the BizCompass. As shown in Figure~\ref{fig:corr-knowledge-actual} (A), model performance on SWE-bench~\citep{jimenez2024swe} is positively correlated with knowledge-based tasks in BizCompass.
This indicates that the cognitive skills involved in complex code reasoning, such as decompositional thinking, consistency checking, and structured information manipulation, also strengthen a model's capacity for factual recall and conceptual integration.
For application-oriented tasks, the correlation remains positive but becomes less pronounced at higher performance levels.
Once a model's SWE-bench accuracy surpasses approximately 60\%, further improvements in code reasoning contribute little to its downstream application performance. 
This suggests that real-world application ability relies not only on symbolic reasoning but also on contextual understanding and the transfer of reasoning to business settings.

\paragraph{Relation to Long-Context Ability.} 
We further analyze how long context understanding relates to model performance on BizCompass. As shown in Figure~\ref{fig:corr-knowledge-actual} (B), performance on LongBench v2~\citep{bai2024longbench} is positively correlated with both knowledge-based and application-based tasks in BizCompass. Models with stronger long-context abilities tend to achieve higher overall scores, but the relationship varies across task types. For knowledge-based tasks, the correlation is stronger yet segmented across models, while application-based tasks display a smoother and more consistent trend. Models such as Gemini-2.5-Pro and Qwen-3 effectively leverage extended context for cross-segment reasoning and knowledge integration, whereas others with similar LongBench accuracy, such as GPT-4o and DeepSeek-R1, show smaller gains. This suggests that long context capacity alone does not guarantee better knowledge reasoning; performance depends on how well a model utilizes extended context to preserve coherence and perform higher-level inference. In contrast, application tasks rely more on reasoning-to-action transfer, where longer context provides modest but more uniform benefits.

\paragraph{Relation to Math Ability.} 
Figure~\ref{fig:corr-knowledge-actual} (C) shows the relationship between model performance on BizCompass-application tasks and their AIME2024 math scores.
The correlation appears weak. Models with stronger mathematical reasoning, such as Gemini-2.5-Pro and DeepSeek-R1, achieve higher scores on business-oriented tasks, but their advantages are limited. While basic mathematical competence provides a foundation for structured reasoning, further improvements in business performance depend on broader capabilities such as combining diverse information, identifying implicit objectives, and handling incomplete or changing contexts. These abilities rely more on pragmatic understanding and adaptive alignment than on numerical skills. Therefore, mathematical and business reasoning represent distinct yet partially overlapping cognitive capacities.

\begin{table}[ht]
\centering
\resizebox{\linewidth}{!}{
\begin{tabular}{lccc}
\toprule
\multirow{2}{*}{Question types} & \multicolumn{3}{c}{Model} \\
\cmidrule(lr){2-4}
& Llama3-8B & SFT\_1\tablefootnote{The model is fine-tuned on the \href{https://huggingface.co/datasets/sujet-ai/Sujet-Finance-Instruct-177k}{sujet-ai/Sujet-Finance-Instruct-177k} dataset.}
& SFT\_2\tablefootnote{The model is fine-tuned on the \href{https://huggingface.co/datasets/Josephgflowers/Finance-Instruct-500k}{Josephgflowers/Finance-Instruct-500k} dataset.} \\
\midrule
SC. (\%) & 38.89 & 51.23 \gain{12.34} & 37.96 \loss{0.93} \\
MC. (\%) & 14.55 & 26.97 \gain{12.42} & 11.21 \loss{3.34} \\
TableQA & 3.24 & 3.28 \gain{0.04} & 3.11 \loss{0.13} \\
GeneralQA & 2.77 & 2.79 \gain{0.02} & 2.56 \loss{0.21} \\
\bottomrule
\end{tabular}}
\caption{Performance comparison of 2 SFT models against the Llama3-8B baseline across different question types in Finance domain. \textcolor{red}{Red} indicates improvement, and \textcolor{darkgreen}{green} indicates decrease. SC. and MC. are evaluated by accuracy (\%), while TableQA and GeneralQA are evaluated by overall scores.} \label{tab:sft_india_finance_comparison}
\label{tab:sft_india_finance_comparison_vertical}
\let\gain\undefined
\let\loss\undefined
\end{table}

\subsection{Domain-specific SFT}
In this section, we address RQ3, which investigates whether domain-specific fine-tuning can effectively enhance the reasoning ability of LLMs, using financial tasks as a representative case.
To ensure our evaluation reflects practical deployment constraints, inference experiments for the fine-tuned models were conducted on $4\times$NVIDIA A40 GPUs.
As shown in Table~\ref{tab:sft_india_finance_comparison}, targeted knowledge injection through supervised fine-tuning improves model performance on simple knowledge-based financial questions.
The SFT model trained on the Sujet-Finance-Instruct dataset achieves substantial gains in single-choice and multi-choice tasks, indicating that domain-specific supervision effectively enhances factual recall and pattern recognition.
However, its improvement on more complex tasks such as TableQA and GeneralQA is marginal, suggesting that fine-tuning primarily benefits surface-level knowledge acquisition rather than deeper reasoning.
These findings imply that while domain instruction tuning strengthens localized knowledge representation, it contributes little to the model's ability to perform compositional or context-dependent reasoning.
Furthermore, to provide practical guidance for real-world deployment, we systematically profiled the computational overhead of the fine-tuned model; detailed performance metrics are provided in Appendix~\ref{app:sft}.

\section{Conclusion}
We present BizCompass, a benchmark that unifies domain knowledge and real-world business reasoning to systematically evaluate LLMs. Extensive experiments reveal that while proprietary models outperform open-source ones, both suffer from reasoning bottlenecks in compositional, multi-step, and cross-domain tasks. Domain-specific fine-tuning improves factual accuracy but offers limited gains in contextual reasoning, highlighting the need to integrate structured knowledge with adaptive reasoning alignment. BizCompass establishes a foundation for future research on model diagnosis, reasoning enhancement, and domain-adaptive alignment in business-critical contexts.

\section{Limitations}
BizCompass currently focuses on text-based reasoning. Future work will extend it toward multimodal business understanding, as many real-world tasks require interpreting images, charts, and other visual materials alongside text. Incorporating these modalities, together with interactive and decision-oriented evaluations, will enable a more comprehensive assessment of business intelligence in large language models.

\bibliography{custom}

@article{llm_deepseek2025r1,
  title={{DeepSeek-R1} incentivizes reasoning in LLMs through reinforcement learning},
  author={{DeepSeek-AI} and Daya Guo and Dejian Yang and Haowei Zhang and Jun-Mei Song and Ruoyu Zhang and Runxin Xu and Qihao Zhu and Shirong Ma and Peiyi Wang and Xiaoling Bi and Xiaokang Zhang and Xingkai Yu and Yu Wu and Z. F. Wu and Zhibin Gou and Zhihong Shao and Zhuoshu Li and Ziyi Gao and Aixin Liu and Bing Xue and Bing-Li Wang and Bochao Wu and Bei Feng and Chengda Lu and Chenggang Zhao and Chengqi Deng and Chenyu Zhang and Chong Ruan and Damai Dai and Deli Chen and Dong-Li Ji and Erhang Li and Fangyun Lin and Fucong Dai and Fuli Luo and Guangbo Hao and Guanting Chen and Guowei Li and H. Zhang and Han Bao and Hanwei Xu and Haocheng Wang and Honghui Ding and Huajian Xin and Huazuo Gao and Hui Qu and Hui Li and Jianzhong Guo and Jiashi Li and Jiawei Wang and JingChang Chen and Jingyang Yuan and Junjie Qiu and Junlong Li and Jiong Cai and Jiaqi Ni and Jian Liang and Jin Chen and Kai Dong and Kai Hu and Kaige Gao and Kang Guan and Kexin Huang and Kuai Yu and Lean Wang and Lecong Zhang and Liang Zhao and Litong Wang and Liyue Zhang and Lei Xu and Leyi Xia and Mingchuan Zhang and Minghua Zhang and M. Tang and Meng Li and Miaojun Wang and Mingming Li and Ning Tian and Panpan Huang and Peng Zhang and Qiancheng Wang and Qinyu Chen and Qiushi Du and Ruiqi Ge and Ruisong Zhang and Ruizhe Pan and Runji Wang and R. J. Chen and Ruiqi Jin and Ruyi Chen and Shanghao Lu and Shangyan Zhou and Shanhuang Chen and Shengfeng Ye and Shiyu Wang and Shuiping Yu and Shunfeng Zhou and Shuting Pan and S. S. Li and Shuang Zhou and Shao-Kang Wu and Tao Yun and Tian Pei and Tianyu Sun and T. Wang and Wangding Zeng and Wanjia Zhao and Wen Liu and Wenfeng Liang and Wenjun Gao and Wen-Xia Yu and Wentao Zhang and Wangding Xiao and Wei An and Xiaodong Liu and Xiaohan Wang and Xiaokang Chen and Xiaotao Nie and Xin Cheng and Xin Liu and Xin Xie and Xingchao Liu and Xinyu Yang and Xinyuan Li and Xuecheng Su and Xuheng Lin and X. Q. Li and Xiangyu Jin and Xi-Cheng Shen and Xiaosha Chen and Xiaowen Sun and Xiaoxiang Wang and Xinnan Song and Xinyi Zhou and Xianzu Wang and Xinxia Shan and Y. K. Li and Y. Q. Wang and Y. X. Wei and Yang Zhang and Yanhong Xu and Yao Li and Yao Zhao and Yaofeng Sun and Yaohui Wang and Yi Yu and Yichao Zhang and Yifan Shi and Yi Xiong and Ying He and Yishi Piao and Yisong Wang and Yixuan Tan and Yiyang Ma and Yiyuan Liu and Yongqiang Guo and Yuan Ou and Yuduan Wang and Yue Gong and Yu-Jing Zou and Yujia He and Yunfan Xiong and Yu-Wei Luo and Yu-mei You and Yuxuan Liu and Yuyang Zhou and Y. X. Zhu and Yanping Huang and Yao Li and Yi Zheng and Yuchen Zhu and Yunxiang Ma and Ying Tang and Yukun Zha and Yuting Yan and Zehui Ren and Zehui Ren and Zhangli Sha and Zhe Fu and Zhean Xu and Zhenda Xie and Zhen-guo Zhang and Zhewen Hao and Zhicheng Ma and Zhigang Yan and Zhiyu Wu and Zihui Gu and Zijia Zhu and Zijun Liu and Zi-An Li and Ziwei Xie and Ziyang Song and Zizheng Pan and Zhen Huang and Zhipeng Xu and Zhongyu Zhang and Zhen Zhang},
  journal={Nature},
  year={2025},
  volume={645},
  pages={633 - 638},
  url={https://api.semanticscholar.org/CorpusID:275789950},
  doi={10.1038/s41586-025-09422-z}
}

@inproceedings{llm_brown2020language,
 author = {Brown, Tom and Mann, Benjamin and Ryder, Nick and Subbiah, Melanie and Kaplan, Jared D and Dhariwal, Prafulla and Neelakantan, Arvind and Shyam, Pranav and Sastry, Girish and Askell, Amanda and Agarwal, Sandhini and Herbert-Voss, Ariel and Krueger, Gretchen and Henighan, Tom and Child, Rewon and Ramesh, Aditya and Ziegler, Daniel and Wu, Jeffrey and Winter, Clemens and Hesse, Chris and Chen, Mark and Sigler, Eric and Litwin, Mateusz and Gray, Scott and Chess, Benjamin and Clark, Jack and Berner, Christopher and McCandlish, Sam and Radford, Alec and Sutskever, Ilya and Amodei, Dario},
 booktitle = {Advances in Neural Information Processing Systems},
 pages = {1877--1901},
 title = {Language Models are Few-Shot Learners},
 url = {https://proceedings.neurips.cc/paper_files/paper/2020/file/1457c0d6bfcb4967418bfb8ac142f64a-Paper.pdf},
 volume = {33},
 year = {2020},
 doi={10.5555/3495724.3495883}
}

@article{llm_openai2023gpt4,
      title={{GPT-4 Technical Report}}, 
      author={{OpenAI} and Josh Achiam and Steven Adler and Sandhini Agarwal and Lama Ahmad and Ilge Akkaya and Florencia Leoni Aleman and Diogo Almeida and Janko Altenschmidt and Sam Altman and Shyamal Anadkat and Red Avila and Igor Babuschkin and Suchir Balaji and Valerie Balcom and Paul Baltescu and Haiming Bao and Mohammad Bavarian and Jeff Belgum and Irwan Bello and Jake Berdine and Gabriel Bernadett-Shapiro and Christopher Berner and Lenny Bogdonoff and Oleg Boiko and Madelaine Boyd and Anna-Luisa Brakman and Greg Brockman and Tim Brooks and Miles Brundage and Kevin Button and Trevor Cai and Rosie Campbell and Andrew Cann and Brittany Carey and Chelsea Carlson and Rory Carmichael and Brooke Chan and Che Chang and Fotis Chantzis and Derek Chen and Sully Chen and Ruby Chen and Jason Chen and Mark Chen and Ben Chess and Chester Cho and Casey Chu and Hyung Won Chung and Dave Cummings and Jeremiah Currier and Yunxing Dai and Cory Decareaux and Thomas Degry and Noah Deutsch and Damien Deville and Arka Dhar and David Dohan and Steve Dowling and Sheila Dunning and Adrien Ecoffet and Atty Eleti and Tyna Eloundou and David Farhi and Liam Fedus and Niko Felix and Simón Posada Fishman and Juston Forte and Isabella Fulford and Leo Gao and Elie Georges and Christian Gibson and Vik Goel and Tarun Gogineni and Gabriel Goh and Rapha Gontijo-Lopes and Jonathan Gordon and Morgan Grafstein and Scott Gray and Ryan Greene and Joshua Gross and Shixiang Shane Gu and Yufei Guo and Chris Hallacy and Jesse Han and Jeff Harris and Yuchen He and Mike Heaton and Johannes Heidecke and Chris Hesse and Alan Hickey and Wade Hickey and Peter Hoeschele and Brandon Houghton and Kenny Hsu and Shengli Hu and Xin Hu and Joost Huizinga and Shantanu Jain and Shawn Jain and Joanne Jang and Angela Jiang and Roger Jiang and Haozhun Jin and Denny Jin and Shino Jomoto and Billie Jonn and Heewoo Jun and Tomer Kaftan and Łukasz Kaiser and Ali Kamali and Ingmar Kanitscheider and Nitish Shirish Keskar and Tabarak Khan and Logan Kilpatrick and Jong Wook Kim and Christina Kim and Yongjik Kim and Jan Hendrik Kirchner and Jamie Kiros and Matt Knight and Daniel Kokotajlo and Łukasz Kondraciuk and Andrew Kondrich and Aris Konstantinidis and Kyle Kosic and Gretchen Krueger and Vishal Kuo and Michael Lampe and Ikai Lan and Teddy Lee and Jan Leike and Jade Leung and Daniel Levy and Chak Ming Li and Rachel Lim and Molly Lin and Stephanie Lin and Mateusz Litwin and Theresa Lopez and Ryan Lowe and Patricia Lue and Anna Makanju and Kim Malfacini and Sam Manning and Todor Markov and Yaniv Markovski and Bianca Martin and Katie Mayer and Andrew Mayne and Bob McGrew and Scott Mayer McKinney and Christine McLeavey and Paul McMillan and Jake McNeil and David Medina and Aalok Mehta and Jacob Menick and Luke Metz and Andrey Mishchenko and Pamela Mishkin and Vinnie Monaco and Evan Morikawa and Daniel Mossing and Tong Mu and Mira Murati and Oleg Murk and David Mély and Ashvin Nair and Reiichiro Nakano and Rajeev Nayak and Arvind Neelakantan and Richard Ngo and Hyeonwoo Noh and Long Ouyang and Cullen O'Keefe and Jakub Pachocki and Alex Paino and Joe Palermo and Ashley Pantuliano and Giambattista Parascandolo and Joel Parish and Emy Parparita and Alex Passos and Mikhail Pavlov and Andrew Peng and Adam Perelman and Filipe de Avila Belbute Peres and Michael Petrov and Henrique Ponde de Oliveira Pinto and Michael and Pokorny and Michelle Pokrass and Vitchyr H. Pong and Tolly Powell and Alethea Power and Boris Power and Elizabeth Proehl and Raul Puri and Alec Radford and Jack Rae and Aditya Ramesh and Cameron Raymond and Francis Real and Kendra Rimbach and Carl Ross and Bob Rotsted and Henri Roussez and Nick Ryder and Mario Saltarelli and Ted Sanders and Shibani Santurkar and Girish Sastry and Heather Schmidt and David Schnurr and John Schulman and Daniel Selsam and Kyla Sheppard and Toki Sherbakov and Jessica Shieh and Sarah Shoker and Pranav Shyam and Szymon Sidor and Eric Sigler and Maddie Simens and Jordan Sitkin and Katarina Slama and Ian Sohl and Benjamin Sokolowsky and Yang Song and Natalie Staudacher and Felipe Petroski Such and Natalie Summers and Ilya Sutskever and Jie Tang and Nikolas Tezak and Madeleine B. Thompson and Phil Tillet and Amin Tootoonchian and Elizabeth Tseng and Preston Tuggle and Nick Turley and Jerry Tworek and Juan Felipe Cerón Uribe and Andrea Vallone and Arun Vijayvergiya and Chelsea Voss and Carroll Wainwright and Justin Jay Wang and Alvin Wang and Ben Wang and Jonathan Ward and Jason Wei and CJ Weinmann and Akila Welihinda and Peter Welinder and Jiayi Weng and Lilian Weng and Matt Wiethoff and Dave Willner and Clemens Winter and Samuel Wolrich and Hannah Wong and Lauren Workman and Sherwin Wu and Jeff Wu and Michael Wu and Kai Xiao and Tao Xu and Sarah Yoo and Kevin Yu and Qiming Yuan and Wojciech Zaremba and Rowan Zellers and Chong Zhang and Marvin Zhang and Shengjia Zhao and Tianhao Zheng and Juntang Zhuang and William Zhuk and Barret Zoph},
      year={2024},
      eprint={2303.08774},
      archivePrefix={arXiv},
      primaryClass={cs.CL},
      url={https://arxiv.org/abs/2303.08774},
      doi={10.48550/arXiv.2303.08774}
}

@article{llm_in_bus_siano2025news,
  title={{The News in Earnings Announcement Disclosures: Capturing Word Context Using LLM Methods}},
  author={Siano, Federico},
  journal={Management Science},
  year={2025},
  publisher={INFORMS},
  doi= {10.1287/mnsc.2024.05417}
}

@article{llm_in_bus_de2025chatgpt,
  title={{ChatGPT for Textual Analysis? How to Use Generative LLMs in Accounting Research}},
  author={de Kok, Ties},
  journal={Management Science},
  year={2025},
  publisher={INFORMS},
  doi={10.1287/mnsc.2023.03253}
}

@article{llm_in_bus_chen2024large,
  title={{Large Language Model in Creative Work: The Role of Collaboration Modality and User Expertise}},
  author={Chen, Zenan and Chan, Jason},
  journal={Management Science},
  volume={70},
  number={12},
  pages={9101--9117},
  year={2024},
  publisher={INFORMS},
  doi={10.1287/mnsc.2023.03014}
}

@article{llm_in_bus_zhou2024can,
  title={{Can ChatGPT Perform a Grounded Theory Approach to Do Risk Analysis? An Empirical Study}},
  author={Zhou, Yaxian and Yuan, Yufei and Huang, Kai and Hu, Xiangpei},
  journal={Journal of Management Information Systems},
  volume={41},
  number={4},
  pages={982--1015},
  year={2024},
  publisher={Taylor \& Francis},
  doi={10.1080/07421222.2024.2415772}
}

@article{inter_eisenhardt1992strategic,
  title={Strategic decision making},
  author={Eisenhardt, Kathleen M and Zbaracki, Mark J},
  journal={Strategic management journal},
  volume={13},
  number={S2},
  pages={17--37},
  year={1992},
  publisher={Wiley Online Library},
  doi={10.1002/smj.4250130904}
}

@article{inter_vahed2025interdisciplinarity,
  title={Interdisciplinarity opens new frontiers for decision science: Interdisciplinary research},
  author={Vahed, Sarah},
  journal={Nature Reviews Psychology},
  volume={4},
  number={8},
  pages={505--505},
  year={2025},
  publisher={Nature Publishing Group US New York},
  doi={10.1038/s44159-025-00468-4}
}

@article{FinQA_2021,
  title={{FinQA: A Dataset of Numerical Reasoning over Financial Data}},
  author={Chen, Zhiyu and Chen, Wenhu and Smiley, Charese and Shah, Sameena and Borova, Iana and Langdon, Dylan and Moussa, Reema and Beane, Matt and Huang, Ting-Hao and Routledge, Bryan and Wang, William Yang},
  journal={Proceedings of EMNLP 2021},
  year={2021},
  doi={10.18653/v1/2021.emnlp-main.300}
}

@article{financebench_2023,
      title={{FinanceBench: A New Benchmark for Financial Question Answering}}, 
      author={Pranab Islam and Anand Kannappan and Douwe Kiela and Rebecca Qian and Nino Scherrer and Bertie Vidgen},
      year={2023},
      journal = {arXiv preprint arXiv:2311.11944},
      primaryClass={cs.CL},
      doi={10.48550/arXiv.2311.11944}
}

@article{FinanceQA_2025,
  title={{FinanceQA: A Benchmark for Evaluating Financial Analysis Capabilities of Large Language Models}},
  author={Mateega, Spencer and Georgescu, Carlos and Tang, Danny},
  journal={arXiv preprint arXiv:2501.18062},
  year={2025},
  doi={10.48550/arXiv.2501.18062}
}

@inproceedings{XFinbench_2025,
  title={{XFinBench: Benchmarking LLMs in Complex Financial Problem Solving and Reasoning}},
  author={Zhang, Zhihan and Cao, Yixin and Liao, Lizi},
  booktitle={Findings of the Association for Computational Linguistics: ACL 2025},
  pages={8715--8758},
  year={2025},
  doi={10.18653/v1/2025.findings-acl.457}
}

@article{fin_agent_bench_2025,
  title={{Finance Agent Benchmark: Benchmarking LLMs on Real-world Financial Research Tasks}},
  author={Bigeard, Antoine and Nashold, Langston and Krishnan, Rayan and Wu, Shirley},
  journal={arXiv preprint arXiv:2508.00828},
  year={2025},
  doi={10.48550/arXiv.2508.00828}
}

@article{FinMMR_2025,
  title={{FinMMR: Make Financial Numerical Reasoning More Multimodal, Comprehensive, and Challenging}},
  author={Tang, Zichen and Liu, Jiacheng and Yang, Zhongjun and Li, Rongjin and Rong, Zihua and He, Haoyang and Hao, Zhuodi and Hu, Xinyang and Ji, Kun and Ma, Ziyan and others},
  journal={arXiv preprint arXiv:2508.04625},
  year={2025},
  doi={10.48550/arXiv.2508.04625}
}

@inproceedings{bizbench_2024,
  title={{BizBench: A Quantitative Reasoning Benchmark for Business and Finance}},
  author={Krumdick, Michael and Koncel-Kedziorski, Rik and Lai, Viet Dac and Reddy, Varshini and Lovering, Charles and Tanner, Chris},
  booktitle={Proceedings of the 62nd Annual Meeting of the Association for Computational Linguistics (Volume 1: Long Papers)},
  pages={8309--8332},
  year={2024},
  doi={10.18653/v1/2024.acl-long.452}
}

@inproceedings{Finben_2024,
  title={{FinBen: A Holistic Financial Benchmark for Large Language Models}},
  author={Xie, Qianqian and Han, Weiguang and Chen, Zhengyu and Xiang, Ruoyu and Zhang, Xiao and He, Yueru and Xiao, Mengxi and Li, Dong and Dai, Yongfu and Feng, Duanyu and others},
  booktitle={Advances in Neural Information Processing Systems},
  volume={37},
  pages={95716--95743},
  year={2024},
  doi = {10.52202/079017-3033}
}

@inproceedings{cfinbench_2025,
  title={{CF}in{B}ench: A Comprehensive {C}hinese Financial Benchmark for Large Language Models},
  author={Nie, Ying and Yan, Binwei and Guo, Tianyu and Liu, Hao and Wang, Haoyu and He, Wei and Zheng, Binfan and Wang, Weihao and Li, Qiang and Sun, Weijian and others},
  booktitle={Proceedings of the 2025 Conference of the Nations of the Americas Chapter of the Association for Computational Linguistics: Human Language Technologies (Volume 1: Long Papers)},
  pages={876--891},
  year={2025},
  doi={10.18653/v1/2025.naacl-long.4}
}

@inproceedings{fineval_2025,
  title={{F}in{E}val: A {C}hinese Financial Domain Knowledge Evaluation Benchmark for Large Language Models},
  author={Guo, Xin and Xia, Haotian and Liu, Zhaowei and Cao, Hanyang and Yang, Zhi and Liu, Zhiqiang and Wang, Sizhe and Niu, Jinyi and Wang, Chuqi and Wang, Yanhui and others},
  booktitle={Proceedings of the 2025 Conference of the Nations of the Americas Chapter of the Association for Computational Linguistics: Human Language Technologies (Volume 1: Long Papers)},
  pages={6258--6292},
  year={2025},
  doi = {10.18653/v1/2025.naacl-long.318}
}

@inproceedings{financemath_2024,
  title={{FinanceMATH: Knowledge-Intensive Math Reasoning in Finance Domains}},
  author={Zhao, Yilun and Liu, Hongjun and Long, Yitao and Zhang, Rui and Zhao, Chen and Cohan, Arman},
  booktitle={Proceedings of the 62nd Annual Meeting of the Association for Computational Linguistics (Volume 1: Long Papers)},
  pages={12841--12858},
  year={2024},
  doi={10.18653/v1/2024.acl-long.693}
}

@article{bbtfin_2023,
  title={{BBT-Fin: Comprehensive Construction of Chinese Financial Domain Pre-trained Language Model, Corpus and Benchmark}},
  author={Lu, Dakuan and Wu, Hengkui and Liang, Jiaqing and Xu, Yipei and He, Qianyu and Geng, Yipeng and Han, Mengkun and Xin, Yingsi and Xiao, Yanghua},
  journal={arXiv preprint arXiv:2302.09432},
  year={2023},
  doi={10.48550/arXiv.2302.09432}
}

@article{finmaster_2025,
  title={{FinMaster: A Holistic Benchmark for Mastering Full-Pipeline Financial Workflows with LLMs}},
  author={Jiang, Junzhe and Yang, Chang and Cui, Aixin and Jin, Sihan and Wang, Ruiyu and Li, Bo and Huang, Xiao and Sun, Dongning and Wang, Xinrun},
  journal={arXiv preprint arXiv:2505.13533},
  year={2025},
  doi={10.48550/arXiv.2505.13533}
}

@inproceedings{pixiu_2023,
  title={{PIXIU: A Comprehensive Benchmark, Instruction Dataset and Large Language Model for Finance}},
  author={Xie, Qianqian and Han, Weiguang and Zhang, Xiao and Lai, Yanzhao and Peng, Min and Lopez-Lira, Alejandro and Huang, Jimin},
  booktitle={Advances in Neural Information Processing Systems},
  volume={36},
  pages={33469--33484},
  year={2023},
  doi={10.5555/3666122.3667576}
}

@inproceedings{Ins_MMBench_2025,
  title={{INS-MMBench: A Comprehensive Benchmark for Evaluating LVLMs' Performance in Insurance}},
  author={Lin, Chenwei and Lyu, Hanjia and Xu, Xian and Luo, Jiebo},
  booktitle={Proceedings of the IEEE/CVF International Conference on Computer Vision},
  pages={9036--9047},
  year={2025}
}

@inproceedings{secque_2025,
  title={{SECQUE: A Benchmark for Evaluating Real-World Financial Analysis Capabilities}},
  author={BenYoash, Noga and Brief, Menachem and Ovadia, Oded and Shenderovitz, Gil and Mishaeli, Moshik and Lemberg, Rachel and Sheetrit, Eitam},
  booktitle={Proceedings of the Fourth Workshop on Generation, Evaluation and Metrics (GEM$^2$)},
  pages={212--230},
  year={2025}
}

@inproceedings{fintextqa_2024,
  title={{FinTextQA: A Dataset for Long-form Financial Question Answering}},
  author={Chen, Jian and Zhou, Peilin and Hua, Yining and Xin, Loh and Chen, Kehui and Li, Ziyuan and Zhu, Bing and Liang, Junwei},
  booktitle={Proceedings of the 62nd Annual Meeting of the Association for Computational Linguistics (Volume 1: Long Papers)},
  pages={6025--6047},
  year={2024},
  doi={10.18653/v1/2024.acl-long.328}
}

@inproceedings{docmath_2023,
  title={{DocMath-Eval: Evaluating Math Reasoning Capabilities of LLMs in Understanding Long and Specialized Documents}},
  author={Zhao, Yilun and Long, Yitao and Liu, Hongjun and Kamoi, Ryo and Nan, Linyong and Chen, Lyuhao and Liu, Yixin and Tang, Xiangru and Zhang, Rui and Cohan, Arman},
  booktitle = {Proceedings of the 62nd Annual Meeting of the Association for Computational Linguistics (Volume 1: Long Papers)},
  year = {2024},
  url = {https://aclanthology.org/2024.acl-long.852/},
  doi = {10.18653/v1/2024.acl-long.852},
  pages = {16103--16120}
}

@article{famma_2025,
  title={{FAMMA: A Benchmark for Financial Multilingual Multimodal Question Answering}},
  author={Xue, Siqiao and Chen, Tingting and Zhou, Fan and Dai, Qingyang and Chu, Zhixuan and Mei, Hongyuan},
  year={2025},
  journal = {arXiv preprint arXiv:2410.04526},
  doi={10.48550/arXiv.2410.04526}
}

@article{teufel2002summarizing,
  title={{Summarizing Scientific Articles: Experiments with Relevance and Rhetorical Status}},
  author={Teufel, Simone and Moens, Marc},
  journal={Computational linguistics},
  volume={28},
  number={4},
  pages={409--445},
  year={2002},
  doi={10.1162/089120102762671936}
}

@article{measure_cobbe2021gsm8k,
  title={{Training Verifiers to Solve Math Word Problems}},
  author={Cobbe, Karl and Kosaraju, Vineet and Bavarian, Mohammad and Chen, Mark and Jun, Heewoo and Kaiser, Lukasz and Plappert, Matthias and Tworek, Jerry and Hilton, Jacob and Nakano, Reiichiro and others},
  journal={arXiv preprint arXiv:2110.14168},
  year={2021},
  doi={10.48550/arXiv.2110.14168}
}

@inproceedings{measure_wei2022chain,
  title={{Chain-of-Thought Prompting Elicits Reasoning in Large Language Models}},
  author={Wei, Jason and Wang, Xuezhi and Schuurmans, Dale and Bosma, Maarten and Xia, Fei and Chi, Ed and Le, Quoc V and Zhou, Denny and others},
  booktitle={Advances in neural information processing systems},
  volume={35},
  pages={24824--24837},
  year={2022},
  doi={10.5555/3600270.3602070}
}

@inproceedings{yang2018hotpotqa,
  title={{HotpotQA: A Dataset for Diverse, Explainable Multi-hop Question Answering}},
  author={Yang, Zhilin and Qi, Peng and Zhang, Saizheng and Bengio, Yoshua and Cohen, William and Salakhutdinov, Ruslan and Manning, Christopher D.},
  booktitle={Proceedings of the 2018 Conference on Empirical Methods in Natural Language Processing},
  pages={2369--2380},
  year={2018},
  organization={Association for Computational Linguistics},
  doi={10.18653/v1/D18-1259}
}

@inproceedings{khashabi2018looking,
  title={{Looking Beyond the Surface: A Challenge Set for Reading Comprehension over Multiple Sentences}},
  author={Khashabi, Daniel and Chaturvedi, Snigdha and Roth, Michael and Upadhyay, Shyam and Roth, Dan},
  booktitle={Proceedings of the 2018 Conference of the North American Chapter of the Association for Computational Linguistics: Human Language Technologies},
  volume={1},
  pages={252--262},
  year={2018},
  organization={Association for Computational Linguistics},
  doi={10.18653/v1/N18-1023}
}

@inproceedings{burges2005learning,
  title={{Learning to rank using gradient descent}},
  author={Burges, Christopher JC and Shaked, Tal and Renshaw, Erin and Lazier, Ari and Deeds, Matt and Hamilton, Nicole and Hullender, Greg},
  booktitle={Proceedings of the 22nd International Conference on Machine Learning},
  pages={89--96},
  year={2005},
  doi={10.1145/1102351.1102363}
}

@inproceedings{wang2025omnieval,
  title={{OmniEval: An Omnidirectional and Automatic RAG Evaluation Benchmark in Financial Domain}},
  author={Wang, Shuting and Tan, Jiejun and Dou, Zhicheng and Wen, Ji-Rong},
  booktitle={Proceedings of the 2025 conference on empirical methods in natural language processing},
  pages={5737--5762},
  year={2025},
  doi={10.18653/v1/2025.emnlp-main.292}
}

@article{stratton2024purposeful,
  title={{Purposeful Sampling: Advantages and Pitfalls}},
  author={Stratton, Samuel J},
  journal={Prehospital and disaster medicine},
  volume={39},
  number={2},
  pages={121--122},
  year={2024},
  publisher={Cambridge University Press},
  doi={10.1017/S1049023X24000281}
}

@article{Biehl2006,
  title={Relationships among the academic business disciplines: A multi-method citation analysis},
  author={Biehl, Markus and Kim, Henry and Wade, Michael},
  journal={Omega},
  volume={34},
  number={4},
  pages={359--371},
  year={2006},
  publisher={Elsevier},
  doi={10.1016/j.omega.2004.12.002}
}

@article{econ_support1,
  author  = {Kalaitzidakis, Pantelis and Mamuneas, Theofanis P. and Stengos, Theodore},
  title   = {An updated ranking of academic journals in economics},
  journal = {Canadian Journal of Economics},
  volume  = {44},
  number  = {4},
  pages   = {1525--1538},
  year    = {2011},
  doi={10.1111/j.1540-5982.2011.01683.x}
}

@article{econ_support2,
  author  = {Heckman, James J. and Moktan, Shruti},
  title   = {{Publishing and promotion in economics: The tyranny of the Top Five}},
  journal = {Journal of Economic Literature},
  volume  = {58},
  number  = {2},
  pages   = {419--470},
  year    = {2020},
  doi     = {10.1257/jel.20191574}
}

@misc{econ_support3,
  title={{Inferring Missing Citations: A Quantitative Multi-Criteria Ranking of all Journals in Economics}},
  author={Combes, Pierre-Philippe and Linnemer, Laurent},
  year={2010},
  howpublished={HAL Working Paper},
  url={https://ideas.repec.org/p/hal/wpaper/halshs-00520325.html}
}

@article{econ_support4,
  author  = {Engemann, Kristie M. and Wall, Howard J.},
  title   = {{A Journal Ranking for the Ambitious Economist}},
  journal = {Federal Reserve Bank of St.~Louis Review},
  volume  = {91},
  number  = {3},
  pages   = {127--139},
  year    = {2009},
  doi={10.20955/R.91.127-140}
}

@article{econ_support5,
url = {https://doi.org/10.1515/1538-0645.1520},
title = {{New Approaches to Ranking Economics Journals}},
author = {Yolanda K. Kodrzycki and Pingkang Yu},
pages = {0000101515153806451520},
volume = {5},
number = {1},
journal = {The B.E. Journal of Economic Analysis \& Policy},
doi = {10.1515/1538-0645.1520},
year = {2006}
}

@article{econ_support6,
  title={Editorial favoritism in the field of laboratory experimental economics},
  author={Cloos, Janis and Greiff, Matthias and Rusch, Hannes},
  journal={Journal of Behavioral and Experimental Economics},
  volume={107},
  pages={102082},
  year={2023},
  publisher={Elsevier},
  doi={10.1016/j.socec.2023.102082}
}

@article{econ_support7,
  author  = {Hamermesh, Daniel S.},
  title   = {Citations in economics: Measurement, uses, and impacts},
  journal = {Journal of Economic Literature},
  volume  = {56},
  number  = {1},
  pages   = {115--156},
  year    = {2018},
  doi     = {10.1257/jel.20161326}
}

@article{fin_support1,
  author  = {Bajo, Edoardo and Barbi, Massimiliano and Hillier, David},
  title   = {Where should I publish to get promoted? A finance journal ranking based on business‐school promotions},
  journal = {Journal of Banking \& Finance},
  volume  = {114},
  year    = {2020},
  pages   = {105780},
  doi={10.1016/j.jbankfin.2020.105780}
}

@article{fin_support2,
  author  = {Currie, Robert and Pandher, G. S.},
  title   = {Finance journal rankings: Active scholar assessment revisited},
  journal = {Journal of Banking \& Finance},
  volume  = {111},
  year    = {2020},
  pages   = {105717},
  doi={10.1016/j.jbankfin.2019.105717}
}

@article{fin_support3,
  author  = {Borokhovich, Kenneth A. and Bricker, Robert J. and Simkins, Betty J.},
  title   = {{An Analysis of Finance Journal Impact Factors}},
  journal = {Journal of Finance},
  volume  = {55},
  number  = {3},
  pages   = {1457--1469},
  year    = {2000},
  doi={10.1111/0022-1082.00254}
}

@article{stats_support1,
  title={{Citation Patterns in the Journals of Statistics and Probability}},
  author={Stigler, Stephen M},
  journal={Statistical Science},
  pages={94--108},
  year={1994},
  publisher={JSTOR},
}

@article{stats_support2,
  author  = {Ruiz-Castillo, Javier and Waltman, Ludo},
  title   = {Field-normalized citation impact indicators using algorithmically constructed classification systems of science},
  journal = {Journal of Informetrics},
  volume  = {9},
  number  = {1},
  year    = {2015},
  pages   = {102--117},
  doi     = {10.1016/j.joi.2014.11.010}
}

@misc{stats_support3,
  title={{Overview of Statistics as a Scientific Discipline and Practical Implications for the Evaluation of Faculty Excellence}},
  author={American Statistical Association and others},
  year={2024},
}

@article{om_support1,
  title={Insights into factors affecting Production and Operations Management (POM) journal evaluation},
  author={Theoharakis, Vasilis and Voss, Chris and Hadjinicola, George C and Soteriou, Andreas C},
  journal={Journal of Operations Management},
  volume={25},
  number={4},
  pages={932--955},
  year={2007},
  publisher={Elsevier},
  doi={10.1016/j.jom.2006.09.002}
}

@article{om_support2,
  title={{How does the UK Academic Journal Guide compare to other journal rating guides?}},
  author={Hudson, Robert},
  journal={Scientometrics},
  pages={1--35},
  year={2025},
  publisher={Springer},
  doi={10.1007/s11192-025-05474-0}
}

@article{om_support3,
  title={{Journal Quality List}},
  author={Edition, Forty-ninth},
  year={2000},
}

@book{om_support4,
  title={{Mathematical programming}},
  author={Vajda, Steven},
  year={2009},
  publisher={Courier Corporation},
}

@article{om_support5,
  title={{Editorial: Perspectives of ISE/OR researchers}},
  author={Ding, Yu},
  journal={IISE Transactions},
  volume={55},
  number={1},
  pages={1--1},
  year={2023},
  publisher={Taylor \& Francis},
  doi={10.1080/24725854.2022.2106391}
}

@inproceedings{shah2023trillion,
  title     = {{Trillion Dollar Words: A New Financial Dataset, Task \& Market Analysis}},
  author    = {Shah, Agam and Paturi, Suvan and Chava, Sudheer},
  booktitle = {Proceedings of the 61st Annual Meeting of the Association for Computational Linguistics (Volume 1: Long Papers)},
  pages     = {6664--6679},
  year      = {2023},
  publisher = {Association for Computational Linguistics},
  doi={10.18653/v1/2023.acl-long.368}
}

@article{Malo2014GoodDO,
  title   = {Good debt or bad debt: Detecting semantic orientations in economic texts},
  author  = {Malo, P. and Sinha, A. and Korhonen, P. and Wallenius, J. and Takala, P.},
  journal = {Journal of the Association for Information Science and Technology},
  year    = {2014},
  volume  = {65},
  doi={10.1002/asi.23062Digital Object Identifier (DOI)}
}

@inproceedings{yang2020generating,
  title     = {{Generating Plausible Counterfactual Explanations for Deep Transformers in Financial Text Classification}},
  author    = {Yang, Linyi and Kenny, Eoin and Ng, Tin Lok James and Yang, Yi and Smyth, Barry and Dong, Ruihai},
  booktitle = {Proceedings of the 28th International Conference on Computational Linguistics (COLING)},
  pages     = {6150--6160},
  year      = {2020},
  doi={10.18653/v1/2020.coling-main.541}
}

@article{feng2023empowering,
  title         = {{Empowering Many, Biasing a Few: Generalist Credit Scoring through Large Language Models}},
  author        = {Feng, Duanyu and Dai, Yongfu and Huang, Jimin and Zhang, Yifang and Xie, Qianqian and Han, Weiguang and Lopez-Lira, Alejandro and Wang, Hao},
  year          = {2023},
  journal = {arXiv preprint arXiv:2310.00566},
  primaryClass  = {cs.LG},
  doi={10.48550/arXiv.2310.00566}
}

@inproceedings{sharma2023financial,
  title     = {{Financial Numeric Extreme Labelling: A Dataset and Benchmarking}},
  author    = {Sharma, Soumya and Khatuya, Subhendu and Hegde, Manjunath and Shaikh, Afreen and Dasgupta, Koustuv and Goyal, Pawan and Ganguly, Niloy},
  booktitle = {Findings of the Association for Computational Linguistics: ACL 2023},
  pages     = {3550--3561},
  year      = {2023},
  doi={10.18653/v1/2023.findings-acl.219}
}

@inproceedings{chen2021finqa,
  title={{FinQA: A Dataset of Numerical Reasoning over Financial Data}},
  author={Chen, Zhiyu and Chen, Wenhu and Smiley, Charese and Shah, Sameena and Borova, Iana and Langdon, Dylan and Moussa, Reema and Beane, Matt and Huang, Ting-Hao and Routledge, Bryan R and others},
  booktitle={Proceedings of the 2021 Conference on Empirical Methods in Natural Language Processing},
  pages={3697--3711},
  year={2021},
  doi={10.18653/v1/2021.emnlp-main.300}
}

@inproceedings{zhu2021tat,
  title={{TAT-QA: A Question Answering Benchmark on a Hybrid of Tabular and Textual Content in Finance}},
  author={Zhu, Fengbin and Lei, Wenqiang and Huang, Youcheng and Wang, Chao and Zhang, Shuo and Lv, Jiancheng and Feng, Fuli and Chua, Tat-Seng},
  booktitle={Proceedings of the 59th annual meeting of the Association for Computational Linguistics and the 11th international joint conference on natural language processing (volume 1: long papers)},
  pages={3277--3287},
  year={2021},
  doi={10.18653/v1/2021.acl-long.254}
}

@article{orlm_2025,
  title={{ORLM: A Customizable Framework in Training Large Models for Automated Optimization Modeling}},
  author={Huang, Chenyu and Tang, Zhengyang and Hu, Shixi and Jiang, Ruoqing and Zheng, Xin and Ge, Dongdong and Wang, Benyou and Wang, Zizhuo},
  journal={Operations Research},
  year={2025},
  publisher={INFORMS},
  doi={10.1287/opre.2024.1233}
}

@inproceedings{alphafin_2024,
  title={{AlphaFin: Benchmarking Financial Analysis with Retrieval-Augmented Stock-Chain Framework}},
  author={Li, Xiang and Li, Zhenyu and Shi, Chen and Xu, Yong and Du, Qing and Tan, Mingkui and Huang, Jun},
  booktitle={Proceedings of the 2024 joint international conference on computational linguistics, language resources and evaluation (LREC-COLING 2024)},
  pages={773--783},
  year={2024}
}

@inproceedings{statQA_2024,
  title={{Are Large Language Models Good Statisticians?}},
  author={Zhu, Yizhang and Du, Shiyin and Li, Boyan and Luo, Yuyu and Tang, Nan},
  booktitle={Advances in Neural Information Processing Systems},
  volume={37},
  pages={62697--62731},
  year={2024},
  doi = {10.52202/079017-2005}
}

@inproceedings{findabench_2025,
  title={{FinDABench: Benchmarking Financial Data Analysis Ability of Large Language Models}},
  author={Liu, Shu and Zhao, Shangqing and Jia, Chenghao and Zhuang, Xinlin and Long, Zhaoguang and Zhou, Jie and Zhou, Aimin and Lan, Man and Chong, Yang},
  booktitle={Proceedings of the 31st International Conference on Computational Linguistics},
  pages={710--725},
  year={2025},
}

@article{supercluefin_2024,
  title={{SuperCLUE-Fin: Graded Fine-Grained Analysis of Chinese LLMs on Diverse Financial Tasks and Applications}},
  author={Xu, Liang and Zhu, Lei and Wu, Yaotong and Xue, Hang},
  journal={arXiv preprint arXiv:2404.19063},
  year={2024},
  doi={10.48550/arXiv.2404.19063}
}

@inproceedings{qrdata_2024,
  title={{Are LLMs Capable of Data-based Statistical and Causal Reasoning? Benchmarking Advanced Quantitative Reasoning with Data}},
  author={Liu, Xiao and Wu, Zirui and Wu, Xueqing and Lu, Pan and Chang, Kai-Wei and Feng, Yansong},
  booktitle={Findings of the Association for Computational Linguistics: ACL 2024},
  pages={9215--9235},
  year={2024},
  doi={10.18653/v1/2024.findings-acl.548}
}

@misc{FinLongEval_2023,
  author = {Xinguang, Jiang and Sihan, Hu and Dingfu, Yu and Yuhao, Zhang and Zhongliang, Yang and Yu, Li and Linna, Zhou and Valuesimplex AI Lab},
  title  = {{FinLongEval}},
  url    = {https://github.com/valuesimplex/FinLongEval},
  year   = {2023},
  month  = {Dec},
}

@inproceedings{convfinqa_2022,
  title={{ConvFinQA: Exploring the Chain of Numerical Reasoning in Conversational Finance Question Answering}},
  author={Chen, Zhiyu and Li, Shiyang and Smiley, Charese and Ma, Zhiqiang and Shah, Sameena and Wang, William Yang},
  booktitle={Proceedings of the 2022 conference on empirical methods in natural language processing},
  pages={6279--6292},
  year={2022},
  doi={10.18653/v1/2022.emnlp-main.421}
}

@article{statllm_2026,
  title={{StatLLM: A Dataset for Evaluating the Performance of Large Language Models in Statistical Analysis}},
  author={Song, Xinyi and Lee, Lina and Xie, Kexin and Liu, Xueying and Deng, Xinwei and Hong, Yili},
  journal={Scientific Data},
  year={2026},
  publisher={Nature Publishing Group UK London},
  doi={10.1038/s41597-026-06731-4}
}

@INPROCEEDINGS{FLUE_2022,
    author = {Shah, Raj Sanjay  and
      Chawla, Kunal and
      Eidnani, Dheeraj and
      Shah, Agam and
      Du, Wendi and
      Chava, Sudheer and
      Raman, Natraj and
      Smiley, Charese and
      Chen, Jiaao and
      Yang, Diyi },
    title = {{When FLUE Meets FLANG: Benchmarks and Large Pretrained Language Model for Financial Domain}},
    booktitle = {Proceedings of the 2022 Conference on Empirical Methods in Natural Language Processing (EMNLP)},
    year = {2022},
    publisher = {Association for Computational Linguistics},
  doi={10.18653/v1/2022.emnlp-main.148}
}

@article{discfinllm_2023,
  title={{DISC-FinLLM: A Chinese Financial Large Language Model based on Multiple Experts Fine-tuning}},
  author={Chen, Wei and Wang, Qiushi and Long, Zefei and Zhang, Xianyin and Lu, Zhongtian and Li, Bingxuan and Wang, Siyuan and Xu, Jiarong and Bai, Xiang and Huang, Xuanjing and others},
  journal={arXiv preprint arXiv:2310.15205},
  year={2023},
  doi={10.48550/arXiv.2310.15205}
}

@inproceedings{sharma2022finred,
  title     = {{FinRED: A Dataset for Relation Extraction in Financial Domain}},
  author    = {Sharma, Soumya and Nayak, Tapas and Bose, Arusarka and Meena, Ajay Kumar and Dasgupta, Koustuv and Ganguly, Niloy and Goyal, Pawan},
  booktitle = {Companion Proceedings of the Web Conference 2022},
  pages     = {595--597},
  year      = {2022},
  doi={10.1145/3487553.3524637}
}

@inproceedings{wu2018hybrid,
  title     = {{Hybrid Deep Sequential Modeling for Social Text-Driven Stock Prediction}},
  author    = {Wu, Huizhe and Zhang, Wei and Shen, Weiwei and Wang, Jun},
  booktitle = {Proceedings of the 27th ACM International Conference on Information and Knowledge Management (CIKM)},
  pages     = {1627--1630},
  year      = {2018},
  publisher = {ACM},
  doi={10.1145/3269206.3269290}
}

@inproceedings{zhou2021trade,
  title     = {{Trade the Event: Corporate Events Detection for News-Based Event-Driven Trading}},
  author    = {Zhou, Zhihan and Ma, Liqian and Liu, Han},
  booktitle = {Findings of the Association for Computational Linguistics: ACL-IJCNLP 2021},
  pages     = {2114--2124},
  year      = {2021},
  publisher = {Association for Computational Linguistics},
  doi={10.18653/v1/2021.findings-acl.186}
}

@article{ham2021new,
  title={{New Rankings of Economics Journals: Documenting and Explaining the Rise of the New Society Journals}},
  author={Ham, John C and Wright, Julian and Ye, Ziqiu},
  journal={Available at SSRN 3606030},
  year={2021},
  doi={10.2139/ssrn.3606030}
}

@article{LoughranMcDonald2011,
  author  = {Loughran, Tim and McDonald, Bill},
  title   = {{When Is a Liability Not a Liability? Textual Analysis, Dictionaries, and 10-Ks}},
  journal = {The Journal of Finance},
  year    = {2011},
  volume  = {66},
  number  = {1},
  pages   = {35--65},
  doi     = {10.1111/j.1540-6261.2010.01625.x}
}

@book{LopezDePrado2018,
  title={{Advances in Financial Machine Learning}},
  author={De Prado, Marcos Lopez},
  year={2018},
  publisher={John Wiley \& Sons}
}

@book{Chan2013,
  author    = {Chan, Ernest P.},
  title     = {{Algorithmic Trading: Winning Strategies and Their Rationale}},
  year      = {2013},
  publisher={John Wiley \& Sons}
}

@article{gebru2021datasheets,
  title={Datasheets for datasets},
  author={Gebru, Timnit and Morgenstern, Jamie and Vecchione, Briana and Vaughan, Jennifer Wortman and Wallach, Hanna and Iii, Hal Daum{\'e} and Crawford, Kate},
  journal={Communications of the ACM},
  volume={64},
  number={12},
  pages={86--92},
  year={2021},
  publisher={ACM New York, NY, USA},
  doi={10.1145/3458723}
}

@inproceedings{Mitchell2019,
  title={{Model Cards for Model Reporting}},
  author={Mitchell, Margaret and Wu, Simone and Zaldivar, Andrew and Barnes, Parker and Vasserman, Lucy and Hutchinson, Ben and Spitzer, Elena and Raji, Inioluwa Deborah and Gebru, Timnit},
  booktitle={Proceedings of the conference on fairness, accountability, and transparency},
  pages={220--229},
  year={2019},
  doi = {10.1145/3287560.3287596}
}

@book{anderson2001taxonomy,
  title={{A Taxonomy for Learning, Teaching, and Assessing: A Revision of {Bloom}'s Taxonomy of Educational Objectives}},
  author={Anderson, Lorin W and Krathwohl, David R},
  year={2001},
  publisher={Addison Wesley Longman, Inc.}
}

@book{bandalos2018measurement,
  title={{Measurement Theory and Applications for the Social Sciences}},
  author={Bandalos, Deborah L},
  year={2018},
  publisher={Guilford Publications}
}

@article{taherdoost2016,
  title={{Validity and Reliability of the Research Instrument; How to Test the Validation of a Questionnaire/Survey in a Research}},
  author={Taherdoost, Hamed},
  journal={International journal of academic research in management (IJARM)},
  volume={5},
  year={2016}
}

@book{lane2006handbook,
  title={{Handbook of Test Development}},
  author={Lane, Suzanne and Raymond, Mark R and Haladyna, Thomas M and others},
  volume={2},
  year={2016},
  publisher={Routledge New York, NY}
}

@article{haynes1995content,
  title={{Content Validity in Psychological Assessment: A Functional Approach to Concepts and Methods}},
  author={Haynes, Stephen N and Richard, David C and Kubany, Edward S},
  journal={Psychological assessment},
  volume={7},
  number={3},
  pages={238},
  year={1995},
  publisher={American Psychological Association},
  doi={10.1037/1040-3590.7.3.238}
}

@article{sloman1993feature,
  title={Feature-based induction},
  author={Sloman, Steven A},
  journal={Cognitive psychology},
  volume={25},
  number={2},
  pages={231--280},
  year={1993},
  publisher={Elsevier},
  doi={10.1006/cogp.1993.1006}
}

@article{he2024opendatalab,
  title={{Opendatalab: Empowering General Artificial Intelligence with Open Datasets}},
  author={He, Conghui and Li, Wei and Jin, Zhenjiang and Xu, Chao and Wang, Bin and Lin, Dahua},
  journal={arXiv preprint arXiv:2407.13773},
  year={2024},
  doi={10.48550/arXiv.2407.13773}
}

@article{wang2024unimernetuniversalnetworkrealworld,
  title={{Unimernet: A Universal Network for Real-world Mathematical Expression Recognition}},
  author={Wang, Bin and Gu, Zhuangcheng and Liang, Guang and Xu, Chao and Zhang, Bo and Shi, Botian and He, Conghui},
  journal={arXiv preprint arXiv:2404.15254},
  year={2024},
  doi={10.48550/arXiv.2404.15254}
}

@article{wang2024cdmreliablemetricfair,
  title={{CDM: A Reliable Metric for Fair and Accurate Formula Recognition Evaluation}},
  author={Wang, Bin and Wu, Fan and Ouyang, Linke and Gu, Zhuangcheng and Zhang, Rui and Xia, Renqiu and Zhang, Bo and He, Conghui},
  journal={arXiv preprint arXiv:2409.03643},
  volume={5},
  number={6},
  year={2024},
  doi={10.48550/arXiv.2409.03643}
}

@article{wang2024mineruopensourcesolutionprecise,
  title={{MinerU: An Open-Source Solution for Precise Document Content Extraction}},
  author={Wang, Bin and Xu, Chao and Zhao, Xiaomeng and Ouyang, Linke and Wu, Fan and Zhao, Zhiyuan and Xu, Rui and Liu, Kaiwen and Qu, Yuan and Shang, Fukai and others},
  journal={arXiv preprint arXiv:2409.18839},
  year={2024},
  doi={10.48550/arXiv.2409.18839}
}

@inproceedings{lewis2020retrieval,
  title={{Retrieval-Augmented Generation for Knowledge-Intensive NLP Tasks}},
  author={Lewis, Patrick and Perez, Ethan and Piktus, Aleksandra and Petroni, Fabio and Karpukhin, Vladimir and Goyal, Naman and K{\"u}ttler, Heinrich and Lewis, Mike and Yih, Wen-tau and Rockt{\"a}schel, Tim and others},
  booktitle={Advances in neural information processing systems},
  volume={33},
  pages={9459--9474},
  year={2020}
}

@inproceedings{sarthi2024raptor,
  title={{RAPTOR: Recursive Abstractive Processing for Tree-Organized Retrieval}},
  author={Sarthi, Parth and Abdullah, Salman and Tuli, Aditi and Khanna, Shubh and Goldie, Anna and Manning, Christopher D},
  booktitle={The Twelfth International Conference on Learning Representations},
  year={2024}
}

@article{han2024retrieval,
  title={{Retrieval-Augmented Generation with Graphs (GraphRAG)}},
  author={Han, Haoyu and Wang, Yu and Shomer, Harry and Guo, Kai and Ding, Jiayuan and Lei, Yongjia and Halappanavar, Mahantesh and Rossi, Ryan A and Mukherjee, Subhabrata and Tang, Xianfeng and others},
  journal={arXiv preprint arXiv:2501.00309},
  year={2024},
  doi={10.48550/arXiv.2501.00309}
}

@article{ozuru2013comparing,
  title={Comparing comprehension measured by multiple-choice and open-ended questions},
  author={Ozuru, Yasuhiro and Briner, Stephen and Kurby, Christopher A and McNamara, Danielle S},
  journal={Canadian Journal of Experimental Psychology/Revue canadienne de psychologie exp{\'e}rimentale},
  volume={67},
  number={3},
  pages={215},
  year={2013},
  publisher={Educational Publishing Foundation},
  doi={10.1037/a0032918}
}

@inproceedings{ganzeboom2010new,
  title={A standard international socio-economic index of occupational status},
  author={Ganzeboom, Harry BG},
  booktitle={annual conference of international social survey programme, Lisbon},
  volume={1},
  year={2010},
  doi={10.1016/0049-089X(92)90017-B}
}

@article{kaniel2006so,
  title={{So What Orders Do Informed Traders Use?}},
  author={Kaniel, Ron and Liu, Hong},
  journal={The Journal of Business},
  volume={79},
  number={4},
  pages={1867--1913},
  year={2006},
  publisher={JSTOR},
  doi={10.1086/503651}
}

@article{carliner2015job,
  title={The job of a performance consultant: a qualitative content analysis of job descriptions},
  author={Carliner, Saul and Castonguay, Chantal and Sheepy, Emily and Ribeiro, Ofelia and Sabri, Hiba and Saylor, Chantal and Valle, Andre},
  journal={European Journal of Training and Development},
  volume={39},
  number={6},
  pages={458--483},
  year={2015},
  publisher={Emerald Group Publishing Limited},
  doi={10.1108/EJTD-01-2015-0006}
}

@article{gemini_2025,
  title={Gemini 2.5: Pushing the frontier with advanced reasoning, multimodality, long context, and next generation agentic capabilities},
  author={Comanici, Gheorghe and Bieber, Eric and Schaekermann, Mike and Pasupat, Ice and Sachdeva, Noveen and Dhillon, Inderjit and Blistein, Marcel and Ram, Ori and Zhang, Dan and Rosen, Evan and others},
  journal={arXiv preprint arXiv:2507.06261},
  year={2025},
  doi={10.48550/arXiv.2507.06261}
}

@misc{gpt_2024,
 author = {OpenAI},
 title = {Hello gpt-4o},
 year = {2024},
 url = {https://openai.com/index/hello-gpt-4o},
 note = {Accessed:2024-05-13}
}

@misc{claude_2025,
 author = {Anthropic},
 title = {Claude3.7 Sonnet and Claude Code},
 year = {2025},
 url = {https://www.anthropic.com/news/claude-3-7-sonnet},
 note = {Accessed:2025-02-25}
}

@misc{llama_2025,
 author = {AI\@Meta},
 title = {The Llama4 Herd: The Beginning of a New Era of Natively Multimodal AI Innovation},
 year = {2025},
 url = {https://ai.meta.com/blog/llama-4-multimodalintelligence},
 note = {Accessed:2025-04-05}
}

@misc{qwen_2024,
 author = {Qwen Team},
 title = {QWQ: Reflect Deeply on the Boundaries of the Unknown},
 year = {2024},
 url = {https://qwenlm.github.io/blog/qwq-32b-preview},
 note = {Accessed:2024-11-28}
}

@article{deepdistill_2025,
  title={{DeepDistill: Enhancing LLM Reasoning Capabilities via Large-Scale Difficulty-Graded Data Training}},
  author={Tian, Xiaoyu and Zhao, Sitong and Wang, Haotian and Chen, Shuaiting and Peng, Yiping and Ji, Yunjie and Zhao, Han and Li, Xiangang},
  journal={arXiv preprint arXiv:2504.17565},
  year={2025},
  doi={10.48550/arXiv.2504.17565}
}

@misc{grok_2025,
 author = {xAI},
 title = {Models},
 year = {2025},
 url = {https://x.ai/news/grok-4},
 note = {Accessed:2025-07-09}
}

@inproceedings{Geval_2023,
  title={{G-Eval: NLG Evaluation using GPT-4 with Better Human Alignment}},
  author={Liu, Yang and Iter, Dan and Xu, Yichong and Wang, Shuohang and Xu, Ruochen and Zhu, Chenguang},
  booktitle={Proceedings of the 2023 conference on empirical methods in natural language processing},
  pages={2511--2522},
  year={2023},
  doi={10.18653/v1/2023.emnlp-main.153}
}

@inproceedings{li2024mediq,
  title={{MediQ: Question-Asking LLMs and a Benchmark for Reliable Interactive Clinical Reasoning}},
  author={Li, Shuyue S and Balachandran, Vidhisha and Feng, Shangbin and Ilgen, Jonathan S and Pierson, Emma and Koh, Pang W and Tsvetkov, Yulia},
  booktitle={Advances in Neural Information Processing Systems},
  volume={37},
  pages={28858--28888},
  year={2024},
  doi={10.52202/079017-0908}
}

@inproceedings{bai2024longbench,
  title={{LongBench: A Bilingual, Multitask Benchmark for Long Context Understanding}},
  author={Bai, Yushi and Lv, Xin and Zhang, Jiajie and Lyu, Hongchang and Tang, Jiankai and Huang, Zhidian and Du, Zhengxiao and Liu, Xiao and Zeng, Aohan and Hou, Lei and others},
  booktitle={Proceedings of the 62nd annual meeting of the association for computational linguistics (volume 1: Long papers)},
  pages={3119--3137},
  year={2024},
  doi={10.18653/v1/2024.acl-long.172}
}

@inproceedings{jimenez2024swe,
    title={{SWE-bench: Can Language Models Resolve Real-world Github Issues?}},
    author={Carlos E Jimenez and John Yang and Alexander Wettig and Shunyu Yao and Kexin Pei and Ofir Press and Karthik R Narasimhan},
    booktitle={The Twelfth International Conference on Learning Representations},
    year={2024},
    url={https://openreview.net/forum?id=VTF8yNQM66}
}

@article{messick1995validity,
  title={{VALIDITY OF PSYCHOLOGICAL ASSESSMENT: VALIDATION OF INFERENCES FROM PERSONS' RESPONSES AND PERFORMANCES AS SCIENTIFIC INQUIRY INTO SCORE MEANING}},
  author={Messick, Samuel},
  journal={American psychologist},
  volume={50},
  number={9},
  pages={741},
  year={1995},
  publisher={American Psychological Association},
  doi={10.1002/j.2333-8504.1994.tb01618.x}
}

@article{shanteau1992competence,
  title={Competence in experts: The role of task characteristics},
  author={Shanteau, James},
  journal={Organizational behavior and human decision processes},
  volume={53},
  number={2},
  pages={252--266},
  year={1992},
  publisher={Elsevier},
  doi={10.1016/0749-5978(92)90064-E}
}

@article{brunswik1955representative,
  title={Representative design and probabilistic theory in a functional psychology.},
  author={Brunswik, Egon},
  journal={Psychological review},
  volume={62},
  number={3},
  pages={193},
  year={1955},
  publisher={American Psychological Association},
  doi={10.1037/h0047470}
}

@article{rush2016impact,
  title={The impact of item-writing flaws and item complexity on examination item difficulty and discrimination value},
  author={Rush, Bonnie R and Rankin, David C and White, Brad J},
  journal={BMC medical education},
  volume={16},
  number={1},
  pages={250},
  year={2016},
  publisher={Springer},
  doi={10.1186/s12909-016-0773-3}
}

@inproceedings{huang2019cosmos,
  title={{Cosmos QA: Machine Reading Comprehension with Contextual Commonsense Reasoning}},
  author={Huang, Lifu and Le Bras, Ronan and Bhagavatula, Chandra and Choi, Yejin},
  booktitle={Proceedings of the 2019 conference on empirical methods in natural language processing and the 9th international joint conference on natural language processing (EMNLP-IJCNLP)},
  pages={2391--2401},
  year={2019},
  doi={10.18653/v1/D19-1243}
}

@inproceedings{judging_2023,
  title={{Judging LLM-as-a-judge with MT-bench and Chatbot Arena}},
  author={Zheng, Lianmin and Chiang, Wei-Lin and Sheng, Ying and Zhuang, Siyuan and Wu, Zhanghao and Zhuang, Yonghao and Lin, Zi and Li, Zhuohan and Li, Dacheng and Xing, Eric and others},
  booktitle={Advances in neural information processing systems},
  volume={36},
  pages={46595--46623},
  year={2023}
}

@article{berthet2022impact,
  title={{The Impact of Cognitive Biases on Professionals’ Decision-Making: A Review of Four Occupational Areas}},
  author={Berthet, Vincent},
  journal={Frontiers in psychology},
  volume={12},
  pages={802439},
  year={2022},
  publisher={Frontiers Media SA},
  doi={10.3389/fpsyg.2021.802439}
}

@article{forster2008failures,
  title={{Failures to Ignore Entirely Irrelevant Distractors: The Role of Load}},
  author={Forster, Sophie and Lavie, Nilli},
  journal={Journal of Experimental Psychology: Applied},
  volume={14},
  number={1},
  pages={73},
  year={2008},
  publisher={American Psychological Association},
  doi={10.1037/1076-898X.14.1.73}
}

@article{gierl2017developing,
  title={{Developing, Analyzing, and Using Distractors for Multiple-Choice Tests in Education: A Comprehensive Review}},
  author={Gierl, Mark J and Bulut, Okan and Guo, Qi and Zhang, Xinxin},
  journal={Review of educational research},
  volume={87},
  number={6},
  pages={1082--1116},
  year={2017},
  publisher={Sage Publications Sage CA: Los Angeles, CA},
  doi={10.3102/0034654317726529}
}

@book{haladyna2004developing,
  title={{Developing and Validating Multiple-Choice Test Items}},
  author={Haladyna, Thomas M},
  year={2004},
  publisher={Routledge}
}

@article{haladyna2002review,
  title={{A Review of Multiple-Choice Item-Writing Guidelines for Classroom Assessment}},
  author={Haladyna, Thomas M and Downing, Steven M and Rodriguez, Michael C},
  journal={Applied measurement in education},
  volume={15},
  number={3},
  pages={309--333},
  year={2002},
  publisher={Taylor \& Francis},
  doi={10.1207/S15324818AME1503_5}
}

@article{radatz1979error,
  title={Error Analysis in Mathematics Education},
  author={Radatz, Hendrik},
  journal={Journal for Research in mathematics Education},
  volume={10},
  number={3},
  pages={163--172},
  year={1979},
  publisher={National Council of Teachers of Mathematics},
  doi={10.2307/748804}
}

@article{wu2025enhancing,
  title={{Enhancing Financial Decision-making under Cyber Threats: a Dual-branch Framework Integrating Bayesian Deep Learning and Explainable AI}},
  author={Wu, Yuhe and Chen, Yuran and Liu, Zhuang and Lin, Wayne},
  journal={Annals of Operations Research},
  pages={1--33},
  year={2025},
  publisher={Springer},
  issn={1572-9338},
  doi={10.1007/s10479-025-06973-2},
  url={https://doi.org/10.1007/s10479-025-06973-2}
}

@article{nickerson1998confirmation,
  title={{Confirmation Bias: A Ubiquitous Phenomenon in Many Guises}},
  author={Nickerson, Raymond S},
  journal={Review of general psychology},
  volume={2},
  number={2},
  pages={175--220},
  year={1998},
  publisher={SAGE Publications Sage CA: Los Angeles, CA},
  doi={10.1037/1089-2680.2.2.175}
}

@inproceedings{wan2024logicasker,
  title={{LogicAsker: Evaluating and Improving the Logical Reasoning Ability of Large Language Models}},
  author={Wan, Yuxuan and Wang, Wenxuan and Yang, Yiliu and Yuan, Youliang and Huang, Jen-tse and He, Pinjia and Jiao, Wenxiang and Lyu, Michael},
  booktitle={Proceedings of the 2024 Conference on Empirical Methods in Natural Language Processing},
  pages={2124--2155},
  year={2024},
  doi={10.18653/v1/2024.emnlp-main.128}
}

@inproceedings{lin2004rouge,
  title={{ROUGE: A Package for Automatic Evaluation of Summaries}},
  author={Lin, Chin-Yew},
  booktitle={Text summarization branches out},
  pages={74--81},
  year={2004}
}

@inproceedings{wei2022chain,
  title={Chain-of-thought prompting elicits reasoning in large language models},
  author={Wei, Jason and Wang, Xuezhi and Schuurmans, Dale and Bosma, Maarten and Xia, Fei and Chi, Ed and Le, Quoc V and Zhou, Denny and others},
  booktitle={Advances in neural information processing systems},
  volume={35},
  pages={24824--24837},
  year={2022}
}

@inproceedings{judgebench_2025,
  author       = {Sijun Tan and
                  Siyuan Zhuang and
                  Kyle Montgomery and
                  William Yuan Tang and
                  Alejandro Cuadron and
                  Chenguang Wang and
                  Raluca A. Popa and
                  Ion Stoica},
  title        = {{JudgeBench: A Benchmark for Evaluating LLM-Based Judges}},
  booktitle    = {The Thirteenth International Conference on Learning Representations,
                  {ICLR} 2025},
  year         = {2025}
}

\newpage
\appendix
\section{Benchmark Comparison}\label{app:comparison}
Table~\ref{tab:com_comparison} compares BizCompass with existing benchmarks across both knowledge coverage and application dimensions. Most prior datasets focus narrowly on finance-related question answering or isolated analytical tasks, lacking broader domain representation and role-specific evaluation. In contrast, BizCompass achieves comprehensive disciplinary coverage spanning finance, economics, statistics, and operations management, while also incorporating three professional perspectives: analysis, trading, and consulting. Furthermore, it unifies multiple question formats, including single-choice, multiple-choice, and open-ended tasks, enabling a more systematic and fine-grained assessment of business reasoning capabilities in LLMs.
\begin{table*}[t]
\centering
\resizebox{\linewidth}{!}{
\setlength{\tabcolsep}{3pt}
\begin{tabular}{l|cccc|ccc|ccc}
\toprule
\multirow{2}{*}{\textbf{Benchmark}} & \multicolumn{4}{c}{\textbf{Knowledge Coverage}} & \multicolumn{3}{c}{\textbf{Application Dimensions}} & \multicolumn{3}{c}{\textbf{Question Types}}\\
\cmidrule(lr){2-5} \cmidrule(lr){6-8} \cmidrule(lr){9-11}
 & Finance & Economics & Statistics & OM & Analysis & Trading & Consulting & SC. & MC. & OQ. \\
\midrule
FinQA \citep{FinQA_2021} & \cmark & \xmark & \xmark & \xmark & \cmark & \xmark & \xmark & \xmark & \xmark & \cmark \\
TAT-QA \citep{zhu2021tat} & \cmark & \xmark & \xmark & \xmark & \cmark & \xmark & \xmark & \xmark & \xmark & \cmark \\
CovFinQA \citep{convfinqa_2022} & \cmark & \xmark & \xmark & \xmark & \cmark & \xmark & \xmark & \xmark & \xmark & \cmark \\
FLUE \citep{FLUE_2022} & \cmark & \cmark & \xmark & \xmark & \cmark & \xmark & \xmark & \cmark & \xmark & \xmark \\
FinanceBench \citep{financebench_2023} & \cmark & \xmark & \xmark & \xmark & \cmark & \xmark & \xmark & \xmark & \xmark & \cmark \\
FinEval \citep{fineval_2025} & \cmark & \cmark & \xmark & \xmark & \cmark & \xmark & \cmark & \cmark & \xmark & \cmark \\
BBT-CFLEB \citep{bbtfin_2023} & \cmark & \xmark & \xmark & \xmark & \cmark & \xmark & \xmark & \xmark & \xmark & \cmark \\
FinLBench \citep{FinLongEval_2023} & \cmark & \cmark & \xmark & \xmark & \cmark & \xmark & \xmark & \cmark & \xmark & \cmark \\
PIXIU \citep{pixiu_2023} & \cmark & \cmark & \xmark & \xmark & \cmark & \cmark & \xmark & \xmark & \xmark & \cmark \\
DISC-FinLLM \citep{discfinllm_2023} & \cmark & \xmark & \xmark & \xmark & \cmark & \xmark & \cmark & \xmark & \xmark & \cmark \\
BBT-Fin \citep{bbtfin_2023} & \cmark & \xmark & \xmark & \xmark & \cmark & \xmark & \xmark & \xmark & \xmark & \cmark \\
DOCMATH-EVAL \citep{docmath_2023} & \cmark & \xmark & \xmark & \xmark & \cmark & \xmark & \xmark & \xmark & \xmark & \cmark \\
StatQA \citep{statQA_2024} & \xmark & \xmark & \cmark & \xmark & \cmark & \xmark & \xmark & \xmark & \xmark & \cmark \\
QRDATA \citep{qrdata_2024} & \xmark & \xmark & \cmark & \xmark & \cmark & \xmark & \xmark & \xmark & \cmark & \cmark \\
BizBench \citep{bizbench_2024} & \cmark & \cmark & \xmark & \xmark & \cmark & \xmark & \xmark & \cmark & \xmark & \cmark\\
FinBen \citep{Finben_2024} & \cmark & \cmark & \cmark & \xmark & \cmark & \cmark & \xmark & \cmark & \cmark & \cmark \\
FinanceMATH \citep{financemath_2024} & \cmark & \cmark & \xmark & \xmark & \cmark & \xmark & \xmark & \xmark & \xmark & \xmark \\
SuperCluE-Fin \citep{supercluefin_2024} & \cmark & \cmark & \cmark & \cmark & \cmark & \xmark & \cmark & \xmark & \xmark & \cmark \\
FinDABench \citep{findabench_2025} & \cmark & \xmark & \cmark & \xmark & \cmark & \xmark & \xmark & \xmark & \xmark & \cmark \\
AlphaFin \citep{alphafin_2024} & \cmark & \xmark & \xmark & \xmark & \cmark & \cmark & \cmark & \xmark & \xmark & \cmark \\
FinTextQA \citep{fintextqa_2024} & \cmark & \xmark & \xmark & \xmark & \cmark & \xmark & \cmark & \xmark & \xmark & \cmark \\
OmniEval \citep{wang2025omnieval} & \cmark & \xmark & \xmark & \cmark & \cmark & \xmark & \cmark & \xmark & \xmark & \cmark \\
INS-MMBench \citep{Ins_MMBench_2025} & \xmark & \xmark & \xmark & \cmark & \cmark & \xmark & \xmark & \xmark & \cmark & \cmark\\
FinanceQA \citep{FinanceQA_2025} & \cmark & \xmark & \xmark & \xmark & \cmark & \xmark & \xmark & \xmark & \xmark & \cmark\\
XFinBench \citep{XFinbench_2025} & \cmark & \cmark & \xmark & \xmark & \cmark & \xmark & \xmark & \cmark & \xmark & \cmark \\
Finance Agent \citep{fin_agent_bench_2025} & \cmark & \xmark & \xmark & \xmark & \cmark & \xmark & \xmark & \cmark & \xmark & \cmark \\
CFinBench \citep{cfinbench_2025} & \cmark & \cmark & \xmark & \xmark & \cmark & \xmark & \xmark & \cmark & \cmark & \xmark\\
FinMaster \citep{finmaster_2025} & \cmark & \cmark & \cmark & \cmark & \cmark & \xmark & \cmark & \xmark & \xmark & \cmark\\
FinMMR \citep{FinMMR_2025} & \cmark & \cmark & \xmark & \xmark & \cmark & \cmark & \xmark & \xmark & \xmark & \xmark \\
StatLLM \citep{statllm_2026} & \cmark & \xmark & \cmark & \xmark & \cmark & \xmark & \xmark & \xmark & \xmark & \cmark \\ 
SECQUE \citep{secque_2025} & \cmark & \xmark & \cmark & \cmark & \cmark & \xmark & \xmark & \xmark & \xmark & \cmark \\
FAMMA \citep{famma_2025} & \cmark & \cmark & \xmark & \xmark & \cmark & \xmark & \xmark & \xmark & \cmark & \cmark \\
IndustryOR \citep{orlm_2025} & \xmark & \xmark & \xmark & \cmark & \cmark & \xmark & \cmark & \xmark & \xmark & \cmark \\

\midrule
\rowcolor{gray!15}
\textbf{BizCompass (ours)} & \textbf{\cmark} & \textbf{\cmark} & \textbf{\cmark} & \textbf{\cmark} & \textbf{\cmark} & \textbf{\cmark} & \textbf{\cmark} & \textbf{\cmark} & \textbf{\cmark} & \textbf{\cmark}\\
\bottomrule
\end{tabular}
}
\caption{Comparison of existing benchmarks with BizCompass. 
Red \cmark\ indicates full coverage, and green \xmark\ indicates lack of coverage. 
"SC." is short for "Single Choice", "MC." is short for "Multiple Choice", and ``OQ." is short for ``Open Question".
BizCompass achieves comprehensive disciplinary coverage and role-aware task design across analysis, trading, and consulting, while prior benchmarks are limited to narrower domains or tasks.}
\label{tab:com_comparison}
\end{table*}

\begin{figure*}[h]
    \centering
    \includegraphics[width=0.98\linewidth]{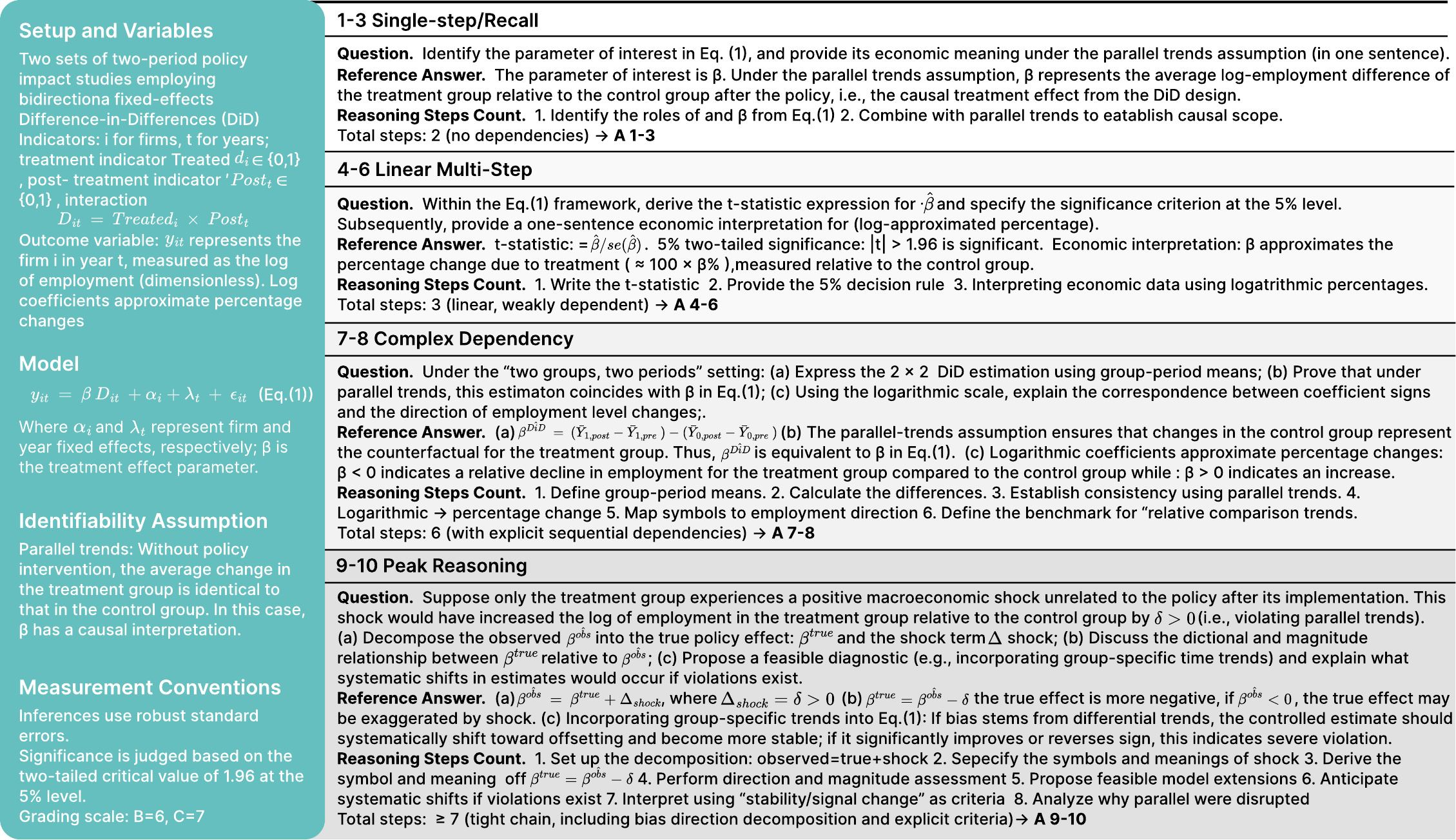}
    \caption{4 examples showing different scores in A (Depth of Reasoning) dimension.}
    \label{fig:A_examples}
\end{figure*}

\section{Data Sources}\label{app:raw-data} 
\paragraph{Economics.}
Economics departments favour outlets that jointly signal methodological rigour, disciplinary reach, and balanced field coverage.  Bibliometric surveys place the “Top Five” general journals—\emph{AER}, \emph{QJE}, \emph{JPE}, \emph{Econometrica}, and \emph{RES}—at the centre of citation influence and promotion decisions \citep{econ_support1,econ_support2}.  
To avoid overweighting these flagships, we add the top‐ranked field titles that remain in the overall top-30 after field normalisation—\emph{JME}, \emph{JIE}, \emph{JDE}, \emph{JHR}, and \emph{JOPE}—as identified by multi-criteria and “ambition-adjusted” rankings \citep{econ_support3,econ_support4,econ_support5}.  
Including \emph{Experimental Economics} secures coverage of laboratory and behavioural methods \citep{econ_support6}.  
Meta-science work shows that papers in these venues receive markedly more citations than the discipline average, ensuring durable visibility \citep{econ_support7}.
The selected core Economics Journal details are shown in Table~\ref{tab:core_econ_journals}.

\begin{table*}[h!]
    \centering
    \small
    \renewcommand{\arraystretch}{1.2} 
    \setlength{\tabcolsep}{4pt} 
    \begin{tabularx}{\textwidth}{
        @{\hspace{1em}} 
        >{\hsize=1.3\hsize}L 
        >{\hsize=0.7\hsize}L 
        @{}
    }
        \toprule
        \bfseries Full title & \bfseries Website \\
        
        \addlinespace[0.5em]
        \pubheader{American Economic Association} \\
        \addlinespace[0.2em]
        American Economic Journal: Applied Economics & \href{https://www.aeaweb.org/journals/app}{\textit{aeaweb.org/journals/app}} \\
        American Economic Journal: Economic Policy & \href{https://www.aeaweb.org/journals/pol}{\textit{aeaweb.org/journals/pol}} \\
        American Economic Journal: Macroeconomics & \href{https://www.aeaweb.org/journals/mac}{\textit{aeaweb.org/journals/mac}} \\
        American Economic Journal: Microeconomics & \href{https://www.aeaweb.org/journals/mic}{\textit{aeaweb.org/journals/mic}} \\
        American Economic Review (incl. Papers \& Proceedings) & \href{https://www.aeaweb.org/journals/aer}{\textit{aeaweb.org/journals/aer}} \\
        
        \addlinespace[0.8em]
        \pubheader{Springer Nature} \\
        \addlinespace[0.2em]
        Experimental Economics & \href{https://www.springer.com/journal/10683}{\textit{springer.com/journal/10683}} \\
        Journal of Economic Growth & \href{https://www.springer.com/journal/10887}{\textit{springer.com/journal/10887}} \\

        \addlinespace[0.8em]
        \pubheader{Oxford University Press} \\
        \addlinespace[0.2em]
        The Economic Journal & \href{https://academic.oup.com/ej}{\textit{academic.oup.com/ej}} \\
        Journal of the European Economic Association & \href{https://academic.oup.com/jeea}{\textit{academic.oup.com/jeea}} \\
        Quarterly Journal of Economics & \href{https://academic.oup.com/qje}{\textit{academic.oup.com/qje}} \\
        Review of Economic Studies & \href{https://academic.oup.com/restud}{\textit{academic.oup.com/restud}} \\

        \addlinespace[0.8em]
        \pubheader{Elsevier} \\
        \addlinespace[0.2em]
        Journal of Development Economics & \href{https://www.sciencedirect.com/journal/journal-of-development-economics}{\textit{sciencedirect.com/j/develop-econ}} \\
        Journal of Economic Theory & \href{https://www.sciencedirect.com/journal/journal-of-economic-theory}{\textit{sciencedirect.com/j/econ-theory}} \\
        Journal of International Economics & \href{https://www.sciencedirect.com/journal/journal-of-international-economics}{\textit{sciencedirect.com/j/intl-econ}} \\
        Journal of Monetary Economics & \href{https://www.sciencedirect.com/journal/journal-of-monetary-economics}{\textit{sciencedirect.com/j/monetary-econ}} \\
        Journal of Econometrics & \href{https://www.sciencedirect.com/journal/journal-of-econometrics}{\textit{sciencedirect.com/j/econometrics}} \\
        Journal of Public Economics & \href{https://www.sciencedirect.com/journal/journal-of-public-economics}{\textit{sciencedirect.com/j/public-econ}} \\
        Review of Economic Dynamics & \href{https://www.sciencedirect.com/journal/review-of-economic-dynamics}{\textit{sciencedirect.com/j/rev-econ-dynam}} \\

        \addlinespace[0.8em]
        \pubheader{University of Wisconsin Press} \\
        \addlinespace[0.2em]
        Journal of Human Resources & \href{https://jhr.uwpress.org}{\textit{jhr.uwpress.org}} \\
        
        \addlinespace[0.8em]
        \pubheader{University of Chicago Press} \\
        \addlinespace[0.2em]
        Journal of Labor Economics & \href{https://www.journals.uchicago.edu/journals/jole}{\textit{journals.uchicago.edu/j/jole}} \\
        Journal of Political Economy & \href{https://www.journals.uchicago.edu/journals/jpe}{\textit{journals.uchicago.edu/j/jpe}} \\

        \addlinespace[0.8em]
        \pubheader{Wiley--Blackwell} \\
        \addlinespace[0.2em]
        Quantitative Economics & \href{https://qeconomics.org}{\textit{qeconomics.org}} \\
        Econometrica & \href{https://www.econometricsociety.org/publications/econometrica}{\textit{econometricsociety.org/econometrica}} \\
        Theoretical Economics & \href{https://econtheory.org}{\textit{econtheory.org}} \\
        RAND Journal of Economics & \href{https://onlinelibrary.wiley.com/journal/17562171}{\textit{wiley.com/journal/17562171}} \\
        International Economic Review & \href{https://onlinelibrary.wiley.com/journal/14682354}{\textit{wiley.com/journal/14682354}} \\
        Journal of Applied Econometrics & \href{https://onlinelibrary.wiley.com/journal/10991255}{\textit{wiley.com/journal/10991255}} \\
        
        \addlinespace[0.8em]
        \pubheader{Cambridge University Press} \\
        \addlinespace[0.2em]
        Econometric Theory & \href{https://www.cambridge.org/core/journals/econometric-theory}{\textit{cambridge.org/j/econometric-theory}} \\

        \addlinespace[0.8em]
        \pubheader{MIT Press} \\
        \addlinespace[0.2em]
        Review of Economics and Statistics & \href{https://direct.mit.edu/rest}{\textit{direct.mit.edu/rest}} \\

        \bottomrule
    \end{tabularx}
    \caption{Core Economics Journals Selected as Sources for the BizCompass: Publishers and URLs}
    \label{tab:core_econ_journals}
\end{table*}

\paragraph{Finance.}
Finance scholarship also values outlets that deliver methodological rigour, cross-field impact, and field balance.  
Citation and promotion studies show that the discipline’s “Top Three” general journals—\emph{Journal of Finance}, \emph{Journal of Financial Economics}, and \emph{Review of Financial Studies}—dominate career-making citations \citep{fin_support1,fin_support2,fin_support3}.  
To provide breadth, we add leading field journals that rank A/A$^{+}$ in active-scholar assessments (e.g.\ \emph{Journal of Corporate Finance}, \emph{Review of Finance}, \emph{Review of Asset Pricing Studies}, \emph{Journal of Financial Intermediation}, \emph{Journal of Financial Stability}, \emph{Journal of Financial Markets}, \emph{Journal of International Money and Finance}) together with quantitative-methods, insurance/actuarial, and real-asset titles.  
Combined, the 30-journal set captures the citation core of contemporary financial economics (details are shown in Table~\ref{tab:core_finance_journals}).

\begin{table*}[h!]
    \centering
    \small
    \renewcommand{\arraystretch}{1.2} 
    \setlength{\tabcolsep}{4pt} 
    \begin{tabularx}{\textwidth}{
        @{\hspace{1em}} 
        >{\hsize=1.3\hsize}L 
        >{\hsize=0.7\hsize}L 
        @{}
    }
        \toprule
        \bfseries Full title & \bfseries Website \\
        
        \addlinespace[0.5em]
        \pubheader{Wiley--Blackwell} \\
        \addlinespace[0.2em]
        European Financial Management & \href{https://onlinelibrary.wiley.com/journal/1468036x}{\textit{wiley.com/journal/1468036x}} \\
        Journal of Finance & \href{https://onlinelibrary.wiley.com/journal/15406261}{\textit{wiley.com/journal/15406261}} \\
        Journal of Risk and Insurance & \href{https://onlinelibrary.wiley.com/journal/15396975}{\textit{wiley.com/journal/15396975}} \\
        Mathematical Finance & \href{https://onlinelibrary.wiley.com/journal/14679965}{\textit{wiley.com/journal/14679965}} \\
        Real Estate Economics & \href{https://onlinelibrary.wiley.com/journal/15406229}{\textit{wiley.com/journal/15406229}} \\
        Financial Review & \href{https://onlinelibrary.wiley.com/journal/15406295}{\textit{wiley.com/journal/15406295}} \\
        International Review of Finance & \href{https://onlinelibrary.wiley.com/journal/14682443}{\textit{wiley.com/journal/14682443}} \\
        Journal of Financial Research & \href{https://onlinelibrary.wiley.com/journal/14756803}{\textit{wiley.com/journal/14756803}} \\
        Journal of Applied Corporate Finance & \href{https://onlinelibrary.wiley.com/journal/17456622}{\textit{wiley.com/journal/17456622}} \\
        Corporate Governance: An International Review & \href{https://onlinelibrary.wiley.com/journal/14678683}{\textit{wiley.com/journal/14678683}} \\
        
        \addlinespace[0.8em]
        \pubheader{Elsevier} \\
        \addlinespace[0.2em]
        Journal of Corporate Finance & \href{https://www.journals.elsevier.com/journal-of-corporate-finance}{\textit{elsevier.com/j/corp-finance}} \\
        Journal of Financial Intermediation & \href{https://www.journals.elsevier.com/journal-of-financial-intermediation}{\textit{elsevier.com/j/fin-intermed}} \\
        Journal of Financial Markets & \href{https://www.journals.elsevier.com/journal-of-financial-markets}{\textit{elsevier.com/j/fin-markets}} \\
        Journal of Financial Stability & \href{https://www.journals.elsevier.com/journal-of-financial-stability}{\textit{elsevier.com/j/fin-stability}} \\
        Insurance: Mathematics and Economics & \href{https://www.journals.elsevier.com/insurance-mathematics-and-economics}{\textit{elsevier.com/j/ins-math-econ}} \\
        Journal of Financial Economics & \href{https://www.journals.elsevier.com/journal-of-financial-economics}{\textit{elsevier.com/j/fin-econ}} \\
        Journal of International Money and Finance & \href{https://www.journals.elsevier.com/journal-of-international-money-and-finance}{\textit{elsevier.com/j/intl-money-fin}} \\

        \addlinespace[0.8em]
        \pubheader{Springer Nature} \\
        \addlinespace[0.2em]
        Journal of Financial Services Research & \href{https://www.springer.com/journal/10693}{\textit{springer.com/journal/10693}} \\
        Journal of Real Estate Finance and Economics & \href{https://www.springer.com/journal/11146}{\textit{springer.com/journal/11146}} \\
        Review of Quantitative Finance and Accounting & \href{https://www.springer.com/journal/11156}{\textit{springer.com/journal/11156}} \\

        \addlinespace[0.8em]
        \pubheader{Taylor \& Francis} \\
        \addlinespace[0.2em]
        Applied Financial Economics & \href{https://www.tandfonline.com/toc/rafe20/current}{\textit{tandfonline.com/toc/rafe20}} \\
        European Journal of Finance & \href{https://www.tandfonline.com/toc/rejf20/current}{\textit{tandfonline.com/toc/rejf20}} \\
        North American Actuarial Journal & \href{https://www.tandfonline.com/toc/uaaj20/current}{\textit{tandfonline.com/toc/uaaj20}} \\
        Scandinavian Actuarial Journal & \href{https://www.tandfonline.com/toc/sajf20/current}{\textit{tandfonline.com/toc/sajf20}} \\
        
        \addlinespace[0.8em]
        \pubheader{Cambridge University Press for the Int. Actuarial Association} \\
        \addlinespace[0.2em]
        ASTIN Bulletin & \href{https://www.cambridge.org/core/journals/astin-bulletin}{\textit{cambridge.org/j/astin-bulletin}} \\
        
        \addlinespace[0.8em]
        \pubheader{Oxford University Press} \\
        \addlinespace[0.2em]
        Review of Financial Studies & \href{https://academic.oup.com/rfs}{\textit{academic.oup.com/rfs}} \\
        Review of Finance & \href{https://academic.oup.com/rof}{\textit{academic.oup.com/rof}} \\
        Review of Asset Pricing Studies & \href{https://academic.oup.com/raps}{\textit{academic.oup.com/raps}} \\
        Review of Corporate Finance Studies & \href{https://academic.oup.com/rcfs}{\textit{academic.oup.com/rcfs}} \\
        Journal of Financial Econometrics & \href{https://academic.oup.com/jfec}{\textit{academic.oup.com/jfec}} \\

        \bottomrule
    \end{tabularx}
    \caption{Core Finance Journals Selected as Sources for the BizCompass: Publishers and URLs}
    \label{tab:core_finance_journals}
\end{table*}

\paragraph{Statistics.}
Citation capital in statistics is highly concentrated.  
Harter and Luby \cite{stats_support1} show that the ten most-cited statistics journals capture just over half ($\approx52\%$) of all within-field citations, while Ruiz-Castillo and Waltman \cite{stats_support2} report a journal-level Gini coefficient of $\approx0.64$—well above economics’ $0.47$.  
The American Statistical Association’s tenure guidelines acknowledge this skew by naming only a handful of core venues \citep{stats_support3}.  
Our eight-journal list—covering theory (\emph{Biometrika}, \emph{JRSS B}), computation/visualisation (\emph{CS\&DA}, \emph{JCGS}), high-dimensional inference (\emph{JMVA}), public-health methods (\emph{Biostatistics}), policy applications (\emph{JRSS C}), and a generalist forum (\emph{Scand.\ J.\ Stat.})—thus captures the field’s reputational and citation nucleus without diluting prestige. The detailed Statistics journal sources are shown in Table~\ref{tab:core_statistics_journals}.

\begin{table*}[h!]
    \centering
    \small
    \renewcommand{\arraystretch}{1.2} 
    \setlength{\tabcolsep}{4pt} 
    \begin{tabularx}{\textwidth}{
        @{\hspace{1em}} 
        >{\hsize=1.3\hsize}L 
        >{\hsize=0.7\hsize}L 
        @{}
    }
        \toprule
        \bfseries Full title & \bfseries Website \\
        
        \addlinespace[0.5em]
        \pubheader{Oxford University Press} \\
        \addlinespace[0.2em]
        Biometrika & \href{https://academic.oup.com/biomet}{\textit{academic.oup.com/biomet}} \\
        Biostatistics & \href{https://academic.oup.com/biostatistics}{\textit{academic.oup.com/biostatistics}} \\
        JRSS Series B (Statistical Methodology) & \href{https://academic.oup.com/jrsssb}{\textit{academic.oup.com/jrsssb}} \\
        JRSS Series C (Applied Statistics) & \href{https://academic.oup.com/jrssc}{\textit{academic.oup.com/jrssc}} \\

        \addlinespace[0.8em]
        \pubheader{Elsevier} \\
        \addlinespace[0.2em]
        Computational Statistics \& Data Analysis & \href{https://www.journals.elsevier.com/computational-statistics-and-data-analysis}{\textit{elsevier.com/j/comp-stat-data}} \\
        Journal of Multivariate Analysis & \href{https://www.journals.elsevier.com/journal-of-multivariate-analysis}{\textit{elsevier.com/j/multivar-anlys}} \\

        \addlinespace[0.8em]
        \pubheader{Taylor \& Francis} \\
        \addlinespace[0.2em]
        Journal of Computational and Graphical Statistics & \href{https://www.tandfonline.com/toc/ucgs20/current}{\textit{tandfonline.com/toc/ucgs20}} \\
        Journal of Business \& Economic Statistics & \href{https://www.tandfonline.com/toc/ubes20/current}{\textit{tandfonline.com/toc/ubes20}} \\

        \addlinespace[0.8em]
        \pubheader{Wiley--Blackwell} \\
        \addlinespace[0.2em]
        Scandinavian Journal of Statistics & \href{https://onlinelibrary.wiley.com/journal/14679469}{\textit{wiley.com/journal/14679469}} \\

        \bottomrule
    \end{tabularx}
    \caption{Core Statistics Journals Selected as Sources for the BizCompass: Publishers and URLs}
    \label{tab:core_statistics_journals}
\end{table*}

\paragraph{Operations Management.}
OM/OR research is organized around a compact core. A Web of Science--based analysis for 2018--2022 by Sodhi and Tang shows that four INFORMS titles, namely \emph{Management Science}, \emph{Operations Research}, \emph{Transportation Science}, and \emph{INFORMS Journal on Applied Analytics} (formerly \emph{Interfaces}), collectively account for $\approx51\%$ of all citations in the domain \citep{om_support1}. These journals hold high-tier AJG/ABS-2021 ratings and feature prominently in Harzing’s Journal Quality List \citep{om_support2,om_support3}.

Our initial source set therefore comprises these INFORMS anchors together with widely recognized OM outlets from Wiley (\emph{Decision Sciences}; \emph{Journal of Operations Management}; \emph{Naval Research Logistics}), Taylor \& Francis (\emph{IISE Transactions}; \emph{International Journal of Production Research}), Springer (\emph{Journal of Optimization Theory and Applications}; \emph{Journal of Scheduling}; \emph{Mathematical Programming}; \emph{Mathematical Programming Computation}), and Elsevier (\emph{Operations Research Letters}). Inclusion is justified by standing in the AJG/ABS-2021 guide (typically $\geq$3 in Operations \& Technology), consistent placement in JQL, and sponsorship by leading professional societies (e.g., the Mathematical Optimization Society for \emph{Mathematical Programming}; the Institute of Industrial and Systems Engineers for \emph{IISE Transactions}) \citep{om_support4,om_support5}. The detailed Operation Management journal sources are shown in Table~\ref{tab:core_om_journals}.

\begin{table*}[h!]
    \centering
    \small
    \renewcommand{\arraystretch}{1.2} 
    \setlength{\tabcolsep}{4pt} 
    \begin{tabularx}{\textwidth}{
        @{\hspace{1em}} 
        >{\hsize=1.3\hsize}L 
        >{\hsize=0.7\hsize}L 
        @{}
    }
        \toprule
        \bfseries Full title & \bfseries Website \\
        
        \addlinespace[0.5em]
        \pubheader{INFORMS} \\
        \addlinespace[0.2em]
        Management Science & \href{https://pubsonline.informs.org/journal/mnsc}{\textit{pubsonline.informs.org/j/mnsc}} \\
        Operations Research & \href{https://pubsonline.informs.org/journal/opre}{\textit{pubsonline.informs.org/j/opre}} \\
        Transportation Science & \href{https://pubsonline.informs.org/journal/trsc}{\textit{pubsonline.informs.org/j/trsc}} \\
        INFORMS Journal on Applied Analytics (formerly \emph{Interfaces}) & \href{https://pubsonline.informs.org/journal/inte}{\textit{pubsonline.informs.org/j/inte}} \\

        \addlinespace[0.8em]
        \pubheader{Wiley} \\
        \addlinespace[0.2em]
        Decision Sciences & \href{https://onlinelibrary.wiley.com/journal/15405915}{\textit{wiley.com/journal/15405915}} \\
        Journal of Operations Management & \href{https://onlinelibrary.wiley.com/journal/18731317}{\textit{wiley.com/journal/18731317}} \\
        Naval Research Logistics & \href{https://onlinelibrary.wiley.com/journal/15206750}{\textit{wiley.com/journal/15206750}} \\

        \addlinespace[0.8em]
        \pubheader{Taylor \& Francis} \\
        \addlinespace[0.2em]
        IISE Transactions (formerly \emph{IIE Transactions}) & \href{https://www.tandfonline.com/toc/uiie20/current}{\textit{tandfonline.com/toc/uiie20}} \\
        International Journal of Production Research & \href{https://www.tandfonline.com/toc/tprs20/current}{\textit{tandfonline.com/toc/tprs20}} \\

        \addlinespace[0.8em]
        \pubheader{Springer} \\
        \addlinespace[0.2em]
        Journal of Optimization Theory and Applications & \href{https://link.springer.com/journal/10957}{\textit{springer.com/journal/10957}} \\
        Journal of Scheduling & \href{https://link.springer.com/journal/10951}{\textit{springer.com/journal/10951}} \\
        Mathematical Programming & \href{https://link.springer.com/journal/10107}{\textit{springer.com/journal/10107}} \\
        Mathematical Programming Computation & \href{https://link.springer.com/journal/12532}{\textit{springer.com/journal/12532}} \\

        \addlinespace[0.8em]
        \pubheader{Elsevier} \\
        \addlinespace[0.2em]
        Operations Research Letters & \href{https://www.journals.elsevier.com/operations-research-letters}{\textit{elsevier.com/j/op-res-letters}} \\

        \bottomrule
    \end{tabularx}
    \caption{Core Operations Management Journals Selected as Sources for the BizCompass: Publishers and URLs}
    \label{tab:core_om_journals}
\end{table*}

\begin{table*}[t]
\centering
\resizebox{1\linewidth}{!}{
\begin{tabular}{@{}l| *{2}{c} | *{3}{c} | *{2}{c} | *{2}{c} | *{2}{c} | c | *{2}{c} | *{2}{c} @{}}
\toprule
\multirow{2}{*}{\textbf{Source Category}}
  & \multicolumn{2}{c|}{\textbf{AP}}
  & \multicolumn{3}{c|}{\textbf{MTA}}
  & \multicolumn{2}{c|}{\textbf{RM}}
  & \multicolumn{2}{c|}{\textbf{SD}}
  & \multicolumn{2}{c|}{\textbf{PM}}
  & \multicolumn{1}{c|}{\textbf{AT}}
  & \multicolumn{2}{c|}{\textbf{FD}}
  & \multicolumn{2}{c}{\textbf{FDA}} \\
\cmidrule(lr){2-3}\cmidrule(lr){4-6}\cmidrule(lr){7-8}\cmidrule(lr){9-10}
\cmidrule(lr){11-12}\cmidrule(lr){13-13}\cmidrule(lr){14-15}\cmidrule(lr){16-17}
  & A & T
  & A & T & C
  & A & C
  & T & C
  & A & C
  & T
  & A & C
  & A & C \\
\midrule
Practitioner-oriented Textbooks  & \cmark & \xmark & \xmark & \xmark & \xmark & \cmark & \cmark & \xmark & \xmark & \cmark & \xmark & \cmark & \xmark & \xmark & \xmark & \xmark \\
Business Documents  & \xmark & \xmark & \xmark & \xmark & \xmark & \xmark & \xmark & \cmark & \cmark & \xmark & \cmark & \xmark & \xmark & \xmark & \xmark & \xmark \\
Existing Benchmarks & \xmark & \cmark & \cmark & \cmark & \cmark & \cmark & \xmark & \xmark & \xmark & \xmark & \xmark & \xmark & \cmark & \cmark & \cmark & \cmark \\
\bottomrule
\end{tabular}
}

\begin{tablenotes}[flushleft]
\scriptsize 
\item \textbf{Sources for ticks:}
AP–T: CIKM18 \citep{wu2018hybrid};
MTA–A: FPB \citep{Malo2014GoodDO};
MTA–T: M\&A \citep{yang2020generating};
MTA–C: FOMC \citep{shah2023trillion};
FD–A: CCFraud \citep{feng2023empowering};
FD–C: Taiwan \citep{feng2023empowering};
FDA–A: FNXL \citep{sharma2023financial}, FinQA \citep{chen2021finqa}, TATQA \citep{zhu2021tat}, FinRED \citep{sharma2022finred};
FDA–C: EDTSUM \citep{zhou2021trade}.
\end{tablenotes}

\caption{Application-Based Tasks across four source categories.
AP=Asset Pricing; MTA=Market Trend Analysis; RM=Risk Management; SD=Strategy Development; PM=Portfolio Management; AT=Algorithmic Trading; FD=Fraud Detection; FDA=Financial Document Analysis.
Subcolumns A/T/C denote Analysis/Trading/Consulting. (\cmark) indicates the task–role pair \emph{uses data} from the given source category; (\xmark) indicates it \emph{does not use} data from that source.}
\label{tab:source_task_role_matrix}
\end{table*}

\section{Benchmark Construction}\label{app:construct}
\subsection{Phase1: Data Preprocessing}
\paragraph{Document Conversion.}
We convert all raw material from PDF to Markdown using MinerU~\citep{wang2024mineruopensourcesolutionprecise,he2024opendatalab}. 
MinerU integrates layout-aware parsing (DocLayout-YOLO) and OCR (UniMERNet) \citep{wang2024unimernetuniversalnetworkrealworld,wang2024cdmreliablemetricfair} to preserve structural elements such as section headings, mathematical formulas, tables, and inline expressions.
\paragraph{Data Cleaning.}
We pre-clean the converted Markdown raw material to obtain the analysis-ready corpus. First, we perform macro deletion: remove sections unrelated to question construction (Abstract, Related Work, Acknowledgements, Funding, Author/Ethics statements, References, and non-core appendices) and delete all figure content and containers.
Second, we sanitize the remaining text: strip in-text citation markers, footnotes, and links while preserving the original wording, equations, tables, and section order; discard paragraphs with insufficient content or pedagogical value.

\subsection{Phase2: Question Construction}
Due to the limited long-context understanding ability of LLMs, our experiments show that directly generating high-quality benchmark samples is impractical, as it exposes three major problems: the question stem and answering prerequisites are not closed, so key conditions are hard to specify in a single pass, largely because models are optimized to answer rather than to question incomplete premises and struggle to maintain reasoning consistency over long documents \citep{li2024mediq, bai2024longbench}; answers are not traceable to explicit evidence, and cross-references can induce out-of-text inference; and the lack of auditable intermediate artifacts makes errors hard to localize and reproduce. Therefore, we adopt a six-step pipeline that decomposes construction into independent and verifiable stages. Each stage addresses one decision point, relies only on explicit evidence from the corpus, and outputs traceable intermediates and logs. This design preserves efficiency while improving item quality.
\paragraph{Section Summarization.}
To convert an unstructured corpus into structured evaluation units, this stage constructs knowledge units. First, we extract only three types of entities that are essential to a paper's core argument: key equations (e.g., identification equations and equilibrium conditions), core propositions (e.g., theorems and hypotheses), and key tables. The extraction rules are explicitly defined; for example, tables must report baseline regression results or key identification test outcomes. Subsequently, the process aggregates these entities according to a Holistic Integrity Principle. When a proposition and its associated key equation together form a self-contained logical unit, they are integrated into a macro knowledge point in order to preserve the complete argumentative structure. Finally, each atomic or macro knowledge point is assigned one or more labels from a predefined set of question types. By combining the pattern recognition capabilities of the LLM with explicit rules, we ensure that the extraction of knowledge units is both traceable and faithful to the original logical structure of the paper.
\paragraph{Question Decomposition.}
To assess integrative reasoning beyond factual recall, the core task at this stage is to synthesize isolated knowledge points into complex questions. We employ a process that uses the Question Type labels of a knowledge point as input to generate a synthetic instruction. This instruction directs the LLM to integrate the analytical dimensions implied by all associated labels and produce an initial question draft. For example, if a knowledge point is tagged with Methodology and Boundary Condition, the resulting instruction guides the LLM to formulate a question that requires explaining the method while also discussing its underlying assumptions. This approach shifts the assessment from factual recall to the ability to integrate knowledge. By anchoring question generation to traceable type labels, we enhance the transparency of intent in the construction of each question.

\paragraph{Retrieval \& Backfilling.}
To ensure each question is self-contained, this stage's core task is to ground the query draft in the source text. 
We employ an in-context retrieval strategy, leveraging the long context window of Gemini 2.5 Pro. This LLM-as-Retriever approach is chosen over traditional chunk-based Retrieval-Augmented Generation (RAG)~\citep{lewis2020retrieval} to mitigate known issues such as information fragmentation and the difficulty in capturing long-range dependencies~\citep{sarthi2024raptor,han2024retrieval}.
By providing the full document as context, the model first locates and extracts verbatim the core passage that directly answers the question, which constitutes the answer draft. 
Second, it synthesizes a comprehensive set of evidence, including all necessary background like variable definitions, model assumptions, and key context. Adherence to a zero-paraphrasing principle is strictly enforced through prompting and automated checks. By constructing a \textit{\textbf{Question-Answer-Evidence}} triplet, this stage ensures the verifiability of every question, establishing a foundation for reliable downstream evaluation.

\paragraph{Restructure.}
To transform discrete \textit{Evidence} sets into structured units, our process begins by mapping each \textit{Evidence} set to a standardized template: descriptive context and variable definitions are organized into the \textit{Background}, core equations and data are placed in the \textit{Model} section, and the initial question draft serves as the basis of the \textit{Question}.

To enrich the processed cases with greater cognitive depth, we design executable textual constraints requiring that each question integrate multiple pieces of evidence and that its sub-questions follow a logical progression. To assess higher-order synthesis, the process further merges complementary cases into a more comprehensive question that forms a complete analytical chain, thereby simulating the multi-step reasoning required in actual research.

\paragraph{Scoring \& Filtering.}
To ensure the quality of the question set, all generated questions must pass a data-driven filtering procedure that evaluates them using a quantitative scoring function across three dimensions: Reasoning Depth, Knowledge Synthesis, and Conceptual Centrality (see Appendix~\ref{subsubsec:rubric}). This filtering step helps maintain quality consistency and reduces the workload of subsequent manual evaluation.

\paragraph{Type Conversion.}z
This stage optimizes the assessment format by matching the most suitable question type to each evaluation objective. A question is converted into a \textit{Choice Question} when its answer space is relatively convergent and common fallacies can be used to construct plausible distractors; otherwise, it is retained as a \textit{QA} problem. For choice questions, we leverage a fallacy knowledge base to construct distractors, as described in Appendix~\ref{subsec:fkl}. This process pairs the depth of QA-style reasoning with the efficiency and objectivity of choice questions~\citep{haladyna2004developing,ozuru2013comparing}.
\subsection{Phase3: Evaluation}
\paragraph{Framework Overview.}
We propose an expert-driven \textit{Dual-Track Evaluation Strategy} for the two task families in the benchmark. 
The strategy focuses on item quality and proceeds in two stages: 
(i) Stratified expert sampling and recruitment; 
(ii) Dual-Track Multi-Dimensional Evaluation and Filtering. 

\paragraph{Expert Recruitment.}
We use stratified purposive sampling to form two non-overlapping panels~\citep{stratton2024purposeful}, each confined to its track:
(i) \textit{Specialist Advisory Panel} for Knowledge-Based items (\(n=40\), equally split across Operations Management, Statistics, Finance, and Economics). Members hold at least a Master's degree, with 57.5\% holding or currently pursuing a Ph.D. in their respective fields. 
(ii) \textit{Core Practitioner Panel} for Application-Based items (\(n=30\), equally split across Analyst, Trader, and Consultant roles).
Members possess a relevant degree and at least three years of verifiable full-time industry experience. Candidates were sourced from universities, professional communities, certification rosters (\eg, CFA/FRM), and alumni networks of leading finance and consulting firms.

\paragraph{Evaluation and Filtering Rules.}
To implement our dual-track validation framework and ensure robustness, we adopted a dual-review coverage design, ensuring that every candidate item was independently adjudicated by at least two experts.
To prevent cognitive fatigue, items were distributed in batches over a 2-week review period.
Specifically, each expert in the knowledge-based track reviewed approximately 400 items, while each expert in the application-based track reviewed approximately 600 items.
During the review, experts provided independent scores against predefined quality dimensions.
To handle disagreements, we implemented a strict \textit{``Review-Discuss-Resolve''} protocol.
Items receiving conflicting decisions or low-quality scores were flagged as ``discrepant items''.
Experts were required to discuss and decide on the final label.
If consensus could not be reached after two rounds of discussion, the item was strictly discarded to maintain the high quality of the final dataset.
The following sections detail the specific evaluation dimensions that underlie this review process:

\paragraph{Track A: Knowledge-Based Tasks.}
This track ensures that knowledge items measure LLMs’ mastery of the theoretical foundations in Operations Management, Statistics, Finance, and Economics, with attention to core concepts.
\begin{itemize}[leftmargin=*,labelsep=5pt]
    \item \textbf{Clarity.} Clarity is the basis of valid assessment. In psychometrics, ambiguity introduces construct-irrelevant variance, meaning scores are influenced by factors unrelated to the target construct (e.g., general language decoding), which harms reliability and validity \citep{bandalos2018measurement}. In LLM benchmarking, an unclear item probes guessing under vague instructions rather than domain knowledge. Ensuring clarity is the first step toward face validity, fixing the semantic ground on which answers are given and allowing us to isolate the model’s domain knowledge \citep{taherdoost2016,wu2025enhancing}. \textit{\textbf{Core Question:} Is the problem statement precise and unambiguous?}

    \item \textbf{Conceptual centrality.} Research in cognitive science shows that domain knowledge forms an interdependent network rather than a set of isolated facts \citep{sloman1993feature}. A concept’s centrality depends on how many other concepts rely on it. Scientific discourse analyses also find that sentences tied to central arguments are more informative and salient \citep{teufel2002summarizing}. Items that focus on core concepts, such as opportunity cost in economics or the central limit theorem in statistics, carrying greater diagnostic value than those focused on peripheral facts. They more effectively assess whether a model has internalized the principles that drive reasoning, rather than merely memorized fragments. \textit{\textbf{Core Question:} Does the item test a core concept of the discipline?}

    \item \textbf{Domain representativeness.} This dimension instantiates content validity: the extent to which an instrument covers the important parts of the domain \citep{haynes1995content, taherdoost2016}. A benchmark with strong content validity samples items that represent the field’s overall structure. This reduces “gaming the benchmark,” where a model overfits narrow subtopics without global understanding. Expert judgments on representativeness ensure breadth and balance so the benchmark serves as a valid proxy for composite competence. \textit{\textbf{Core Question:} Does the item reflect knowledge typically expected of practitioners in this field?}

    \item \textbf{Cognitive complexity.} Cognitive complexity evaluates the depth and level of processes needed to solve the item. Following Bloom’s Revised Taxonomy, processes span remember, understand, apply, analyze, evaluate, and create \citep{anderson2001taxonomy}. A high-quality benchmark includes a sufficient share of items that require higher-order processes (e.g., analysis, evaluation) to differentiate reasoning ability \citep{rush2016impact}. This dimension filters recall-only items and ensures the bank elicits and measures multi-step reasoning, concept integration, and critical judgment rather than mere information retrieval. \textit{\textbf{Core Question:} What level of cognition is required, from recall to multi-step reasoning?}
\end{itemize}

\paragraph{Track B: Application-Based Tasks.}
This track ensures that application items reflect realistic business conditions and align with the core competencies of specific roles.
\begin{itemize}[leftmargin=*,labelsep=5pt]
    \item \textbf{Clarity.} As in Track A, clarity is essential. An unclear business scenario forces the model to assume missing conditions, turning the task from a reasoning test into a guessing exercise. This undermines reliability—the reproducibility of results. Clear items allow us to attribute performance differences to underlying ability rather than arbitrary interpretations \citep{taherdoost2016}. \textit{\textbf{Core Question:} Are the scenario and task clearly described with explicit intent?}

    \item \textbf{Realism.} Realism evaluates consistency with actual business practice. It secures ecological validity, i.e., the degree to which results generalize to real, non-research decision contexts \citep{brunswik1955representative}. An application item that ignores business logic, data constraints, or decision pressure cannot demonstrate practical utility even if a model produces a correct answer. Realism links benchmark performance to real-world potential. \textit{\textbf{Core Question:} Does the item reflect a credible real-world business scenario?}

    \item \textbf{Role-relevance.} This dimension assesses alignment with role responsibilities using standardized Job/Task Analysis (JTA) to establish content validity \citep{lane2006handbook}. JTA is the gold standard in professional testing and certification: subject-matter experts identify and rate the tasks and knowledge essential to each role. Having role-matched experts judge relevance provides evidence for construct validity that the benchmark measures the claimed role-specific abilities. \textit{\textbf{Core Question:} Is the item closely aligned with the core duties of the specified role?}

    \item \textbf{Difficulty.} Difficulty captures the depth of domain knowledge, the length and complexity of the reasoning chain, and the breadth of integration. Assessment theory requires sufficient discriminatory power to separate ability levels \citep{lane2006handbook}. Expert difficulty judgments help build a well-graded bank, reducing ceiling and floor effects and enabling more precise ranking and evaluation across models. \textit{\textbf{Core Question:} What level of combined knowledge, reasoning, and information integration is required?}
\end{itemize}

\section{Question Quality Filter}
\label{subsubsec:rubric}

\subsection{Sample Quality Evaluation Dimensions}

We argue that a good benchmark sample should exhibit sufficient reasoning depth, integrate information comprehensively, and remain closely aligned with the primary claim of raw data. Correspondingly, we propose three dimensions to systematically evaluate sample quality, namely A (Depth of Reasoning), B (Degree of Synthesis), and C (Conceptual Centrality), which we define as follows.

\paragraph{Depth of Reasoning.} Depth of Reasoning captures whether an example goes beyond surface recall to require multi-step inference, making it crucial for distinguishing genuine reasoning ability in LLMs. Classic benchmarks such as GSM8K probe this capacity through stepwise mathematical reasoning~\citep{measure_cobbe2021gsm8k}, and subsequent work on chain-of-thought prompting has shown that explicit intermediate reasoning steps are essential for solving complex problems, underscoring the importance of evaluating reasoning depth ~\citep{measure_wei2022chain}.
The criteria for different scoring bands are classified in Table~\ref{tab:bands}.
To elucidate the distinctions among the four different scoring bands, Figure~\ref{fig:A_examples} illustrates representative questions and corresponding reasoning examples derived from the same background information.

\paragraph{Degree of Synthesis.} 
Degree of Synthesis captures the extent to which an example requires integrating multiple pieces of information into a coherent whole. 
This dimension is essential because real-world problems often demand the fusion of knowledge across domains or sources, making synthesis a key determinant of benchmark quality. 
For instance, HotpotQA requires reasoning across multiple documents to synthesize evidence~\citep{yang2018hotpotqa}, while MultiRC involves combining information scattered across sentences~\citep{khashabi2018looking}.

\paragraph{Conceptual Centrality.} We define conceptual centrality as the closeness of a text unit to the primary claim or contribution of a paper. Prior work in scientific discourse analysis has shown that sentences aligned with central arguments are typically more information-dense and salient than peripheral content~\citep{teufel2002summarizing}, and thus are better suited for generating higher-quality samples.

\paragraph{Operational Definitions.}

\noindent\textit{Information source.}
A source used directly by the question and the answer that determines the conclusion.
Count a source only if removing it would change the answer.
One or more of the following six types:
\begin{itemize}[leftmargin=*,labelsep=5pt]
  \item \textit{Equations or formal conditions} (model relations, constraints, optimality conditions)
  \item \textit{Numerical tables} (coefficients, standard errors, sample sizes, $R^2$)
  \item \textit{Identification or validity assumptions} (explicitly stated conditions)
  \item \textit{Variable and parameter definitions} (meaning, units, transformations)
  \item \textit{Context and sample design} (treatment definitions, timing, scope)
  \item \textit{Estimation and design details} (procedures, settings, data construction)
\end{itemize}

\noindent\textit{Reasoning step.}
An inference step is a necessary operation that yields a new intermediate or final conclusion from the premises. Count one step if and only if at least one of the following conditions holds. Splitting is allowed only when an intermediate statement is necessary and substantive; do not count mere transcription, reading a single table entry, or mechanically breaking one algebraic manipulation into micro-steps:

\begin{itemize}[leftmargin=*,labelsep=5pt]
  \item Deriving a new property or directional claim from a formal condition
  \item Combining two different information sources to obtain a joint conclusion
  \item Performing one statistical decision or interval judgment
  \item Conducting one comparative static, counterfactual, or required simplification within the model
\end{itemize}

\subsection{Sample Quality Scoring}
\label{sec:sample_quality_scoring}
To ensure the quality of benchmark samples, we adopt a two-step filtering framework. It consists of \textit{Multi-dimensional Evaluation}, aggregating the three quality dimensions into a weighted score to capture overall adequacy, 
and \textit{Dimension-Wise Filtering}, enforcing minimum thresholds on each dimension to prevent samples with severe deficiencies from being retained. 
We next describe how these two scoring procedures are aligned with expert annotations.

\paragraph{Operational Scoring Rules and Bands.}
Using the operational definitions of information source and reasoning step, we assign per-example scores on three dimensions \textit{Depth of Reasoning} (A), \textit{Degree of Synthesis} (B) and \textit{Conceptual Centrality} (C). For each dimension, choose an integer from 1–10 by selecting the highest band fully satisfied by the example and provide a one-sentence justification. The band criteria are summarized in Table~\ref{tab:bands}. The overall score is Eq.~\ref{eq:scoring}; weight estimation follows.

\begin{table*}[t]
\small
\centering
\renewcommand{\arraystretch}{1.18}

\begin{tabularx}{\linewidth}{@{}p{1.6cm}Y@{}}
\toprule
\multicolumn{2}{@{}l@{}}{\textbf{A. Depth of Reasoning}}\\
\midrule
1–3  & \textit{Steps:} $\leq$2; \textit{Dependence:} none; \textit{Ops:} no comparative statics or counterfactuals.\\
\rowcolor{gray!5}
4–6  & \textit{Steps:} 3–4; \textit{Dependence:} weak; \textit{Ops:} at least one statistical decision or necessary calculation.\\
\rowcolor{gray!10}
7–8  & \textit{Steps:} 5–6; \textit{Dependence:} later conclusions rely on earlier results; \textit{Ops:} at least one of comparative statics, counterfactual, or multi-stage derivation; \textit{Sources:} connects at least two information sources.\\
\rowcolor{gray!15}
9–10 & \textit{Steps:} $\geq$7 tightly linked; \textit{Dependence:} strong chain; \textit{Ops:} substantive setup change or bias direction/decomposition under violated assumptions with explicit criteria.\\
\midrule
\multicolumn{2}{@{}l@{}}{\textbf{B. Degree of Synthesis}}\\
\midrule
1–3  & Single source only.\\
\rowcolor{gray!5}
4–6  & Two distinct sources, both necessary and used substantively.\\
\rowcolor{gray!10}
7–8  & At least three sources used substantively, including an equation/formal condition and a numerical table.\\
\rowcolor{gray!15}
9–10 & More than three sources used substantively with non-obvious correspondence or mutual constraints; each source is necessary for the main conclusion.\\
\midrule
\multicolumn{2}{@{}l@{}}{\textbf{C. Conceptual Centrality}}\\
\midrule
1–3  & Peripheral to the paper’s main claims.\\
\rowcolor{gray!5}
4–6  & Supportive but not core.\\
\rowcolor{gray!10}
7–8  & Directly serves a key identification condition, core mechanism, or baseline interpretation.\\
\rowcolor{gray!15}
9–10 & Core contribution (e.g., main theorem, baseline headline result, decisive mechanism evidence).\\
\bottomrule
\end{tabularx}
\caption{Operational scoring bands for A–C dimensions.}
\label{tab:bands}
\end{table*}

\paragraph{Multi-dimensional Evaluation.}
Absolute human scores are often subjective and inconsistent across annotators, making them unreliable as direct supervision. 
To address this, we adopt a preference-based approach: experts were asked to compare triplets of examples and provide pairwise preferences 
(e.g., example$_1 \succ$ example$_2$). This yields a set of pairwise constraints $\mathcal{P} = \{(i,j)\}$, 
where $(i,j)$ indicates that example $i$ is preferred over $j$. 

Given the weighted scoring function 
\begin{equation}
\label{eq:scoring}
   S(x) = \alpha A(x) + \beta B(x) + \gamma C(x),
\end{equation}
we require that $S(x_i) > S(x_j)$ whenever $(i,j) \in \mathcal{P}$. 
To enforce this, we adopt a pairwise ranking loss from the learning-to-rank literature~\citep{burges2005learning}:
{\small
\[
\mathcal{L}(\alpha,\beta,\gamma) 
= \sum_{(i,j) \in \mathcal{P}} \log\big(1 + \exp(-(S(x_i) - S(x_j)))\big).
\]
}
Minimizing this loss is equivalent to maximizing the likelihood under a Bradley--Terry preference model, 
which assumes that the probability of preferring $i$ over $j$ increases with $S(x_i)-S(x_j)$. 

In practice, we normalize the coefficients $(\alpha,\beta,\gamma)$ to sum to one, 
and optimize them using gradient-based methods over preference annotations. 
Specifically, 10 human experts provided pairwise preference labels on 500 triplets of benchmark examples, 
yielding a set of training constraints for the ranking model. 
This process yields the final weights $(0.4, 0.4, 0.2)$, 
indicating that \textit{Depth of Reasoning} and \textit{Degree of Synthesis} contribute most strongly to perceived sample quality, 
while \textit{Conceptual Centrality} plays a supportive but non-negligible role.

\paragraph{Dimension-Wise Filtering.}
While the weighted score reflects relative sample quality, it does not directly guarantee the rejection of particularly poor cases. 
We therefore impose a filtering threshold. 
Let $\mathcal{R}$ denote the set of samples rejected by human experts, 
and let $\hat{\mathcal{R}}(T)$ denote the set of samples filtered out by the automatic rule with threshold $T$. 
The rejection recall is defined as:
\[
\text{Recall}_{\text{reject}}(T) = 
\frac{|\mathcal{R} \cap \hat{\mathcal{R}}(T)|}{|\mathcal{R}|}.
\]
This measures the proportion of expert-rejected samples that are also eliminated by the automatic threshold. 
When $T=6.0$, the recall reaches 90\%, showing that our criterion effectively recovers the majority of samples deemed unacceptable by human experts. 
In practice, we additionally require that each dimension score is at least $6.0$, 
preventing examples with severe deficiencies in any single aspect from being retained.



\section{Assessment Type Conversion}
\label{subsec:fkl}
\subsection{Aligning Assessment Formats with Cognitive Constructs}

\paragraph{Rationale for the Dual-Format Assessment Strategy.}
The choice of an assessment format is a fundamental issue of construct validity, centered on the alignment between the evaluation tool and the cognitive processes being measured, rather than mere scoring efficiency \citep{messick1995validity}. Psychometric literature clearly distinguishes between two primary formats: \textit{Constructed-Response (CR)} and \textit{Selected-Response (SR)} \citep{lane2006handbook}. As a typical CR format, the open-ended Question-Answer (QA) assesses active, generative cognitive processes. In contrast, the Multiple-Choice Question (MCQ), a typical SR format, evaluates recognition-based processes that rely more heavily on prior knowledge and familiarity \citep{ozuru2013comparing}.

The core objective of BizCompass is to evaluate the advanced cognitive capabilities of LLMs within professional domains. Expertise in these domains, whether in academic research or business analysis, requires both generative abilities (e.g., composing in-depth analyses) and discriminative abilities (e.g., making precise judgments among viable options) \citep{shanteau1992competence}. Consequently, a benchmark relying on a single format would lack ecological validity, failing to comprehensively measure a model's integrated performance on the diverse tasks required in authentic, complex professional settings \citep{brunswik1955representative}.

This distinction forms the theoretical cornerstone of our protocol. We treat the decision to convert a QA item to an MCQ as a \textit{validity claim}: we posit that, for the specific assessment target, the core cognitive construct is more appropriately measured through discrimination among options rather than through unprompted generation \citep{messick1995validity}. By strategically deploying both formats, our framework simulates a broader spectrum of real-world cognitive demands, thereby providing a more comprehensive and reliable assessment of an LLM's potential in professional domains.

\paragraph{Type Conversion Evaluation Dimensions.}
We argue that a rigorous format conversion decision requires a systematic evaluation of two core issues: the ability to losslessly capture the question's core construct in a multiple-choice format, and the potential for constructing distractors with high value. We therefore propose two dimensions to guide this evaluation.

\begin{itemize}[leftmargin=*,labelsep=5pt]
\item \textbf{Conceptual Clarity.}
Conceptual Clarity is defined as the degree to which a question's core answer is convergent, determining if it can be losslessly captured by a single option. The clarity of the answer space is a cornerstone of assessment objectivity. If the answer space is divergent and permits multiple valid interpretations, the evaluation's focus shifts to the reasoning process rather than the final conclusion, which contradicts the fundamental purpose of multiple-choice questions to measure precise judgment \citep{rush2016impact}.

\item \textbf{Misconception Potential.}
Misconception Potential captures whether a set of common fallacies exists for constructing distractors. The core value of this potential lies in enhancing a choice question's discriminability: its ability to effectively differentiate between examinees of varying proficiency levels. Commonsense reasoning benchmarks such as COSMOS QA demonstrate that high-quality distractors based on common errors are critical for building challenging multiple-choice questions \citep{huang2019cosmos}. If a question does not permit the systematic construction of functional distractors, its multiple-choice version tends to degenerate into a simple pattern-matching test, thereby losing its diagnostic value \citep{taherdoost2016}.

\end{itemize}

\paragraph{Type Conversion Scoring.}
To aggregate the scores from the two dimensions above into a single conversion suitability score, we employ the same preference-based learning methodology of Appendix~\ref{sec:sample_quality_scoring}. This ensures a consistent, data-driven approach to weight determination. The resulting decision function is:
\[
\text{Total Score} = 0.5 \times A + 0.5 \times B
.\]
where A and B represent the scores for Conceptual Clarity and Misconception Potential, respectively. An item is converted to the multiple-choice format only if its total score is at least 9.0; otherwise, it is retained in its original QA format. This high threshold ensures that only items demonstrating exceptional suitability on both dimensions are converted, thereby prioritizing the retention of QA problems that assess deep reasoning.
\subsection{Framework for Distractor Design}
The quality of distractors is an important factor of a multiple-choice question's measurement precision and diagnostic power. To mitigate the validity threats posed by arbitrarily created distractors, we adopt a evidence-based framework for distractor design. This framework is grounded in a core tenet of modern test theory that validity is not something checked after a test is complete, but rather something built into the instrument through a principled development process \citep{haladyna2004developing}.
It categorizes common errors into five core cognitive domains: conceptual, procedural, calculation, logic, and cognitive bias, as detailed in Table~\ref{tab:ddm}.

\begin{table*}[t]

\setlength{\tabcolsep}{4pt}
\renewcommand\arraystretch{1.08}
\begin{tabularx}{\textwidth}{@{} >{\RaggedRight\arraybackslash}p{0.30\textwidth} >{\RaggedRight\arraybackslash}X @{}}
\toprule
\textbf{Error Type} & \textbf{Definition and Cognitive Target} \\
\midrule

\rowcolor{gray!10}
\multicolumn{2}{@{}p{\dimexpr\textwidth-2\tabcolsep\relax}@{}}{\RaggedRight\textit{\textbf{Conceptual Errors}}} \\
Common Misconception &
Based on intuitive beliefs common among learners but contrary to scientific facts \citep{gierl2017developing, radatz1979error}. \textit{Target: Probe stable alternative concepts.} \\

Almost Correct &
Mostly correct but missing one key detail \citep{haladyna2002review}. \textit{Target: Separate precise understanding from surface recall.} \\

Opposite Idea &
States the opposite of the correct idea \citep{gierl2017developing}. \textit{Target: Detect inverted understanding of a core idea.} \\

\midrule
\rowcolor{gray!10}
\multicolumn{2}{@{}p{\dimexpr\textwidth-2\tabcolsep\relax}@{}}{\RaggedRight\textit{\textbf{Procedural Errors}}} \\
Missing Step &
Skips a required step; the distractor is the result after skipping \citep{radatz1979error}. \textit{Target: Expose gaps in procedure or working memory.} \\

Wrong Order &
All steps are present but in the wrong order \citep{radatz1979error}. \textit{Target: Check control of procedure sequence.} \\

Rule Misuse &
Uses the right procedure but misapplies a sub-rule or formula \citep{radatz1979error}. \textit{Target: Locate weak links inside the procedure.} \\

\midrule
\rowcolor{gray!10}
\multicolumn{2}{@{}p{\dimexpr\textwidth-2\tabcolsep\relax}@{}}{\RaggedRight\textit{\textbf{Calculation Errors}}} \\
Sign Error &
Flips the sign during calculation \citep{radatz1979error}. \textit{Target: Check attention to sign rules.} \\

Unit Error &
Uses the wrong unit or converts units incorrectly \citep{radatz1979error}. \textit{Target: Check understanding of units and conversion.} \\

Order Error &
Ignores the standard order of operations \citep{radatz1979error}. \textit{Target: Find gaps in basic rules.} \\

\midrule
\rowcolor{gray!10}
\multicolumn{2}{@{}p{\dimexpr\textwidth-2\tabcolsep\relax}@{}}{\RaggedRight\textit{\textbf{Logical Errors}}} \\
Treating Correlation as Cause &
Takes correlation as causation \citep{wan2024logicasker}. \textit{Target: Distinguish association from cause.} \\

Hasty Generalization &
Concludes from too little or biased evidence \citep{wan2024logicasker}. \textit{Target: Check grasp of representativeness and inference.} \\

Invalid Logic &
Uses an invalid form (e.g., affirming the consequent) \citep{wan2024logicasker}. \textit{Target: Evaluate logical soundness.} \\

\midrule
\rowcolor{gray!10}
\multicolumn{2}{@{}p{\dimexpr\textwidth-2\tabcolsep\relax}@{}}{\RaggedRight\textit{\textbf{Cognitive Bias Errors}}} \\
True but Irrelevant &
Factually correct but irrelevant to the question \citep{forster2008failures}. \textit{Target: Test filtering of relevant information.} \\

Anchor Trap &
A number looks plausible due to an unrelated anchor in the stem \citep{berthet2022impact}. \textit{Target: Check influence of anchoring.} \\

Confirmation Trap &
Fits prior belief but conflicts with given evidence \citep{nickerson1998confirmation, berthet2022impact}. \textit{Target: Test ability to update belief.} \\

\bottomrule

\end{tabularx}
\caption{Error Types Underlying Multiple-Choice Distractors}
\label{tab:ddm}
\end{table*}

\paragraph{Formatting Rules.}
In addition to the content guidelines provided by the framework, all distractors must adhere to strict formatting rules to ensure fairness and prevent cueing. These include maintaining grammatical parallelism, ensuring options are mutually exclusive for single-choice questions, and avoiding the use of "all of the above" or "none of the above".

\section{Hyperparameter Selection}\label{app:hyper}
We conduct hyperparameter optimization for the two different question types to ensure optimal model performance. 
Specifically, we perform grid search on two key sampling parameters: temperature (ranging from 0 to 0.8) and top-p (ranging from 0.6 to 0.95). 

For single-choice questions, we evaluate performance using accuracy as the metric, while for general QA tasks, we employ LLM evaluation scores. 
The grid search results are presented in Table~\ref{tab:hyper}, where the optimal hyperparameter combinations for each question type are highlighted. 
Based on these results, we select temperature=0.8 and top-p=0.95 for single-choice questions, and temperature=0.8 and top-p=0.6 for general QA tasks in our experiments.

\begin{table*}[h]
\centering
\resizebox{\linewidth}{!}{
\begin{tabular}{c|ccccc|ccccc}
\toprule
\multicolumn{6}{c|}{Single Choice (Metric: Accuracy)} & \multicolumn{5}{c}{General QA (Metric: LLM Eval)} \\
\midrule
& \text{Temp.=0} & \text{Temp.=0.2} & \text{Temp.=0.4} & \text{Temp.=0.6} & \text{Temp.=0.8} &\text{Temp.=0} & \text{Temp.=0.2} & \text{Temp.=0.4} & \text{Temp.=0.6} & \text{Temp.=0.8} \\
\midrule
Top-p=0.6 & \multirow{4}{*}{32.73\%} &32.67\% &32.67\% &33.16\% &34.92\% & \multirow{4}{*}{3.16} &2.60 &2.53 &2.48 & \cellcolor{orange!70}3.06 \\
Top-p=0.8 & &33.16\% &33.21\% &34.92\% &34.65\% & & 2.49 &2.56 &2.52 &2.98 \\
Top-p=0.9& &33.80\% &33.80\% &34.17\% &34.01\% & & 2.53 &2.53 &2.55 &3.02 \\
Top-p=0.95& &32.67\% &34.54\% &33.96\% &\cellcolor{orange!70}35.93\% & & 2.59 &2.57 &2.49 &3.01 \\
\bottomrule
\end{tabular}
}
\caption{Hyperparameter grid search results. The selected hyperparameter combinations for Choice and QA question types have been highlighted.}
\label{tab:hyper}
\end{table*}

\section{Practical Performance Metrics for Domain-Specific SFT}\label{app:sft}
To assess the practical feasibility of deploying fine-tuned LLMs in business workflows, we profiled the computational cost of the fine-tuned model across different problem formats in BizCompass. The evaluated model was deployed on a server equipped with four NVIDIA A40 GPUs.
We tracked four key performance indicators: Average Runtime, Time to First Token (TTFT), Token Generation Speed (Tokens/s), and Average Memory Usage (both GPU VRAM and system RAM).
As summarized in Table~\ref{tab:sft_p}, tasks requiring extensive context processing and generative reasoning, such as Table QA and General QA, demand significantly higher runtimes compared to discriminative tasks (Single and Multiple Choice).
However, the underlying memory footprint remains largely stable across all question types, constrained primarily by the model's static weight allocation.

\begin{table*}[ht]
\centering
\resizebox{\linewidth}{!}{
\begin{tabular}{cccccc}
\toprule
\textbf{Question Type} & \textbf{Avg Runtime (s)} & \textbf{Avg TTFT (s)} & \textbf{Tokens/s} & \textbf{Avg GPU Memory (MB)}& \textbf{Avg RAM (MB)}\\
\midrule
Table QA & 80.238 & 0.018 & 6.014 & 15325.648 & 2434.830 \\
General QA & 83.441 & 0.011 & 6.031 & 15325.646 & 2542.343 \\
Single Choice & 0.872 & 0.008 & 5.238 & 15325.644 & 2615.969 \\
Multiple Choice & 1.079 & 0.008 & 4.233 & 15325.645 & 2575.021 \\
\bottomrule
\end{tabular}}
\caption{Computational cost and practical performance metrics of the fine-tuned model deployed on $4\times$ NVIDIA A40 GPUs across different question types.}
\label{tab:sft_p}
\end{table*}

\section{Evaluation Results}\label{app:eva_result}
\begin{table*}[h]
    \centering
    \resizebox{\linewidth}{!}{
    \begin{tabular}{lcccccccccc}
    \toprule
    \multirow{2}{*}{Model} & \multirow{2}{*}{Size} & \multirow{2}{*}{Thinking} & \multicolumn{4}{c}{FIN} & \multicolumn{4}{c}{ECON}\\
    \cmidrule(lr){4-7} \cmidrule(lr){8-11}
        &  &  & SC. & MC. & TableQA & GeneralQA & SC. & MC. & TableQA & GeneralQA \\
    \midrule
    GPT-4o-2024-05-13 & N/D & \xmark & 69.14\% & 58.18\% / 86.57\% & 4.841 & 4.736 & 58.79\% & 40.44\% / 79.45\% & 4.741 & 4.665 \\
    GPT-4.1-2025-04-14 & N/D & \xmark & 76.55\% & 63.94\% / 89.39\% & 4.950 & 4.957 & 67.88\% & 52.04\% / 85.32\% & 4.950 & 4.955 \\
    GPT-5-2025-08-07 & N/D & \cmark & 89.20\% & 76.97\% / 93.68\% & 4.982 & 4.986 & 88.79\% & 71.79\% / 91.65\% & 4.967 & 4.992 \\
    Gemini-2.5-Flash-thinking & N/D & \cmark & 89.82\% & 74.24\% / 92.95\% & 4.942 & 4.903 & 83.33\% & 65.20\% / 89.81\% & 4.907 & 4.919 \\
    Gemini-2.5-Pro-thinking & N/D & \cmark & 91.05\% & 84.55\% / 95.74\% & 4.954 & 4.948 & 88.79\% & 75.24\% / 93.30\% & 4.953 & 4.940 \\
    Llama3-8B & 8B & \xmark & 38.89\% & 14.55\% / 72.21\% & 3.240 & 2.770 & 30.61\% & 14.11\% / 69.86\% & 2.961 & 2.452 \\
    Llama-3-70B-Instruct & 70B & \xmark & 65.13\% & 53.33\% / 83.41\% & 4.603 & 4.432 & 56.36\% & 37.93\% / 77.73\% & 4.443 & 4.272 \\
    Llama-3.3-70B-Instruct & 70B & \xmark & 71.61\% & 54.24\% / 85.00\% & 4.799 & 4.688 & 62.42\% & 42.63\% / 80.06\% & 4.693 & 4.616 \\
    Llama-4-Scout & 17B & \xmark & 76.24\% & 56.97\% / 85.81\% & 4.128 & 3.656 & 60.00\% & 41.69\% / 81.46\% & 3.905 & 3.295 \\
    Claude 3.7 Sonnet-20250219 & N/D & \xmark & 74.08\% & 62.73\% / 87.88\% & 4.942 & 4.906 & 66.36\% & 48.59\% / 82.86\% & 4.919 & 4.885 \\
    Claude 4 Sonnet-20250514 & N/D & \cmark & 88.27\% & 74.85\% / 93.23\% & 4.939 & 4.896 & 82.12\% & 66.77\% / 90.52\% & 4.895 & 4.870 \\
    Claude 4 Opus-20250514 & N/D & \cmark & 90.43\% & 81.21\% / 94.98\% & 4.937 & 4.904 & 85.76\% & 77.74\% / 93.72\% & 4.904 & 4.886 \\
    Claude 3.7 Sonnet-20250219-thinking & N/D & \cmark & 90.13\% & 79.09\% / 94.72\% & 4.942 & 4.892 & 85.15\% & 69.91\% / 91.04\% & 4.925 & 4.882 \\
    Qwen3-235B-A22B & 235B & \cmark & 87.66\% & 75.76\% / 93.80\% & 4.876 & 4.868 & 84.24\% & 72.73\% / 92.05\% & 4.859 & 4.854 \\
    Qwen2.5-72B-Instruct & 72B & \xmark & 73.15\% & 53.64\% / 87.89\% & 4.757 & 4.625 & 59.09\% & 42.32\% / 81.03\% & 4.673 & 4.529 \\
    Qwen/QwQ-32B & 32B & \cmark & 88.27\% & 74.85\% / 93.63\% & 4.870 & 4.801 & 82.12\% & 65.52\% / 90.26\% & 4.848 & 4.688 \\
    Grok4-0709 & N/D & \xmark & 86.11\% & 80.00\% / 94.85\% & 4.950 & 4.915 & 82.42\% & 70.53\% / 92.22\% & 4.935 & 4.941 \\
    Grok3 & N/D & \xmark & 74.08\% & 65.45\% / 89.35\% & 4.955 & 4.902 & 64.85\% & 50.47\% / 85.32\% & 4.909 & 4.884 \\
    DeepSeek-V3-250324 & 671B & \xmark & 79.02\% & 64.24\% / 90.58\% & 4.891 & 4.828 & 68.49\% & 52.35\% / 85.93\% & 4.846 & 4.776 \\
    DeepSeek-R1-0528 & 671B & \cmark & 87.04\% & 76.36\% / 93.53\% & 4.930 & 4.893 & 81.82\% & 65.52\% / 90.32\% & 4.901 & 4.914 \\
    DeepSeek-R1-Distill-Qwen-7B & 7B & \cmark & 72.23\% & 50.61\% / 84.61\% & 3.973 & 3.942 & 58.18\% & 35.42\% / 77.04\% & 3.611 & 3.639 \\
    Deepseek-R1-Distill-Llama-70B & 70B & \cmark & 86.73\% & 73.33\% / 92.63\% & 4.677 & 4.684 & 78.79\% & 61.13\% / 87.16\% & 4.584 & 4.524 \\
    DeepSeek-R1-Distill-Qwen-32B & 32B & \cmark & 82.41\% & 71.21\% / 92.44\% & 4.622 & 4.584 & 78.79\% & 56.74\% / 86.50\% & 4.596 & 4.510 \\
    \midrule
    \multirow{2}{*}{Model} & \multirow{2}{*}{Size} & \multirow{2}{*}{Thinking} & \multicolumn{4}{c}{OM} & \multicolumn{4}{c}{STAT}\\
    \cmidrule(lr){4-7} \cmidrule(lr){8-11}
        &  &  & SC. & MC. & TableQA & GeneralQA & SC. & MC. & TableQA & GeneralQA \\
    \midrule
    GPT-4o-2024-05-13 & N/D & \xmark & 56.35\% & 48.30\% / 83.49\% & 4.665 & 4.603 & 58.49\% & 50.62\% / 83.55\% & 4.816 & 4.693 \\
    GPT-4.1-2025-04-14 & N/D & \xmark & 66.88\% & 56.97\% / 86.92\% & 4.945 & 4.962 & 69.50\% & 62.04\% / 89.55\% & 4.972 & 4.980 \\
    GPT-5-2025-08-07 & N/D & \cmark & 86.38\% & 72.14\% / 92.25\% & 4.972 & 4.986 & 90.88\% & 76.85\% / 93.34\% & 4.985 & 4.992 \\
    Gemini-2.5-Flash-thinking & N/D & \cmark & 83.90\% & 70.28\% / 91.21\% & 4.913 & 4.851 & 89.62\% & 71.30\% / 92.16\% & 4.955 & 4.918 \\
    Gemini-2.5-Pro-thinking & N/D & \cmark & 87.62\% & 77.71\% / 93.45\% & 4.940 & 4.932 & 92.14\% & 79.32\% / 94.84\% & 4.955 & 4.975 \\
    Llama3-8B & 8B & \xmark & 31.27\% & 21.05\% / 73.25\% & 2.916 & 2.729 & 29.87\% & 16.36\% / 69.77\% & 3.351 & 2.693 \\
    Llama-3-70B-Instruct & 70B & \xmark & 56.66\% & 50.77\% / 83.47\% & 4.384 & 4.253 & 59.43\% & 45.37\% / 83.42\% & 4.564 & 4.384 \\
    Llama-3.3-70B-Instruct & 70B & \xmark & 64.71\% & 56.35\% / 85.93\% & 4.657 & 4.591 & 66.98\% & 48.77\% / 85.00\% & 4.775 & 4.665 \\
    Llama-4-Scout & 17B & \xmark & 58.52\% & 45.20\% / 83.18\% & 3.486 & 3.230 & 65.09\% & 49.07\% / 85.29\% & 3.850 & 3.385 \\
    Claude 3.7 Sonnet-20250219 & N/D & \xmark & 60.68\% & 56.04\% / 85.29\% & 4.882 & 4.873 & 64.15\% & 54.01\% / 86.24\% & 4.954 & 4.912 \\
    Claude 4 Sonnet-20250514 & N/D & \cmark & 82.35\% & 75.23\% / 93.81\% & 4.894 & 4.860 & 89.62\% & 71.30\% / 92.47\% & 4.938 & 4.873 \\
    Claude 4 Opus-20250514 & N/D & \cmark & 86.07\% & 74.30\% / 93.10\% & 4.872 & 4.896 & 90.25\% & 79.01\% / 93.87\% & 4.938 & 4.916 \\
    Claude 3.7 Sonnet-20250219-thinking & N/D & \cmark & 84.52\% & 73.68\% / 92.65\% & 4.920 & 4.861 & 90.57\% & 73.77\% / 92.76\% & 4.936 & 4.899 \\
    Qwen3-235B-A22B & 235B & \cmark & 85.14\% & 74.92\% / 93.42\% & 4.887 & 4.863 & 90.25\% & 74.07\% / 93.68\% & 4.935 & 4.896 \\
    Qwen2.5-72B-Instruct & 72B & \xmark & 60.68\% & 52.63\% / 86.66\% & 4.635 & 4.612 & 63.21\% & 49.69\% / 86.88\% & 4.774 & 4.626 \\
    Qwen/QwQ-32B & 32B & \cmark & 83.90\% & 72.76\% / 92.48\% & 4.758 & 4.648 & 89.94\% & 68.83\% / 92.10\% & 4.900 & 4.771 \\
    Grok4-0709 & N/D & \xmark & 81.43\% & 71.83\% / 92.33\% & 4.930 & 4.887 & 86.48\% & 70.68\% / 92.83\% & 4.964 & 4.957 \\
    Grok3 & N/D & \xmark & 64.40\% & 56.35\% / 87.57\% & 4.907 & 4.883 & 69.50\% & 60.19\% / 88.79\% & 4.961 & 4.909 \\
    DeepSeek-V3-250324 & 671B & \xmark & 67.18\% & 56.35\% / 88.86\% & 4.835 & 4.807 & 76.73\% & 56.48\% / 88.52\% & 4.863 & 4.856 \\
    DeepSeek-R1-0528 & 671B & \cmark & 72.45\% & 69.66\% / 92.00\% & 4.894 & 4.887 & 70.44\% & 71.30\% / 91.63\% & 4.941 & 4.949 \\
    DeepSeek-R1-Distill-Qwen-7B & 7B & \cmark & 68.73\% & 44.27\% / 83.27\% & 3.825 & 4.045 & 70.75\% & 40.43\% / 83.05\% & 4.111 & 4.156 \\
    Deepseek-R1-Distill-Llama-70B & 70B & \cmark & 78.64\% & 65.63\% / 89.32\% & 4.590 & 4.521 & 82.70\% & 61.42\% / 88.45\% & 4.769 & 4.657 \\
    DeepSeek-R1-Distill-Qwen-32B & 32B & \cmark & 79.57\% & 65.02\% / 90.76\% & 4.542 & 4.534 & 85.53\% & 61.11\% / 89.42\% & 4.694 & 4.647 \\
    \bottomrule
    \end{tabular}
    }
    \caption{Complete evaluation results for knowledge-based tasks in BizCompass. N/D means not disclosed.}
    \label{tab:full-knowledge-result}
\end{table*}

{
\onecolumn
\scriptsize
\setlength{\tabcolsep}{2pt}
\begin{longtable}{lccccccccc@{}}
    \toprule
    \multirow{2}{*}{Model} & AP\_A & CIKM18 & FPB & M\&A & FOMC & FRM & CCFraud & Taiwan & RM\_C \\
    \cmidrule(lr){2-9} \cmidrule(lr){10-10}
        & \multicolumn{8}{c}{Accuracy} & GPT Eval\\
    \midrule
    \endfirsthead 
    
    \toprule
    \multirow{2}{*}{Model} & AP\_A & CIKM18 & FPB & M\&A & FOMC & FRM & CCFraud & Taiwan & RM\_C \\
    \cmidrule(lr){2-9} \cmidrule(lr){10-10}
        & \multicolumn{8}{c}{Accuracy} & GPT Eval\\
    \midrule
    \endhead 

    GPT-4o-2024-05-13 & 57.98\% & 47.00\% & 97.20\% & 60.80\% & 63.31\% & 66.67\% & 67.40\% & 3.80\% & 4.272 \\
    GPT-4.1-2025-04-14 & 60.61\% & 46.60\% & 98.20\% & 57.80\% & 63.71\% & 67.81\% & 71.20\% & 3.60\% & 4.896 \\
    GPT-5-2025-08-07 & 96.16\% & 57.40\% & 99.40\% & 59.20\% & 66.33\% & 89.04\% & 85.00\% & 40.20\% & 4.940 \\
    Gemini-2.5-Flash-thinking & 86.87\% & 51.40\% & 98.20\% & 67.20\% & 64.11\% & 70.55\% & 96.60\% & 37.20\% & 4.622 \\
    Gemini-2.5-Pro-thinking & 84.85\% & 48.40\% & 95.00\% & 84.60\% & 65.32\% & 76.94\% & 80.20\% & 5.80\% & 4.798 \\
    Llama3-8B & 24.65\% & 58.40\% & 87.00\% & 75.60\% & 59.27\% & 36.53\% & 6.60\% & 3.60\% & 2.998 \\
    Llama-3-70B-Instruct & 65.25\% & 58.80\% & 97.40\% & 72.60\% & 68.35\% & 56.39\% & 7.60\% & 19.00\% & 3.902 \\
    Llama-3.3-70B-Instruct & 72.73\% & 55.60\% & 93.20\% & 57.60\% & 69.96\% & 60.27\% & 12.80\% & 5.00\% & 4.182 \\
    Llama-4-Scout & 61.21\% & 54.60\% & 98.00\% & 59.00\% & 63.91\% & 28.54\% & 87.00\% & 29.20\% & 3.028 \\
    Claude 3.7 Sonnet-20250219 & 61.82\% & 53.00\% & 93.40\% & 68.00\% & 67.14\% & 63.93\% & 85.60\% & 10.40\% & 4.768 \\
    Claude 4 Sonnet-20250514 & 93.13\% & 50.60\% & 98.60\% & 62.00\% & 66.94\% & 84.25\% & 96.20\% & 13.80\% & 4.638 \\
    Claude 4 Opus-20250514 & 94.75\% & 44.80\% & 88.60\% & 74.60\% & 65.73\% & 84.70\% & 99.40\% & 7.40\% & 4.602 \\
    Claude 3.7 Sonnet-20250219-thinking & 94.14\% & 54.80\% & 98.00\% & 53.40\% & 67.74\% & 85.39\% & 63.60\% & 3.60\% & 4.806 \\
    DeepSeek-V3-250324 & 95.56\% & 48.80\% & 96.80\% & 83.60\% & 67.34\% & 77.63\% & 55.00\% & 3.60\% & 4.508 \\
    DeepSeek-R1-0528 & 95.15\% & 51.80\% & 97.40\% & 60.00\% & 66.53\% & 87.21\% & 46.80\% & 4.20\% & 4.608 \\
    DeepSeek-R1-Distill-Qwen-7B & 81.21\% & 54.70\% & 93.52\% & 68.21\% & 54.44\% & 58.22\% & 22.20\% & 80.80\% & 3.222 \\
    Deepseek-R1-Distill-Llama-70B & 91.90\% & 51.80\% & 58.60\% & 58.60\% & 64.31\% & 74.21\% & 7.00\% & 50.20\% & 2.570 \\
    DeepSeek-R1-Distill-Qwen-32B & 89.70\% & 54.20\% & 95.60\% & 72.60\% & 59.68\% & 73.97\% & 12.20\% & 4.80\% & 3.830 \\
    Qwen3-235B-A22B & 92.73\% & 47.60\% & 97.20\% & 54.60\% & 67.34\% & 84.02\% & 20.00\% & 18.60\% & 4.636 \\
    Qwen2.5-72B-Instruct & 28.28\% & 45.40\% & 98.20\% & 55.80\% & 62.10\% & 52.51\% & 90.60\% & 14.20\% & 3.994 \\
    Qwen/QwQ-32B & 93.54\% & 54.00\% & 95.80\% & 56.80\% & 60.69\% & 80.59\% & 15.80\% & 10.80\% & 3.222 \\
    Grok4-0709 & 77.78\% & 65.20\% & 96.60\% & 80.20\% & 65.12\% & 71.69\% & 80.00\% & 42.60\% & 4.160 \\
    Grok3 & 70.51\% & 52.20\% & 97.60\% & 80.80\% & 65.73\% & 66.44\% & 76.60\% & 26.00\% & 4.766 \\
    \midrule
    \multirow{2}{*}{Model} & \multicolumn{3}{c}{SECQUE} & \multicolumn{3}{c}{EDTSUM} & \multicolumn{3}{c}{AT\_T\_Algorithm} \\
    \cmidrule(lr){2-4} \cmidrule(lr){5-7} \cmidrule(lr){8-10} 
        & R-1 & R-2 & R-L & R-1 & R-2 & R-L & R-1 & R-2 & R-L \\
    \midrule
    GPT-4o-2024-05-13 & 31.05\% & 14.50\% & 19.46\% & 24.76\% & 11.31\% & 19.06\% & 47.25\% & 19.52\% & 26.34\% \\
    GPT-4.1-2025-04-14 & 24.59\% & 10.59\% & 14.71\% & 23.81\% & 9.91\% & 17.77\% & 52.90\% & 21.46\% & 28.87\% \\
    GPT-5-2025-08-07 & 29.54\% & 10.18\% & 16.33\% & 31.81\% & 13.57\% & 25.03\% & 46.39\% & 16.11\% & 27.67\% \\
    Gemini-2.5-Flash-thinking & 33.60\% & 14.27\% & 21.03\% & 26.86\% & 11.83\% & 20.46\% & 52.57\% & 23.80\% & 34.26\% \\
    Gemini-2.5-Pro-thinking & 35.01\% & 14.52\% & 21.19\% & 36.83\% & 17.25\% & 29.52\% & 51.57\% & 21.45\% & 30.29\% \\
    Llama3-8B & 37.60\% & 15.79\% & 23.47\% & 22.49\% & 10.26\% & 17.47\% & 50.36\% & 18.40\% & 26.90\% \\
    Llama-3-70B-Instruct & 34.78\% & 14.93\% & 21.36\% & 30.09\% & 14.44\% & 24.48\% & 51.50\% & 19.50\% & 27.60\% \\
    Llama-3.3-70B-Instruct & 32.86\% & 14.41\% & 20.32\% & 35.16\% & 17.09\% & 28.72\% & 47.27\% & 17.97\% & 25.54\% \\
    Llama-4-Scout & 39.00\% & 16.85\% & 24.11\% & 22.15\% & 10.33\% & 17.31\% & 47.33\% & 18.13\% & 27.80\% \\
    Claude 3.7 Sonnet-20250219 & 33.15\% & 13.34\% & 19.22\% & 29.14\% & 13.49\% & 23.42\% & 51.09\% & 21.85\% & 29.40\% \\
    Claude 4 Sonnet-20250514 & 32.72\% & 12.36\% & 18.81\% & 20.45\% & 8.66\% & 15.54\% & 52.05\% & 20.53\% & 29.46\% \\
    Claude 4 Opus-20250514 & 35.08\% & 13.70\% & 20.20\% & 40.17\% & 18.79\% & 33.73\% & 52.92\% & 21.22\% & 29.13\% \\
    Claude 3.7 Sonnet-20250219-thinking & 34.26\% & 13.18\% & 19.61\% & 36.27\% & 17.02\% & 30.21\% & 50.90\% & 20.17\% & 28.47\% \\
    DeepSeek-V3-250324 & 29.90\% & 10.24\% & 16.45\% & 21.66\% & 8.81\% & 16.13\% & 51.44\% & 19.77\% & 27.94\% \\
    DeepSeek-R1-0528 & 27.59\% & 9.70\% & 15.42\% & 17.19\% & 7.06\% & 12.71\% & 50.49\% & 18.33\% & 26.43\% \\
    DeepSeek-R1-Distill-Qwen-7B & 32.34\% & 11.51\% & 18.23\% & 20.37\% & 8.36\% & 15.43\% & 47.32\% & 16.81\% & 24.57\% \\
    Deepseek-R1-Distill-Llama-70B & 33.95\% & 11.76\% & 19.66\% & 29.83\% & 13.34\% & 23.60\% & 42.10\% & 15.11\% & 23.51\% \\
    DeepSeek-R1-Distill-Qwen-32B & 33.92\% & 13.62\% & 19.98\% & 29.19\% & 13.61\% & 23.57\% & 50.84\% & 19.42\% & 26.91\% \\
    Qwen3-235B-A22B & 30.04\% & 11.42\% & 17.00\% & 18.64\% & 7.37\% & 13.98\% & 20.25\% & 9.17\% & 11.78\% \\
    Qwen2.5-72B-Instruct & 23.13\% & 10.93\% & 14.47\% & 21.25\% & 9.14\% & 15.96\% & 45.45\% & 19.31\% & 25.77\% \\
    Qwen/QwQ-32B & 14.05\% & 6.28\% & 8.77\% & 6.02\% & 2.62\% & 4.83\% & 21.12\% & 9.66\% & 12.37\% \\
    Grok4-0709 & 31.15\% & 12.94\% & 18.01\% & 23.93\% & 10.81\% & 18.76\% & 48.35\% & 19.60\% & 27.50\% \\
    Grok3 & 25.49\% & 11.65\% & 15.60\% & 24.00\% & 10.88\% & 18.77\% & 47.33\% & 20.22\% & 26.57\% \\
    \midrule
    \multirow{2}{*}{Model} & FNXL & FinQA & TATQA & FinRED & SD\_T & SD\_C & PM\_A & PM\_C & AT\_T\_Strategy \\
    \cmidrule(lr){2-5} \cmidrule(lr){6-10} 
        & \multicolumn{4}{c}{GPT Eval (Label)} & \multicolumn{5}{c}{GPT Eval (Score)}\\
    \midrule
    GPT-4o-2024-05-13 & 28.00\% & 52.00\% & 90.40\% & 33.80\% & 3.547 & 3.994 & 4.584 & 4.762 & 4.852 \\
    GPT-4.1-2025-04-14 & 16.40\% & 59.20\% & 91.20\% & 23.30\% & 4.330 & 4.325 & 4.765 & 4.822 & 4.955 \\
    GPT-5-2025-08-07 & 23.00\% & 71.80\% & 92.40\% & 18.40\% & 4.660 & 4.692 & 4.876 & 4.880 & 4.943 \\
    Gemini-2.5-Flash-thinking & 19.20\% & 68.20\% & 92.40\% & 20.30\% & 4.032 & 3.903 & 4.794 & 4.872 & 4.909 \\
    Gemini-2.5-Pro-thinking & 6.00\% & 65.60\% & 88.20\% & 18.90\% & 3.764 & 3.873 & 4.813 & 4.836 & 4.841 \\
    Llama3-8B & 1.30\% & 32.80\% & 77.40\% & 33.10\% & 3.119 & 2.790 & 2.765 & 3.946 & 4.273 \\
    Llama-3-70B-Instruct & 6.60\% & 48.60\% & 87.20\% & 24.00\% & 3.447 & 3.440 & 4.289 & 4.732 & 4.670 \\
    Llama-3.3-70B-Instruct & 2.50\% & 47.40\% & 85.40\% & 19.10\% & 3.513 & 3.544 & 4.270 & 4.800 & 4.795 \\
    Llama-4-Scout & 7.50\% & 58.40\% & 88.60\% & 17.20\% & 2.937 & 2.573 & 3.000 & 3.918 & 3.182 \\
    Claude 3.7 Sonnet-20250219 & 27.40\% & 64.40\% & 90.20\% & 19.60\% & 4.053 & 4.135 & 4.724 & 4.926 & 4.932 \\
    Claude 4 Sonnet-20250514 & 33.00\% & 74.20\% & 92.20\% & 17.90\% & 3.846 & 3.696 & 4.759 & 4.906 & 4.841 \\
    Claude 4 Opus-20250514 & 46.20\% & 69.20\% & 91.40\% & 21.60\% & 3.884 & 3.829 & 4.794 & 4.880 & 4.898 \\
    Claude 3.7 Sonnet-20250219-thinking & 50.00\% & 70.00\% & 89.80\% & 21.10\% & 4.151 & 4.429 & 4.759 & 4.912 & 4.886 \\
    DeepSeek-V3-250324 & 9.70\% & 63.20\% & 88.60\% & 14.20\% & 3.805 & 3.569 & 4.778 & 4.780 & 4.864 \\
    DeepSeek-R1-0528 & 8.80\% & 61.40\% & 90.40\% & 19.10\% & 3.796 & 4.026 & 4.835 & 4.822 & 4.875 \\
    DeepSeek-R1-Distill-Qwen-7B & 0.30\% & 47.60\% & 74.60\% & 19.90\% & 2.731 & 2.879 & 4.317 & 4.206 & 4.238 \\
    Deepseek-R1-Distill-Llama-70B & 3.50\% & 60.00\% & 75.60\% & 19.90\% & 3.050 & 3.604 & 3.443 & 2.224 & 2.080 \\
    DeepSeek-R1-Distill-Qwen-32B & 4.10\% & 63.60\% & 88.00\% & 18.90\% & 3.440 & 3.774 & 4.635 & 4.698 & 4.693 \\
    Qwen3-235B-A22B & 23.30\% & 63.80\% & 89.60\% & 19.60\% & 3.173 & 4.119 & 4.787 & 4.812 & 4.500 \\
    Qwen2.5-72B-Instruct & 17.90\% & 49.40\% & 88.60\% & 23.80\% & 3.484 & 3.679 & 4.546 & 4.834 & 4.818 \\
    Qwen/QwQ-32B & 5.30\% & 59.60\% & 88.40\% & 19.10\% & 2.969 & 3.085 & 4.527 & 4.408 & 4.011 \\
    Grok4-0709 & 29.90\% & 61.20\% & 87.20\% & 21.80\% & 3.503 & 3.903 & 4.743 & 4.780 & 4.716 \\
    Grok3 & 15.10\% & 51.80\% & 85.00\% & 19.60\% & 3.840 & 3.800 & 4.762 & 4.872 & 4.932 \\
    \bottomrule

    \caption{Complete evaluation results for application-based tasks in BizCompass.}
    \label{tab:full-application-result}
\end{longtable}
}

\twocolumn

\section{LLM Eval Prompts}\label{app:llmeval}

Following G-Eval's setting~\cite{Geval_2023}, we select GPT-4o as the evaluation model.

\begin{tcolorbox}[title=LLM Eval Prompt for general QA,fonttitle=\bfseries\large,breakable,code={\setstretch{1.1}}]
You are an expert evaluator. Your task is to score the candidate's answer based on both the correctness of the final answer and the logical soundness of its reasoning.

\vspace{0.8ex}
\textbf{IMPORTANT INSTRUCTION}

If the question contains multiple sub-questions (\eg, 1., 2(a).), you MUST score each sub-question INDEPENDENTLY.

Evaluate the [Conclusion Correctness] and [Reasoning Soundness] for each sub-question in the candidate's answer, and strictly apply the following criteria to score EACH sub-question:

- 5 (Completely Correct):
  - Evaluation Points: The conclusion is correct, concepts are applied precisely, and the reasoning is clear, rigorous, and flawless.

- 4 (Incorrect Conclusion, Sound Reasoning - Minor Deviation):
  - Evaluation Points: The overall framework and method of reasoning are correct, but the conclusion is flawed due to minor calculation errors, data misinterpretations, or slight misunderstandings of boundary conditions. The core logic is sound.

- 3 (Incorrect Conclusion, Sound Reasoning - Major Deviation):
  - Evaluation Points: The general direction of the reasoning is correct, but a key assumption, formula application, or core step contains a significant error, leading to a major deviation in the conclusion.

- 2 (Correct Conclusion, Flawed Reasoning - Minor Flaws):
  - Evaluation Points: The conclusion happens to be correct, but the reasoning is not rigorous, showing minor flaws such as logical leaps, insufficient justification, or confusion of secondary concepts.

- 1 (Correct Conclusion, Flawed Reasoning - Severe Flaws):
  - Evaluation Points: The conclusion happens to be correct, but the reasoning has fundamental errors, such as reversed causality or the use of a completely wrong model/theory. The connection between the reasoning and the conclusion is purely coincidental.

- 0 (Completely Incorrect):
  - Evaluation Points: Both the conclusion and the reasoning process have severe errors, or the answer is completely off-topic and provides little to no valid reasoning.

\vspace{0.8ex}
\textbf{OUTPUT FORMAT:}

Return ONLY a JSON object. The object must contain two fields:
1.  `score`: The arithmetic mean of all sub-question scores.
2.  `scores\_per\_question`: An object containing the independent score for each sub-question.
The format must be exactly as follows:

\{

    ``score": $<$arithmetic mean score$>$,
    
    ``scores\_per\_question": \{
    
        ``$<$question\_number\_1$>$": $<$integer from 0 to 5$>$,
        
        ``$<$question\_number\_2$>$": $<$integer from 0 to 5$>$ \}
    
\}
\end{tcolorbox}

\section{More Results}
\subsection{Statistical Analysis of Business Scenarioss}
\label{sec:sta_ana_of_bus}
To explore the potential applications of LLMs in business, we begin by defining three representative business scenarios—analysis, trading, and consulting—based on authoritative external sources and expert consultation. Specifically, analysis follows the definition provided in the International Labour Organization’s ISCO-08 classification \citep{ganzeboom2010new}, trading is defined according to the description published by Indeed UK \citep{kaniel2006so}, and consulting is defined using the official description provided by LinkedIn Talent Solutions \citep{carliner2015job}. 
\begin{tcolorbox}[colback=lightgray!20, colframe=lightgray!20, sharp corners=all, 
boxrule=0mm, boxsep=0.5mm, left=1.5mm, right=1.5mm, top=1.5mm, bottom=1.5mm]
\begin{definition}[Analysis / Business Analysis]
\label{def:anlysis}
Analysis involves assisting clients in evaluating information that influences investment programmes in both public and private institutions. Professionals in this role are commonly referred to as financial analysts.
\end{definition}
\end{tcolorbox}
\begin{tcolorbox}[colback=lightgray!20, colframe=lightgray!20, sharp corners=all, 
boxrule=0mm, boxsep=0.5mm, left=1.5mm, right=1.5mm, top=1.5mm, bottom=1.5mm]
\begin{definition}[Trading]
\label{def:trading}
Trading involves buying and selling stocks and shares in financial markets with other participants who hold an interest in the stock market and finance.
\end{definition}
\end{tcolorbox}
\begin{tcolorbox}[colback=lightgray!20, colframe=lightgray!20, sharp corners=all, 
boxrule=0mm, boxsep=0.5mm, left=1.5mm, right=1.5mm, top=1.5mm, bottom=1.5mm]
\begin{definition}[Consulting]
\label{def:consulting}
Consulting involves providing specialized expertise and knowledge—either independently or through a consulting firm—to help businesses achieve goals and solve problems.
\end{definition}
\end{tcolorbox}

In order to validate the above definitions, we conduct a systematic analysis of O*NET. The O*NET database is organized in a hierarchical structure: at the top level are career clusters, which group related occupational domains (e.g., financial services). Each cluster is further divided into sub-clusters that capture more specific functional areas (e.g., financial strategy \& investments). At the lowest level are occupations, each documented with a job title, a definition, and a set of task descriptions. Building on this structure, our analysis proceeds in three steps:

\paragraph{Step One: Cluster Selection and Occupation Collection.}
We begin by selecting three career clusters most relevant to business practice—management \& entrepreneurship, marketing \& sales, and financial services. Together, these clusters comprise 14 unique sub-clusters and 156 occupations in total. Of these, 126 occupations fall within the scope of our three scenario definitions. The distribution of in-scope occupations is relatively balanced, with 59 from management \& entrepreneurship, 41 from financial services, and 26 from marketing \& sales.

\paragraph{Step Two: Occupation-to-Scenario Mapping.}
Each occupation is assigned to one of the three business scenarios by matching its related information against \cref{def:anlysis,def:consulting,def:trading}.
The classification is performed by an LLM, with the mapping guided by the following prompt:

\paragraph{Step Three: Scenario Coverage Calculation.}
In the final step, we merge the occupations collected and remove duplicates, thereby yielding a unified occupation set $\tilde{J}$.
For each occupation $j \in \tilde{J}$, we record its assigned category $c_j \in \{\text{consulting}, \text{analysis}, \text{trading}, \text{none}\}$, along with two associated quantities: the number of employed persons $e_j \ge 0$ and the projected number of job openings $o_j \ge 0$. Based on this setup, our interest lies in the relative importance of the three core business scenarios
\(
C = \{\text{consulting}, \text{analysis}, \text{trading}\}.
\)
To capture this, we define two proportional measures. The first is the employment share, which quantifies the proportion of total employment represented by a given category:
\[
S^{(\mathrm{emp})}(c) =
\frac{\sum_{j \in \tilde{J}} e_j \, 1\{c_j = c\}}
     {\sum_{j \in \tilde{J}} e_j} \times 100\%, 
\quad c \in C,
\]
where the denominator aggregates employment across all categories (including \texttt{none}), while the numerator sums only employment in category $c$. Here, $1\{\cdot\}$ denotes the indicator function. The second is the job-opening share, which quantifies the proportion of projected job openings represented by a given category:
\[
S^{(\mathrm{open})}(c) =
\frac{\sum_{j \in \tilde{J}} o_j \, 1\{c_j = c\}}
     {\sum_{j \in \tilde{J}} o_j} \times 100\%, 
\quad c \in C,
\]
where the denominator aggregates openings across all categories (including \texttt{none}), while the numerator sums only openings in category $c$. The employment share reflects the current labor market distribution, whereas the job-opening share captures future demand.  

\begin{table}[ht]
    \centering 
    \resizebox{\linewidth}{!}{
    \begin{tabular}{l *{4}{S[table-format=2.2]}}
        \toprule
        \textbf{Metric} & \textbf{Consulting} & \textbf{Analysis} & \textbf{Trading} & \textbf{Total} \\
        \midrule
        $S^{(\mathrm{emp})}(c)$ (\%)  & 38.76 & 33.60 & 23.20 & 95.56 \\
        $S^{(\mathrm{open})}(c)$ (\%) & 35.01 & 30.72 & 29.52 & 95.25 \\
        \bottomrule
    \end{tabular}}
    \caption{Employment and job-opening shares of three representative business scenarios.}
    \label{tab:role-coverage}
\end{table}

The statistical results are summarized in Table~\ref{tab:role-coverage}, which demonstrates that the three representative business scenarios together cover the vast majority of both employment and job openings.



\begin{table*}[ht]
    \centering
    \setlength{\tabcolsep}{4pt}
    \begin{threeparttable}
    \resizebox{\linewidth}{!}{%
    \begin{tabular}{@{}p{0.29\linewidth} S[table-format=2.2] p{0.29\linewidth} S[table-format=2.2] p{0.29\linewidth} S[table-format=2.2]@{}}
        \toprule
        \multicolumn{2}{c}{Analysis} & \multicolumn{2}{c}{Consulting} & \multicolumn{2}{c}{Trading} \\
        \cmidrule(lr){1-2}\cmidrule(lr){3-4}\cmidrule(lr){5-6}
        Task & {Coverage (\%)} & Task & {Coverage (\%)} & Task & {Coverage (\%)} \\
        \midrule
        asset pricing                   & 36.52 & market trend analysis        & 31.05 & asset pricing         & 30.40 \\
        market trend analysis           & 44.76 & risk management              & 74.55 & market trend analysis & 30.09 \\
        risk management                 & 76.12 & strategy development         & 66.84 & strategy development  & 39.51 \\
        portfolio management            & 43.91 & portfolio management         & 29.14 & algorithmic trading   & 12.16 \\
        fraud detection                 & 48.41 & fraud detection              & 25.77 & \multicolumn{1}{c}{}  & \multicolumn{1}{c}{} \\
        financial document analysis     & 55.90 & financial document analysis  & 31.57 & \multicolumn{1}{c}{}  & \multicolumn{1}{c}{} \\
        \addlinespace[2pt]\midrule
        Overall                         & 88.20 & Overall                      & 91.13 & Overall               & 70.52 \\
        \bottomrule
    \end{tabular}%
    }
    \caption{Task coverage by role.\label{tab:task-coverage-3col}}
    \begin{tablenotes}[flushleft]
        \scriptsize
        \item \textit{Note.} Tasks may match multiple representative tasks; \emph{Overall} is not a sum of rows but the share of tasks matched at least once within each role.
    \end{tablenotes}
    \end{threeparttable}
\end{table*}

\subsection{Task Coverage Validation}

We now assess whether the three roles, Analysis, Trading, and Consulting, align with the observed work content. Using \textsc{O*NET}, we collect all tasks from the 126 in-scope occupations and, for each role, check whether each task matches that role's representative tasks. Our goal is to measure how much of the work of each role is captured by its representative set and to ensure the internal consistency of the design.

\paragraph{Step One: Collect and summarize tasks.}
Starting from the 126 in-scope occupations, we extract all tasks from \textsc{O*NET} and deduplicate them. For each role $c\in\{\text{Analysis},\ \text{Trading},\ \text{Consulting}\}$, let $K_c$ be the set of unique tasks associated with $c$, and let $T_c = |K_c|$. In our data: $T_{\text{Analysis}}=1068$, $T_{\text{Trading}}=329$, and $T_{\text{Consulting}}=947$.

\paragraph{Step Two: Test alignment with real tasks.}
We evaluate a set of 16 task--role pairs. We assign the following subsets to each role:
\begin{itemize}[leftmargin=*,nosep]
  \item \textbf{Analysis}: \emph{asset pricing}, \emph{market trend analysis}, \emph{risk management}, \emph{portfolio management}, \emph{fraud detection}, \emph{financial document analysis}.
  \item \textbf{Trading}: \emph{asset pricing}, \emph{market trend analysis}, \emph{strategy development}, \emph{algorithmic trading}.
  \item \textbf{Consulting}: \emph{market trend analysis}, \emph{risk management}, \emph{strategy development}, \emph{portfolio management}, \emph{fraud detection}, \emph{financial document analysis}.
\end{itemize}

Let $S_c$ denote the subset for role $c$. For each task $k\in K_c$ and each representative task $\tau\in S_c$, we use an LLM (\emph{LLM}, temperature $=0$) to decide whether $k$ is related to $\tau$ under a fixed yes/no rule. The decision is recorded as
\[
f_{\tau,c}(k)\in\{0,1\}\quad (1=\text{relevant},\ 0=\text{not relevant}).
\]

\paragraph{Step Three: Task Coverage Calculation.}
In this step, we assess how completely the representative tasks account for the collected tasks within each role. Let \(K_c\), \(T_c\), and \(S_c\) be as defined above, and let \(f_{\tau,c}(k)\in\{0,1\}\) denote the relevance decision.

Based on this setup, we define two measures. The first is the share of each representative task within its role:
\[
H_{\tau,c}=\frac{1}{T_c}\sum_{k\in K_c} f_{\tau,c}(k)\times 100\%.
\]
This reports the percentage of tasks in role \(c\) that are covered by \(\tau\).

The second is the overall coverage of a role. A task counts as covered if it matches at least one representative task:

$$U_c=\{\,k\in K_c:\exists\,\tau\in S_c\ \text{with}\ f_{\tau,c}(k)=1\,\},$$
$$\mathrm{Coverage}(c)=\frac{|U_c|}{T_c}\times 100\%.$$

Because a task may match more than one representative task, the shares \(H_{\tau,c}\) do not sum to \(100\%\). By contrast, \(\mathrm{Coverage}(c)\) counts each task once.

The statistical results are summarized in Table~\ref{tab:task-coverage-3col}, which show that the representative task sets cover the large majority of observed tasks within each role. Within roles, individual representative tasks already account for substantial portions—for example, \emph{risk management} covers 76.12\% of Analysis tasks and 74.55\% of Consulting; \emph{market trend analysis} covers 44.76\% in Analysis and 31.05\% in Consulting; and in Trading the main contributors are \emph{strategy development} (39.51\%) and \emph{asset pricing} (30.40\%). Aggregating within roles, the \emph{Overall} row reaches 88.20\% for Analysis, 91.13\% for Consulting, and 70.52\% for Trading. Taken together, these results indicate broad coverage: most in-scope tasks can be matched to at least one representative task, suggesting that the benchmark reflects real work and is suitable for a wide range of practical applications.

\newpage
\onecolumn

\begin{tcolorbox}[title=Prompt for Statistic-GeneralQA,fonttitle=\bfseries\large,breakable,code={\setstretch{1.1}}]
\textbf{ROLE:}

You are an expert researcher in statistics, specializing in statistical theory and methodology. You should provide a rigorous and complete answer to the question. The answer must be structured in a step-by-step format that directly corresponds to each sub-question from the input, maintaining the original numbering and lettering (e.g., 1., 2(a)., 2(b).).
\vspace{1.1ex}

\textbf{OUTPUT FORMAT:}

Return ONLY the answer as a single string. The string must be a complete, step-by-step solution. The answer must use the exact same notation and LaTeX expressions as provided in the question.
\vspace{1.6ex}

\textbf{EXAMPLE}

\vspace{0.5ex}
\textbf{INPUT:}

Background:

Research Question. How does the homoscedasticity assumption simplify variance in mixed-effects models, and what are its practical limitations?

Setting:

Response $y$ decomposes into orthogonal effects: $y = \sum a_G$, with fixed factors $J$ and random factors. Conditional variance is $var(y|J) = \sum_{G\notin J} E(a_G^2 | J) + \sum_{G1 \neq G2 \notin J} E(a_{G1} a_{G2} | J)$.

Under homoscedasticity (F3), this simplifies to $var(y|J) = \sum_{G\notin J} \sigma_G^2$.

Questions:

1. Derive the simplified variance under (F3) and show it implies conditionally uncorrelated random effects.

2. Critique (F3) with a clinical trial example where it fails, and discuss consequences for inference if violated but the simplified model is used.

\vspace{1.5ex}
\textbf{OUTPUT:}

1. Under (F3), $E(a_G^2 | J) = \sigma_G^2$ (constant) for random $G$, and $E(a_{G1} a_{G2} | J) =$ constant for $G1 \neq G2$. Since unconditional orthogonality implies $E(a_{G1} a_{G2}) =0$, the constant must be $0$, so conditional covariances vanish. Thus, $var(y|J) = \sum_{G\notin J} \sigma_G^2$, with effects conditionally uncorrelated.

2. In a trial with fixed factor Treatment (Placebo, Low Dose, High Dose) and random factor Patient, high-dose responses may vary more due to side effects, violating constant $\sigma_R^2$ across treatments. If violated but simplified model used, variance estimates are wrong, leading to invalid p-values, confidence intervals, and hypothesis tests for treatment effects.
\end{tcolorbox}

\begin{tcolorbox}[title=Prompt for Economics-Single Choice,fonttitle=\bfseries\large,breakable,code={\setstretch{1.1}}]
\textbf{ROLE:}

You are an expert economic researcher specializing in econometrics and empirical economics. You should carefully analyze the given question and options, determine the single correct choice, and output only the letter of the correct option.
\vspace{1.6ex}

\textbf{OUTPUT FORMAT:}

Return ONLY the answer as a single string containing the chosen letter. The string must not include any other text or explanations.

\vspace{1.6ex}

\textbf{EXAMPLE}
\vspace{0.3ex}

\textbf{INPUT:}

Background:

In ordinary least squares (OLS) regression, a key assumption is homoskedasticity: the error terms have constant variance across all observations. If violated (heteroskedasticity), standard errors may be biased, leading to invalid inference.

\vspace{0.2ex}
Question:

In which of the following scenarios is heteroskedasticity most likely to occur in a regression of income on education?

\vspace{0.2ex}
Options: 

A) Errors are randomly distributed and independent of education level.

B) All individuals report income accurately, with no measurement error.

C) Variance in income is larger for highly educated individuals due to diverse job opportunities.

D) The sample size is small, but errors are normally distributed.
\vspace{1.5ex}

\textbf{OUTPUT:}

C
\end{tcolorbox}

\begin{tcolorbox}[title=Prompt for Finance-Multiple Choice,fonttitle=\bfseries\large,breakable,code={\setstretch{1.1}}]
\textbf{ROLE:}

You are an expert financial researcher specializing in financial theory and quantitative methods. You should carefully analyze the given question and options, determine all correct choices, and output only the letters of the correct options separated by commas (e.g., ``A,B"). 
\vspace{0.1ex}

\textbf{OUTPUT FORMAT:}

Return ONLY the answer as a string containing the chosen letters separated by commas (e.g., ``A,B"). The string must not include any other text or explanations. If no options are correct, return an empty string.
\vspace{1.8ex}

\textbf{EXAMPLE}

\vspace{0.6ex}
\textbf{INPUT:}

Background:

The Capital Asset Pricing Model (CAPM) describes the relationship between systematic risk and expected return for assets, assuming investors are rational and markets are efficient.

Question:

Select all assumptions of the CAPM that are essential for its derivation.

Options:

A: Investors have homogeneous expectations about asset returns.

B: There are no taxes or transaction costs.

C: All investors can borrow and lend at the risk-free rate.

D: Assets are infinitely divisible.
\vspace{1.5ex}

\textbf{OUTPUT:}

A,B,C,D
\end{tcolorbox}

\begin{tcolorbox}[title=Prompt for OM-TableQA,fonttitle=\bfseries\large,breakable,code={\setstretch{1.1}}]
\textbf{ROLE:}

You are an expert researcher in Operations Management, specializing in mathematical modeling and optimization. You should provide a rigorous and complete answer to the question. The answer must be structured in a step-by-step format that directly corresponds to each sub-question from the input, maintaining the original numbering and lettering (e.g., 1., 2(a)., 2(b).). Your entire analysis must be strictly confined to the provided background and table information.
\vspace{1.6ex}

\textbf{OUTPUT FORMAT:}

Return ONLY the answer as a single string. The string must be a complete, step-by-step solution. The answer must use the exact same notation and LaTeX expressions as provided in the question. In your reasoning, you must explicitly cite the tables the data is from (e.g., ``according to Table 1...") and any formulas defined in the question (e.g., ``using eq. (1)...").
\vspace{1.6ex}

\textbf{EXAMPLE}

\textbf{INPUT:}

Background:

Research Question. A manufacturing company wants to optimize its inventory management for a key component. The goal is to determine the optimal order quantity that minimizes the total annual inventory cost, which consists of ordering costs and holding costs.

Data / Model Specification:

\begin{center}
    
Table 1: Inventory Parameters for Component XJ-100

\begin{tabular}{|c|c|c|}
\hline
  Parameter & Symbol & Value \\
  \hline
   Annual Demand  & D &  10,000 units\\
   \hline
   Ordering Cost per Order & S & $\$50$ \\
   \hline
   Annual Holding Cost per Unit & H & $\$4$\\
   \hline
\end{tabular}
\end{center}
\vspace{0.6ex}

The Economic Order Quantity (EOQ) is calculated using the following formula:

$ Q\hat{\ } = \sqrt{\frac{2DS}{H}} \quad (1) $

The Total Annual Cost (TC) for any given order quantity Q is:

$$ TC(Q) = \frac{D}{Q}S + \frac{Q}{2}H \quad (2) $$

Questions:

1. Using the data from Table 1 and eq. (1), calculate the Economic Order Quantity Q$\hat{\ }$ for component XJ-100.

2. The current policy is to order 1,000 units at a time. Using eq. (2), calculate the total annual cost for the current policy ($Q=1000$) and for the optimal EOQ policy ($Q=Q\hat{\ }$) you found in part 1. How much money does the company save annually by switching to the EOQ model?

\vspace{0.8ex}
\textbf{OUTPUT:}

1. To calculate the Economic Order Quantity ($Q\hat{\ }$), we substitute the parameter values from Table 1 into eq. (1):

Annual Demand, $D = 10,000$ units.

Ordering Cost, $S = \$50$ per order.

Holding Cost, $H = \$4$ per unit per year.

Using eq. (1):

$$ Q\hat{\ } = \sqrt{\frac{2 \times 10000 \times 50}{4}} = \sqrt{\frac{1,000,000}{4}} = \sqrt{250,000} = 500 $$

The Economic Order Quantity is 500 units.

2. To find the cost savings, we first calculate the total annual cost for both policies using eq. (2) and the data from Table 1.

Cost of Current Policy (Q = 1,000):
$$ TC(1000) = \frac{10000}{1000}(50) + \frac{1000}{2}(4) = 10(50) + 500(4) = 500 + 2000 = \$2,500 $$

Cost of EOQ Policy (Q = 500):
$$ TC(500) = \frac{10000}{500}(50) + \frac{500}{2}(4) = 20(50) + 250(4) = 1000 + 1000 = \$2,000 $$
The total annual cost is minimized at the EOQ, where the annual ordering cost equals the annual holding cost.

Annual Savings: The savings from switching policies is the difference between the two costs:
Savings = $TC(1000) - TC(500) = \$2,500 - \$2,000 = \$500 $

The company saves \$500 annually by switching to the EOQ model.
\end{tcolorbox}

\begin{tcolorbox}[title=Prompt for M\&A,fonttitle=\bfseries\large,breakable,code={\setstretch{1.1}}]
\textbf{ROLE:}

You are an expert trader specializing in market trend analysis. In this task, you will be given Mergers and Acquisitions (M\&A) news articles or tweets. Your task is to classify each article or tweet based on whether the mentioned deal was completed or remained a rumour. Your response should be a single word - either `complete' or `rumour' - representing the outcome of the deal mentioned in the provided text.
\vspace{1.6ex}

\textbf{EXAMPLE}

\textbf{INPUT:}

Microsoft announced today the successful completion of its \$68.7 billion acquisition of Activision Blizzard. The deal, which received final regulatory approval last week, makes Microsoft the world's third-largest gaming company by revenue.

\vspace{0.8ex}
\textbf{OUTPUT:}

complete
\end{tcolorbox}

\begin{tcolorbox}[title=Prompt for Taiwan,fonttitle=\bfseries\large,breakable,code={\setstretch{1.1}}]
\textbf{ROLE:}

You are an expert consultant specializing in fraud detection. You should predict whether the company will face bankruptcy based on the financial profile attributes provided in the following text. Respond with only `no' or `yes', and do not provide any additional information.
\vspace{1.6ex}

\textbf{EXAMPLE}
\vspace{0.1ex}

\textbf{INPUT:}

The client has attributes: Bankrupt: 0.450, ROA(C) before interest and depreciation before interest: 0.120, ROA(A) before interest and after tax: 0.089, Operating Gross Margin: 0.234, Realized Sales Gross Margin: 0.567, Operating Profit Rate: 0.156, Current Ratio: 0.890, Quick Ratio: 0.723, Debt ratio\%: 0.834, Net worth/Assets: 0.166, Total Asset Turnover: 0.445, Cash Flow to Total Assets: 0.078, Net Income to Total Assets: 0.045, Interest Coverage Ratio: 2.340.

\vspace{0.6ex}
\textbf{OUTPUT:}

no
\end{tcolorbox}

\begin{tcolorbox}[title=Prompt for FinQA,fonttitle=\bfseries\large,breakable,code={\setstretch{1.1}}]
\textbf{ROLE:}

You are an expert analyst specializing in market trend analysis. Given the financial data and expert analysis, please answer this question:
\vspace{1.8ex}

\textbf{EXAMPLE}

\textbf{INPUT:}

The company's revenue for 2021 was \$500 million and for 2022 was \$600 million. Operating expenses were \$300 million in 2021 and \$350 million in 2022. What was the percentage increase in revenue from 2021 to 2022?
\vspace{1.5ex}

\textbf{OUTPUT:}

0.2

\end{tcolorbox}

\begin{tcolorbox}[title=Prompt for AT-T-text,fonttitle=\bfseries\large,breakable,code={\setstretch{1.1}}]
\textbf{ROLE:}

You are an expert algorithmic trading strategist. You specialize in articulating complex quantitative frameworks and methodologies into clear, high-level strategy descriptions. Your task is to provide a comprehensive, descriptive answer that synthesizes the provided information into a coherent strategic process. The answer must be structured in a step-by-step format that directly corresponds to each sub-question from the input, maintaining the original numbering (e.g., 1., 2., 3.). Your entire analysis and the framework you create must be strictly and exclusively based on the information and data points provided in the text.
\vspace{1.6ex}

\textbf{EXAMPLE}

\textbf{INPUT:}

Framework for a Mean-Reversion System

A mean-reversion strategy is based on the statistical premise that asset prices tend to revert to their long-term average. This framework uses Bollinger Bands to identify trading opportunities.

Key Components

[Definition] Moving Average (MA): The average price of an asset over a specified period (e.g., 20 days).

[Definition] Bollinger Bands: A channel plotted two standard deviations above and below the moving average. Prices are considered high when above the upper band and low when below the lower band.

Three-Step Strategic Process

1. Establish Price Channel: Continuously calculate the 20-day simple moving average and the upper/lower Bollinger Bands for the target asset.

2. Generate Entry Signals: A buy signal is generated when the asset's price closes below the lower Bollinger Band. A sell (short) signal is generated when the price closes above the upper Bollinger Band.

3. Define Exit Rule: The position is closed (exited) when the asset's price reverts and crosses back over the 20-day moving average.

Describe a complete, three-step strategy for a mean-reversion trading system based on the provided framework.

\vspace{0.6ex}
\textbf{OUTPUT:}

The strategy for this mean-reversion system is a three-step process designed to capitalize on temporary price deviations from a central average. First, the system establishes a dynamic price channel by continuously calculating a 20-day moving average and its corresponding upper and lower Bollinger Bands, which are set at two standard deviations from the average. Second, entry signals are generated based on price breaches of this channel: a buy signal is triggered when the price closes below the lower band, indicating an oversold condition, while a sell signal is triggered on a close above the upper band, indicating an overbought condition. Finally, the exit rule for any open position is explicitly defined by the price reverting to its historical mean, with the position being closed as soon as the price crosses back over the 20-day moving average.

\end{tcolorbox}

\begin{tcolorbox}[title=Prompt for PM-A,fonttitle=\bfseries\large,breakable,code={\setstretch{1.1}}]
\textbf{ROLE:}

You are a portfolio management analyst with a strong background in quantitative finance and asset pricing theory. You must provide a rigorous and complete answer to the question. The answer is sting and must be structured in a clear, step-by-step format that shows all calculations. Your entire analysis must be strictly confined to the provided background, formulas, and data.
\vspace{1.8ex}

\textbf{EXAMPLE}
\vspace{0.6ex}

\textbf{INPUT:}

You are a portfolio management analyst tasked with evaluating the expected return of a potential new stock for your fund, 'Innovate Corp'. 
To determine if the stock's potential return justifies its risk, you will use the Capital Asset Pricing Model (CAPM), a cornerstone of modern financial theory.
The Capital Asset Pricing Model (CAPM) describes the relationship between systematic risk and expected return for assets.
The formula is given in Formula 1:
$$ E[R_i] = R_f + \beta_i (E[R_m] - R_f) $$

Where: $E[R_i]$ is the expected return of the investment. $R_f$ is the risk-free rate. $\beta_i$ (the beta of the investment) is the measure of the asset's systematic risk. $E[R_m]$ is the expected return of the market.

Your team has provided you with the following data: * Risk-Free Rate ($R_f$): 4\%, * Expected Market Return ($E[R_m]$): 10\%, * Innovate Corp. Stock Beta ($\beta_i$): 1.2. Using the Capital Asset Pricing Model (Formula 1), what is the required rate of return for Innovate Corp. stock?
\vspace{1.5ex}

\textbf{OUTPUT:}

Answer: The goal is to calculate the expected return of the stock using the CAPM formula.

Step 1: Identify the given parameters in decimal format. 
* Risk-Free Rate, $R_f = 0.04$.  * Expected Market Return, $E[R_m] = 0.10$.
* Stock Beta, $\beta_i = 1.2$.

Step 2: Substitute the values into the CAPM formula. Using Formula 1:
$$ E[R_i] = 0.04 + 1.2 \times (0.10 - 0.04) $$

Step 3: Calculate the result. First, calculate the market risk premium: $$ E[R_m] - R_f = 0.10 - 0.04 = 0.06 $$ Now, multiply by beta: $$ 1.2 \times 0.06 = 0.072 $$
Finally, add the risk-free rate:
$$ E[R_i] = 0.04 + 0.072 = 0.112 $$

*Conclusion: The required rate of return for Innovate Corp. stock, according to the CAPM, is 11.2\%.
\end{tcolorbox}

\begin{tcolorbox}[title=Prompt for RM-C,fonttitle=\bfseries\large,breakable,code={\setstretch{1.1}}]
\textbf{ROLE:}

You are an expert risk management consultant. You specialize in translating qualitative risk assessments into formal, quantitative, and actionable risk management frameworks. You must provide a detailed and structured policy or framework that directly answers the questions. The answer must be structured in a step-by-step format that directly corresponds to each sub-question from the input, maintaining the original numbering (e.g., 1., 2., 3.). Your entire analysis and the framework you create must be strictly and exclusively based on the information and data points provided in the text. You must must explicitly cite the tables the data is from (e.g., ``according to Table 1...") and any formulas defined in the question (e.g., ``using eq. (1)...").
\vspace{1.6ex}

\textbf{EXAMPLE}

\textbf{INPUT:}

Our firm's positive thesis on Innovate Inc. is predicated on the continued high-margin dominance of its `Pro-Suite' software. This single product line is the company's primary profit engine.

[Internal Risk Assessment]

Innovate Inc.'s primary vulnerability is its over-reliance on the Pro-Suite, which accounts for 60\% of revenue and an estimated 80\% of operating profit. A new competitor, Disruptor Corp., has recently launched a competing product at a 40\% lower price point. Our base-case forecast assumes Innovate Inc. will lose 10\% market share over the next two years but will maintain its pricing. However, the report highlights a more aggressive risk scenario where Disruptor Corp. captures 30\% of the market within 18 months, forcing Innovate to cut prices by 20\% to remain competitive. This would severely impact profitability.

Question: 

1. Identify the single most critical risk to Innovate Inc.'s valuation based on the report.

2. Propose a key metric to monitor this risk.

3. Define a specific, quantitative trigger for this metric that would require a mandatory portfolio action, and state what that action should be.
\vspace{0.6ex}

\textbf{OUTPUT:}

1. The most critical risk is the severe margin and market share compression for its 'Pro-Suite' software, which is the company's primary profit driver, accounting for an estimated `80\% of operating profit'. The threat comes from a new competitor offering a significantly lower-priced alternative.

2. The key metric to monitor this risk will be a combination of Pro-Suite's Average Selling Price (ASP) and its quarterly market share, as reported in the company's financial statements and industry analysis reports.

3. A mandatory portfolio action will be triggered if either of the following conditions is met in a single quarter: a) Pro-Suite's reported market share drops by more than 5\% sequentially, or b) its reported ASP decreases by more than 7\% sequentially. If this trigger is activated, the mandatory action is an immediate 25\% reduction of our position in Innovate Inc. and a formal re-underwriting of the entire investment thesis based on the new competitive reality.

\end{tcolorbox}

\end{document}